\documentclass[a4paper,11pt]{article}
\pdfoutput=1

\usepackage{cancel}
\usepackage{dsfont}
\usepackage[T1]{fontenc}
\usepackage{float}
\usepackage{siunitx}
\usepackage{jheppub}
\usepackage{amsmath}
\usepackage{mathtools}
\usepackage{graphicx}
\usepackage[dvipsnames]{xcolor}

\newcommand*{\xRed}[1]{{\color{BrickRed}#1}}

\title{A Surface Integrand for the Inverse KLT Kernel}

\author{Christoph Bartsch$^1$,}\emailAdd{christoph.bartsch@mff.cuni.cz}
\author{Karol Kampf$^1$,}\emailAdd{karol.kampf@mff.cuni.cz}
\author{Ji\v{r}\'i Novotn\'y$^1$,}\emailAdd{jiri.novotny@mff.cuni.cz}
\author{Jaroslav Trnka$^{2}$,}\emailAdd{trnka@ucdavis.edu}
\affiliation{$^1$Institute for Particle and Nuclear Physics, Charles University, Prague, Czech Republic}
\affiliation{$^2$Center for Quantum Mathematics and Physics (QMAP), University of California, Davis, California, USA}

\abstract{
We propose a loop-level generalization of the inverse string theory Kawai-Lewellen-Tye (KLT) kernel: the planar \textit{inverse KLT integrand}. The integrand is defined constructively via a novel Berends-Giele-like recursion that exposes the inverse KLT kernel as the simplest toy model of a ``stringy amplitude''. We show that, to all loop orders, the inverse KLT integrand is structurally equivalent to integrands in the cubic scalar tr$\phi^3$ theory. This simplicity is obscured in the conventional Feynman diagram approach, where the inverse KLT integrand receives contributions from an infinity of infinite towers of contact interactions.
The inverse KLT integrand is a rational function of stringified kinematic variables and is naturally defined on the kinematic surface proposed by Arkani-Hamed et al. \cite{Arkani-Hamed:2023lbd,Arkani-Hamed:2023mvg}. It provides an elementary analogue of the surfacehedron integrand for the tr$\phi^3$ theory involving only scalar resonances and unifies the scattering of cubic scalars and pions in the non-linear sigma model (NLSM) to all loop orders via kinematic $\alpha'$-shifts.
}

\begin{document}

\maketitle

\section{Introduction}
One of the challenges of theoretical physics is to identify toy models that capture some aspect of a physical system while stripping away much of its complexity. In a more realistic setting, certain properties of a given model may be obscured by the intricacy of the problem or by the prohibitive effort required to compute its consequences. Toy models allow to focus on the essence of the problem while keeping just enough structure to make progress. Time and again this approach has led to new insights and informed our understanding of problems that were previously thought intractable.

The aim of this article is to bring attention to one such toy model for the study of string-like scattering amplitudes, the \textit{inverse string theory Kawai-Lewellen-Tye (KLT) kernel},
\begin{align*}
m^{\alpha'}_n \!\equiv (S_n^{\alpha'})^{-1}.
\end{align*}
First studied thoroughly in \cite{Mizera:2016jhj,Mizera:2017cqs}, the stringy matrix elements $m_n^{\alpha'}$ of the inverse KLT kernel were recognized to be simple functions of the external kinematics at finite $\alpha'$. As such, they retain many of the properties of string amplitudes such as an infinite resonance structure or the monodromy relations, a characteristic feature of the worldsheet formulation of string theory.
They provide an interesting trade-off between the field theory limit where string-like features of amplitudes are often lost or incomplete and, on the other side of the complexity spectrum, bosonic string amplitudes, or $Z$-theory functions \cite{Broedel:2013tta,Mafra:2016mcc,Carrasco:2016ygv,Carrasco:2016ldy}, which are guaranteed to exhibit stringy behavior but often implement it in an intransparent way, since closed expressions for these amplitudes are not known beyond five points \cite{Arkani-Hamed:2024nzc}.

In fact, we will argue that the inverse KLT kernel might possibly be the \textit{simplest} toy model of stringy amplitudes, not just at tree-level but to all loop orders. We support this claim by formulating a new kind of recursion relation, called the \textit{cubic Berends-Giele recursion}, for the tree-level matrix elements $m_n^{\alpha'}$ that exposes an underlying equivalence of the inverse KLT kernel and amplitudes in the cubic bi-adjoint scalar (BAS) or tr$\phi^3$ field theory \cite{Mafra:2016ltu}. 
This structural similarity reveals a previously unknown simplicity of the inverse KLT kernel which, in its conventional formulation in terms of Feynman diagrams, is obscured by an infinite tower of interaction terms \cite{Mizera:2016jhj}. 

More importantly, by generalizing the cubic Berends-Giele recursion to loop integrands, we reveal the existence of a loop-level analogue of the inverse KLT kernel which we call the \textit{inverse KLT integrand}. This integrand represents a simple stringy analogue of the surface integrand for the $\text{tr}\phi^3$ theory recently studied in \cite{Arkani-Hamed:2023lbd,Arkani-Hamed:2023mvg} involving only scalar resonances. We will show how the inverse KLT integrand naturally implements the recently observed close connection between amplitudes in the $\text{tr}\phi^3$ theory and pions in the non-linear sigma model (NLSM) \cite{Kampf:2013vha} via kinematic $\alpha'$-shifts similar to those described in \cite{Bartsch:2025loa,Bartsch:2025mvy} for the tree-level inverse KLT kernel.

 \, \\
The paper is organized as follows.
In Section \ref{review} we provide a short review of the cubic bi-adjoint scalar theory and how its amplitudes can be efficiently computed using the well-known Berends-Giele recursion relations. We then introduce the inverse string theory KLT kernel and discuss some of its relevant features as well as its connection to stringy pion amplitudes via the kinematic $\alpha'$-shift. Section \ref{KLTtree} is then concerned with formulating the novel Berends-Giele recursion relations for the matrix elements $m_n^{\alpha'}$, exposing the cubic simplicity of the inverse string theory KLT kernel. In Section \ref{KLTloop} we use the cubic Berends-Giele recursion relation to define the \textit{inverse KLT integrand} $m_{L,n}^{\alpha'}$ at $L$ loops and $n$ points. We first focus on the one-loop case where we touch on the connection to the surface integrand \cite{Arkani-Hamed:2023lbd,Arkani-Hamed:2023mvg} defined for the $\text{tr}\phi^3$ field theory. The recursion is then extended to the general $L$ loop case. Finally, in Section \ref{piAlphaLoop} we briefly outline how the $\alpha'$-shift connects the inverse KLT integrand to a suitably stringified version of the ``perfect'' NLSM integrand for pions constructed in \cite{Arkani-Hamed:2024nhp,Arkani-Hamed:2024yvu}.

\section{Review}\label{review}

In this section we introduce some notational conventions used in this article. We also review the general structure of Berends-Giele recursion relations \cite{Berends:1987me} and their application to the tr$\phi^3$ theory \cite{Mafra:2016ltu}. Further, we provide a short introduction to the inverse string theory KLT kernel \cite{Mizera:2016jhj} and discuss its connection to pion amplitudes in the NLSM \cite{Bartsch:2025loa}.

\subsection*{Notation}
Throughout this article we will only consider ordered amplitudes $A_n(p_1\,p_{\sigma(2)}p_{\sigma(3)}\dots p_{\sigma(n)})$ where $p_i$ are on-shell, external momenta $p_i^2\,{=}\,0$ and $\sigma \,{\in}\, S_{n-1}$ denotes some ordering of particle labels $\lbrace 2\dots n\rbrace$. In particular, we will be studying amplitudes of scalar particles exclusively, which are naturally functions just of momentum invariants $p_i\cdotp p_j$. A convenient basis for these invariants at $n$ points (disregarding Gram determinant relations) is given by the planar Mandelstam variables
\begin{align}\label{defX}
    X_{ij} = (p_i+p_{i+1}+\dots + p_{j-1})^2, \hspace{0.5cm} \text{for } 1\le i < j \le n,
\end{align}
satisfying $X_{ij}=X_{ji}$ and on-shell conditions $X_{i,i+1}=0$. All indices are treated as cyclic, identifying labels $n+1\equiv 1$.
The advantage of using a Mandelstam basis is that a given scalar amplitude $A_n$ has a unique representation in terms of the variables \eqref{defX},
\begin{align}\label{defAX}
    A_n\equiv A_n(X_{ij}) \equiv A_n(X).
\end{align}

Given this unique $X$-\textit{representation}, it will be convenient for us to adopt a notation where we consider the amplitude as a formal function of the \textit{$X$-labels} appearing as subscripts in the $X_{ij}$ variables. That is, for some abstract set of labels $\lbrace a_1\dots a_n \rbrace$ we define the $X$-\textit{label representation} of an amplitude by appropriately plugging the labels $a_i$ into the $X$-representation \eqref{defAX} of the amplitude,
\begin{align}\label{defXlabNot}
A_n(a_1a_2\dots a_n)\equiv A_n(X_{a_i\,a_j}).
\end{align}
To give an example, we can define a function of four labels
\begin{align}
    m_4(a_1a_2a_3a_4) = \frac{1}{X_{a_1a_3}} +\frac{1}{X_{a_2a_4}}.
\end{align}
Evaluated on the specific set of labels $\lbrace a_1a_2a_3a_4\rbrace \,{=}\, \lbrace 1234\rbrace$ this function yields the four-point bi-adjoint scalar amplitude (cf. \eqref{BASex}),
\begin{align}
    m_4(1234) = \frac{1}{X_{13}}+\frac{1}{X_{24}} \equiv \frac{1}{(p_1+p_2)^2} + \frac{1}{(p_2+p_3)^2}\,,
\end{align}
once the abstract $X$-variables are identified with on-shell momenta according to \eqref{defX}.
Note that in principle we allow $X$-labels to take any values such that e.g. $m_4(*19\beta)=X_{*,9}^{-1} + X_{1,\beta}^{-1}$ makes sense as a formal function of $X$-labels but not necessarily as an amplitude.

The main reason to use the $X$-label representations \eqref{defAX} and \eqref{defXlabNot} rather than the usual on-shell momentum representation is that it allows to canonically ``off-shellize'' an amplitude. Starting with an amplitude $A(X_{ij})$ as a function of $X$-variables (on-shell amplitudes depend only on $X_{ij}$ with $j\ge i+2$) we may parameterize an off-shell variant of it by adding terms proportional to $X_{i,i+1}\equiv p_i^2$,
\begin{align}\label{paramOffShell}
    \tilde{A}(X_{ij},X_{i,i+1}) = A(X_{ij}) + \sum_{k=1}^n X_{k,k+1}\, F_k(X_{ij},X_{i,i+1}),
\end{align}
Assuming that the functions $F_k$ do not blow up as any $X_{i,i+1}\to 0$, this allows to canonically lift an on-shell amplitude to an off-shell function. Concretely, we demand that the off-shell function, like the on-shell amplitude, should remain independent of the off-shellness variables $X_{i,i+1}$. This can be achieved trivially by setting all $F_k\equiv 0$ in \eqref{paramOffShell} and defining
\begin{align}\label{canonOffShell}
    \tilde{A}(X_{ij},X_{i,i+1}) = \tilde{A}(X_{ij}) \equiv A(X_{ij}).
\end{align}
In other words, we identify the on-shell amplitude and its canonical ``off-shellization'' as functions of $X_{ij}$. While this construction is certainly not unique, it will turn out to be the right choice for the purposes of defining recursion relations for the bi-adjoint scalar theory as well as the inverse KLT kernel discussed subsequently. Because of \eqref{canonOffShell} we will usually not distinguish between the amplitude and its canonical off-shell function and denote both as $A(X_{ij})$ or just $A(X)$.

\subsection{Bi-Adjoint Scalar Theory (BAS)}
Since we will refer to it throughout this article, we give a very brief introduction to the so-called bi-adjoint scalar theory (BAS). Conventionally, it can be defined in terms of the Lagrangian
\begin{align}\label{defBASLag}
    \mathcal{L}_{\text{BAS}} = \frac{1}{2}\partial\phi^{a\bar{a}}\cdot\partial\phi^{a\bar{a}} - \frac{1}{3!} f_{abc}f_{\bar{a}\bar{b}\bar{c}}\phi^{a\bar{a}}\phi^{b\bar{b}}\phi^{c\bar{c}},
\end{align}
where $\phi^{a\bar{a}}$ is a doubly-charged scalar under two non-abelian groups, which, for concreteness, we will take to be $U(N)\times U(\bar{N})$. Amplitudes in this theory can be color-ordered with respect to both groups,
\begin{align}\label{defBASstripped}
    m_n^{a_1\bar{a}_1 \dots a_n\bar{a}_n} =\!\!\!\!\! \sum_{\sigma,\rho \in S_{n-1}}\!\!\!\! \text{tr}(t^{a_1}t^{a_{\sigma(2)}}\dots t^{a_{\sigma(n)}})\text{tr}(t^{\bar{a}_1}t^{\bar{a}_{\rho(2)}}\dots t^{\bar{a}_{\rho(n)}})\, m_n(1 \sigma| 1\rho),
\end{align}
where $t^a,t^{\bar{a}}$ denote the generators of $U(N)$ and $U(\bar{N})$ respectively. The doubly stripped amplitudes $m_n(1 \sigma| 1\rho)$ can be computed from the Lagrangian \eqref{defBASLag} by summing over all trivalent Feynman graphs which are compatible with both orderings $\lbrace{1\sigma \rbrace}$ and $\lbrace 1\rho\rbrace$ as described in \cite{Cachazo:2013iea}. For example, at $n=3$ we have $m_3(\mathds{1}|\mathds{1}) = -m_3(\mathds{1}|132) = 1$ while for $n=4,5$ we can compute
\begin{align}\label{BASex}
\begin{split}
    m_4(\mathds{1}|\mathds{1}) &= \frac{1}{X_{13}} \!+\! \frac{1}{X_{24}}, \hspace{0.3cm} m_4(\mathds{1}|1243) = -\frac{1}{X_{13}}, \hspace{0.3cm} m_4(\mathds{1}|1324) = -\frac{1}{X_{24}}, \\
    m_5(\mathds{1}|\mathds{1}) &= \frac{1}{X_{13}X_{14}} + \text{cyc.}, \hspace{0.3cm} m_5(\mathds{1}|13245) = \frac{1}{X_{14}}\left(\! \frac{1}{X_{13}} \!+\! \frac{1}{X_{24}}\!\right), 
\end{split}
\end{align}
where $\mathds{1}=\lbrace 1\dots n \rbrace$ always denotes the identity permutation of $n$ labels and the momentum invariants $X_{ij}$ are as defined in \eqref{defX}. In the following we will refer to matrix elements $m_n\!\left(\mathds{1}|\mathds{1}\right)$ as \textit{diagonal} whereas \textit{off-diagonal} matrix elements are those $m_n\!\left(\mathds{1}|\rho\right)$ with $\rho\neq\mathds{1}$.

However, throughout this article we will exclusively study the properties of diagonal matrix elements $m_n \,{\equiv}\, m_n(\mathds{1}|\mathds{1})$. Since both orderings $\sigma\,{=}\,\rho$ agree for diagonal matrix elements, they can be equivalently computed as amplitudes of a singly charged scalar in the so-called $\text{tr}\phi^3$ theory,
\begin{align}\label{defLagTrPhi3}
    \mathcal{L}_{\text{tr}\phi^3} = \frac{1}{2} \langle \partial \phi \cdot \partial \phi \rangle - \frac{1}{3!}\langle \phi^3 \rangle,
\end{align}
where $\langle \dots \rangle \equiv \text{tr}(\dots)$ denotes the fundamental trace over the fields $\phi=\phi^{a}t^{a}$. For this reason we will refer to the matrix elements $m_n$ as diagonal BAS or $\text{tr}\phi^3$ amplitudes interchangeably.

\subsection{Berends-Giele Recursion for diagonal BAS amplitudes}
While diagonal BAS amplitudes $m_n$ can be calculated directly from the Lagrangian \eqref{defBASLag} as sums over cubic Feynman graphs, more efficient methods have been devised to facilitate their computation. Of particular relevance for the present article are the Berends-Giele recursion relations derived in \cite{Mafra:2016ltu}. In this section we will recast some of the results obtained there into the $X$-label representation \eqref{defXlabNot} in terms of which the recursion for the matrix elements $m_n$ takes a particularly simple form. This will also allow us to introduce \textit{Berends-Giele diagrams} as useful notation to facilitate later generalizations to stringy matrix elements and loop integrands.

\subsubsection*{The Basics}
Before getting into the specifics of amplitudes in the BAS theory, let us briefly remark on the general idea behind the Berends-Giele approach \cite{Berends:1987me}. At the most basic level, the Berends-Giele recursion can be understood as an efficient regrouping, or partitioning, of Feynman diagrams into recursively definable building blocks, called \textit{Berends-Giele currents}. 

\begin{figure}[t]
    \centering
    \includegraphics[width=12.0cm]{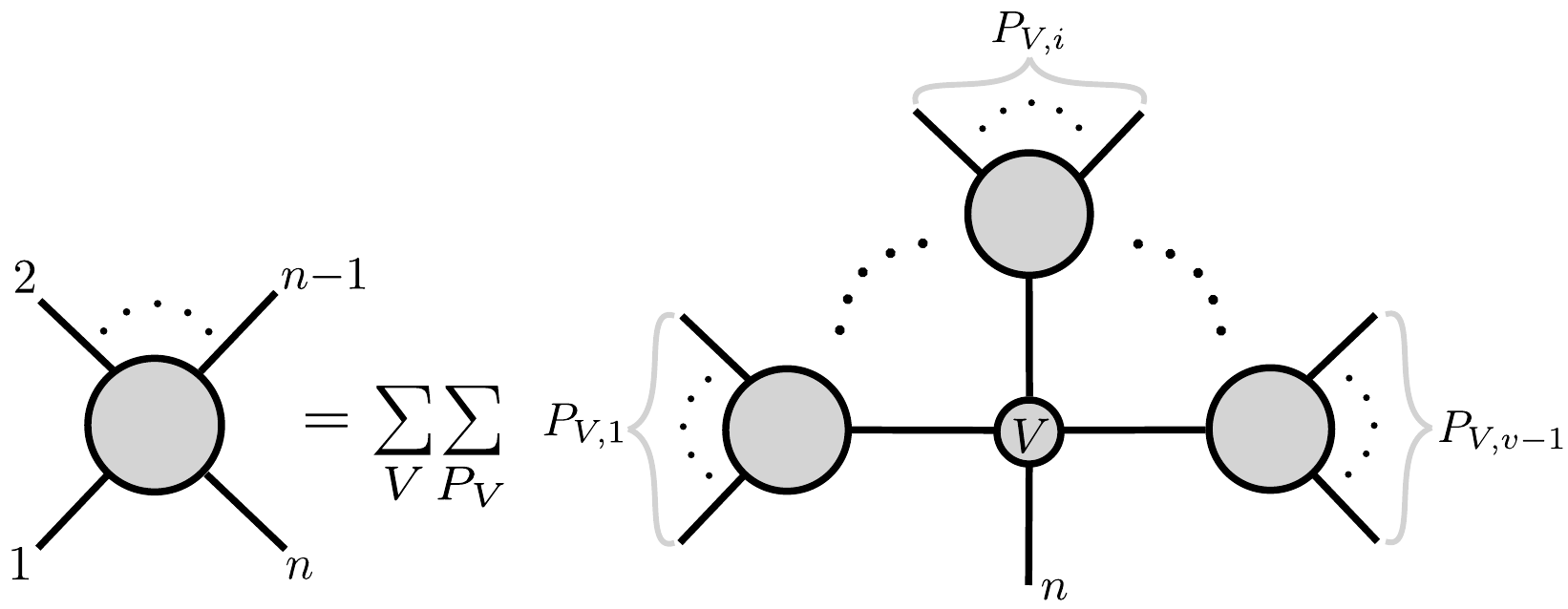}
    \caption{A generic Berends-Giele diagram is specified by a root vertex $V$ (valency $v$) and a compatible partition $P_V$ of external legs. The grey blobs can be identified with lower-point semi-on-shell amplitudes or Berends-Giele currents. The $n$-point amplitude is obtained recursively by summing over all Berends-Giele diagrams.}
    \label{fig:bgdiaggengen}
\end{figure}

This partitioning is accomplished as follows. First, a particular (say the $n$-th) external line is fixed in each Feynman diagram, designating the so-called \textit{root leg}. We then follow this line into the diagram until it first meets a vertex, which we will correspondingly call the \textit{root vertex} $V$. Supposing the vertex has valency $v$ then the remaining external lines can be partitioned into $v{-}1$ non-overlapping sets $\lbrace 1\dots n{-}1\rbrace = \lbrace P_{V,1}|P_{V,2}|\dots|P_{V,v-1} \rbrace$ depending on where they connect to the root vertex. Generically there will be multiple partitions $P_V$ compatible with a given root vertex $V$. The sum of all Feynman diagrams with a particular root vertex $V$ and a fixed partition $P_V$ of external momenta can then be associated with a \textit{Berends-Giele diagram} as shown on the right-hand side of Figure \ref{fig:bgdiaggengen}.

Since a given Berends-Giele diagram represents the sum over all Feynman diagrams compatible with a partition $P_V$, each gray blob in Figure \ref{fig:bgdiaggengen} can essentially be identified with a $(|P_{V,i}|{+}1)$-point amplitude where the line connecting to the root vertex is off-shell. These are the Berends-Giele currents in \cite{Mafra:2016ltu} for which the recursion is conventionally defined.

The full amplitude is obtained by summing over all vertices $V$ in the theory (as, for example, specified by a given Lagrangian) and compatible partitions $P_V$ of the external lines. This amounts to a sum over all Berends-Giele diagrams (again, see Figure \ref{fig:bgdiaggengen}).

Since, throughout this article, we are only considering scalar functions, we will always identify Berends-Giele currents (i.e. amplitudes with one or more legs off-shell) with the on-shell amplitudes themselves via the canonical \textit{off-shellization} \eqref{canonOffShell} of their $X$-label representation. Omitting this (in our case trivial) distinction allows us to talk about recursion relations for amplitudes and matrix elements directly without further complicating the presentation with jargon.

\subsubsection*{Berends-Giele for $m_n(\mathds{1}|\mathds{1})$}
Let us now specify the discussion to diagonal bi-adjoint scalar (or tr$\phi^3$) amplitudes $m_n\equiv m_n(\mathds{1}|\mathds{1})$ with canonical ordering $\mathds{1}{=}\lbrace 1\dots n\rbrace$ and show how this dramatically simplifies the generic form of the Berends-Giele recursion in Figure \ref{fig:bgdiaggengen}.

Since the BAS theory is cubic, there is only a single trivalent root vertex from the Lagrangian \eqref{defBASLag} entering the recursion which is given by
\begin{align}\label{rootVBAS}
    \begin{matrix}
         \hspace{-0.0cm}\vspace{-0.25cm}\includegraphics[width=2.5cm]{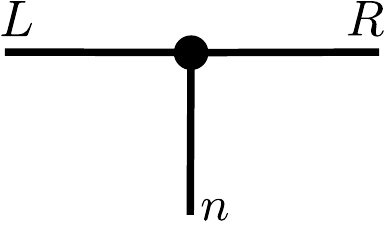}
    \end{matrix}
    \,=\, 1.
\end{align}
Consequently, any partition $P$ of external lines compatible with the root vertex \eqref{rootVBAS} must be into two sets $\lbrace 1\dots n{-}1 \rbrace = \lbrace P_L|P_R\rbrace$. Moreover, since we are considering an ordered amplitude, each set $P_L,P_R$ must itself be ordered. This only leaves a total of $n{-}2$ possible partitions of the form $\lbrace P_{j,L}|P_{j,R}\rbrace = \lbrace 1\dots j{-}1|j\dots n{-}1\rbrace$ where $j=2,\dots,n{-}1$. Summing over the corresponding Berends-Giele diagrams we obtain the $n$-point diagonal matrix element,
\begin{align}\label{BASBGdiag}
\begin{split}
    m_n^{\alpha'} &=
    \begin{matrix}
         \hspace{-0.0cm}\vspace{-0.15cm}\includegraphics[width=3.8cm]{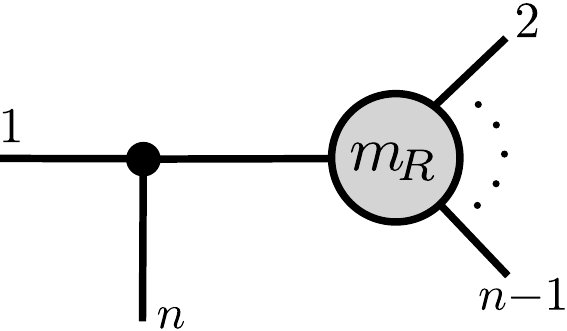}
    \end{matrix}
    +\hspace{-0.2cm}
    \begin{matrix}
         \hspace{-0.0cm}\vspace{-0.15cm}\includegraphics[width=4.0cm]{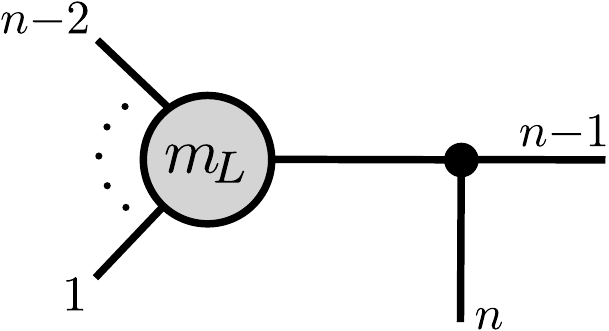}
    \end{matrix}
    \\
    &\hspace{0.45cm} + \sum_{j=3}^{n-2} \hspace{0.1cm}
    \begin{matrix}
         \hspace{-0.0cm}\vspace{-0.15cm}\includegraphics[width=5.7cm]{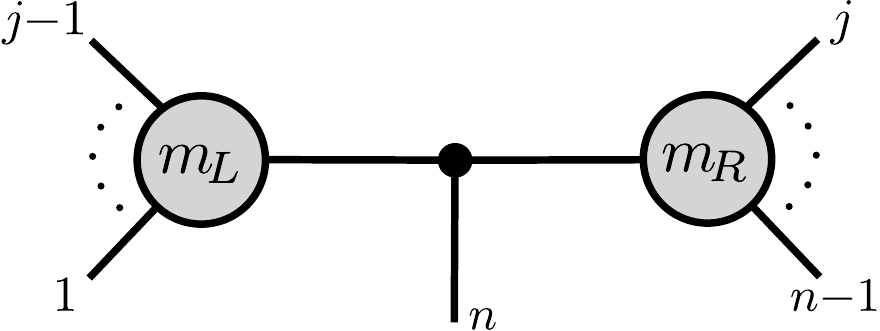}
    \end{matrix}\,\,.
\end{split}
\end{align}
Since the theory is cubic, there are only up to two sub-amplitudes $m_{L/R}$ appearing in each Berends-Giele diagram. This fact makes the Berends-Giele recursion highly efficient, as in each recursive step only $n{-}2$ terms need to be summed to obtain $m_n$.

We can easily read off the algebraic expression for the recursion in $X$-label representation from the Berends-Giele diagrams \eqref{BASBGdiag},
\begin{align}\label{BGBAS}
\begin{split}
    m_n(1\dots n) &= \frac{1}{X_{2n}}\, m_{n-1}(2\dots n) + \frac{1}{X_{1n-1}}\, m_{n-1}(1\dots n{-}1)\\ &+  \sum_{j=3}^{n-2} \frac{1}{X_{1j}X_{jn}}\, m_j(1\dots j)m_{n-j+1}(j\dots n).
\end{split}
\end{align}
Note that for the first two contributions there is only one explicit propagator as either the left- or right-hand momentum corresponds to an on-shell external line $X_{L/R} \,{=}\, 0$. Furthermore, the recursion generates matrix elements $m_n$ in a form that is manifestly independent of the off-shellness variables $X_{i,i+1}$, meaning they are already canonically off-shellized according to \eqref{canonOffShell}. We can therefore directly use matrix elements obtained through \eqref{BGBAS} as input for the next iteration of the recursion without having to distinguish between on-shell amplitudes and (semi-on-shell) Berends-Giele currents.

The Berends-Giele recursion \eqref{BGBAS} computes amplitudes simply as sums of products involving lower point amplitudes. This makes the recursion particularly amenable for implementation on a computer since there are no non-trivial kinematic shifts to keep track of as is the case for a BCFW recursion. Adding to its efficiency, the Berends-Giele recursion generates amplitudes without introducing spurious poles. In fact, no cancellations between different Berends-Giele diagrams occur.

\subsubsection*{A short example ($n\,{=}\,5$)}
To get comfortable with the notation, let us illustrate the recursion \eqref{BASBGdiag} and compute the diagonal matrix element $m_5$. At $n=5$ only three Berends-Giele diagrams contribute,
\begin{align}\label{BASBGex5diag}
    \hspace{-0.6cm} m_5 = \begin{matrix}
         \hspace{-0.0cm}\vspace{-0.3cm}\includegraphics[width=3.2cm]{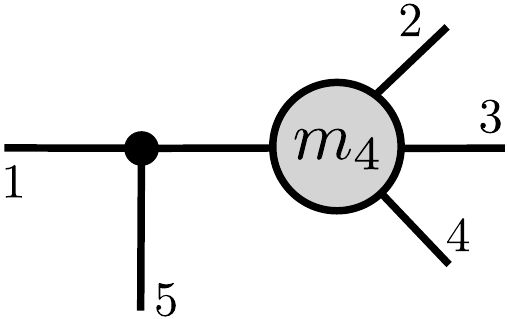}
    \end{matrix}
    \,\,+\,\,
    \begin{matrix}
         \hspace{-0.0cm}\vspace{-0.42cm}\includegraphics[width=3.2cm]{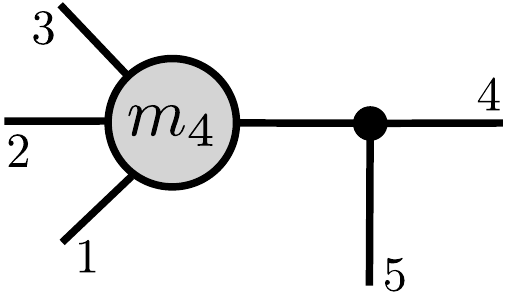}
    \end{matrix}
    \,\,+
    \begin{matrix}
         \hspace{-0.0cm}\vspace{-0.3cm}\includegraphics[width=4.2cm]{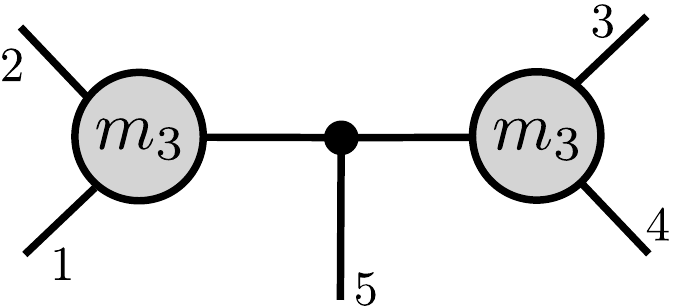}
    \end{matrix}\,,
\end{align}
corresponding to partitions $P= \lbrace 1 | 234\rbrace, \lbrace12|34\rbrace,\lbrace 123| 4\rbrace$. From the diagrams it is straightforward to read off the expressions,
\begin{align}\label{BASBGex5expr}
\begin{split}
    m_5(\mathds{1}) &= \frac{1}{X_{25}}m_4(2345) + \frac{1}{X_{14}}m_4(1234) + \frac{1}{X_{13}X_{35}}m_3(123)m_3(345)\\
    &= \frac{1}{X_{25}}\!\left(\!\frac{1}{X_{24}} + \frac{1}{X_{35}}\!\right) + \frac{1}{X_{14}}\!\left(\!\frac{1}{X_{13}} + \frac{1}{X_{24}}\! \right) + \frac{1}{X_{13}X_{35}},
\end{split}
\end{align}
and correctly recover the five-point matrix element in \eqref{BASex}.

\subsection{The inverse String Theory KLT Kernel}
In this section we briefly review some of the properties of the inverse string theory Kawai-Lewellen-Tye (KLT) kernel $m_n^{\alpha'}\!(\sigma|\rho)$ first studied in \cite{Mizera:2016jhj,Mizera:2017cqs}. Our discussion here will largely follow the initial work, and is mainly intended to introduce some of the notations and conventions used throughout this article. 

The string theory KLT kernel $S_n^{\alpha'}$ was first introduced in a seminal paper by Kawai, Lewellen, and Tye \cite{Kawai:1985xq} who observed that closed string amplitudes $M^{\text{cl}}$ could be written as a ``square'' of ordered open string amplitudes $A^{\text{o}}(\sigma)$ when appropriately sewn together by a scalar function $S^{\alpha'}_n\!(\sigma | \rho)$,
\begin{align}
    M_n^{\text{cl}} = \!\!\! \sum_{\sigma, \rho\in S_{n-3}}\!\!\! A_n^{\text{o}}(1\, \sigma \,n{-}1\,n)\,S^{\alpha'}_n\!(\sigma | \rho)\, A_n^{\text{o}}(1\, \rho\, n\,n{-}1),
\end{align}
where the sum runs over all orderings $\sigma,\rho$ of $n{-}3$ particle labels $\lbrace2\dots n{-}2\rbrace$. Importantly, the KLT kernel $S^{\alpha'}_n(\sigma | \rho)$ admits a simple closed-form expression \cite{Bjerrum-Bohr:2010pnr} involving trigonometric functions of the external kinematics.

In \cite{Mizera:2016jhj,Mizera:2017cqs} Mizera initiated the study of the matrix \textit{inverse} of the KLT kernel,
\begin{align}\label{defInvKLT}
m^{\alpha'}_n \!\equiv (S_n^{\alpha'})^{-1},
\end{align}
and showed that its matrix elements $m^{\alpha'}\!\!\left( \sigma | \rho \right)$ can similarly be written as simple rational functions of the stringy planar Mandelstam variables,
\begin{align}\label{tsVar}
    t_{ij} = \tan(\pi\alpha' X_{ij}), \hspace{0.5cm} s_{ij} = \sin(\pi\alpha' X_{ij}).
\end{align}
These variables inherit all the properties of the field theory $X$-variables \eqref{defX}, i.e. $t_{ij}=t_{ji}$, $t_{i,i+1}=0$ and similar for $s_{ij}$. Expanding for small $\alpha'$ we have $t_{ij}\simeq s_{ij} = \pi\alpha' X_{ij} + \mathcal{O}(\alpha'^3)$.

Some of the simplest examples for the matrix elements of the inverse string theory KLT kernel include the three-point functions $m_3^{\alpha'}\left(\mathds{1}|\mathds{1}\right) = -m_3^{\alpha'}\left(\mathds{1}|132\right) = 1$ as well as
\begin{align}
\begin{split}\label{invKLTex}
     &m_4^{\alpha'}\!\!\left(\mathds{1}|\mathds{1}\right) = \frac{1}{t_{13}}+\frac{1}{t_{24}}, \hspace{0.5cm} m_4^{\alpha'}\!\!\left(\mathds{1}|1243\right) = -\frac{1}{s_{13}}, \hspace{0.5cm} m_4^{\alpha'}\!\!\left(\mathds{1}|1324\right) = -\frac{1}{s_{24}},
    \\
    &m_5^{\alpha'}\!\!\left(\mathds{1}|\mathds{1}\right) = \left(\!\frac{1}{t_{13}t_{14}} \!+ \text{cyc.}\!\right) \!+\!1, \hspace{1.1cm}
    m_5^{\alpha'}\!\!\left(\mathds{1}|13245\right) =\frac{1}{s_{14}}\!\left(\!\frac{1}{t_{13}} \!+\! \frac{1}{t_{24}}\!\right)\!,
    \\
    &m_6^{\alpha'}\!\!\left(\mathds{1}|\mathds{1}\right) = \frac{1}{t_{13}t_{14}t_{15}} \!+\! \frac{1}{2}\frac{1}{t_{13}t_{14}t_{46}} \!+\! \frac{1}{2}\frac{1}{t_{13}t_{36}t_{46}} \!+\! \frac{1}{3}\frac{1}{t_{13}t_{15}t_{35}} \!+\! \frac{1}{t_{13}} \!+ \text{cyc.} ,
\end{split}
\end{align}
at four, five and six points. From \eqref{invKLTex} we see that diagonal matrix elements $m_n^{\alpha'}\!\left(\mathds{1}|\mathds{1}\right)$ depend on $X_{ij}$ only through tangents $t_{ij}$, whereas off-diagonal elements $m_n^{\alpha'}\!\left(\mathds{1}|\rho\right)$, $\rho\neq\mathds{1}$ generically involve $t_{ij}$ and $s_{ij}$. Again, only diagonal matrix elements $m_n^{\alpha'}\equiv m_n^{\alpha'}\!(\mathds{1}|\mathds{1})$ will be considered subsequently.

Diagonal matrix elements $m_n^{\alpha'}$ can be systematically computed using diagrammatic rules \cite{Mizera:2016jhj}. In particular, they can be obtained as sums over all tree-level Feynman diagrams built from stringy propagators and vertices
\begin{align}\label{KLTFeynRules}
    \begin{matrix}
        \hspace{-0.0cm}\vspace{-0.45cm}\includegraphics[width=1.0cm]{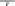}
    \end{matrix} \,=\, \frac{1}{\tan(\pi\alpha' p^2)}, \hspace{1.5cm} \begin{matrix}
        \hspace{-0.0cm}\vspace{-0.1cm}\includegraphics[width=2.0cm]{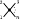}
    \end{matrix} \hspace{-0.7cm}\equiv \kappa_n = 
    \begin{cases}
        C_{\frac{n-3}{2}}   \hspace{0.8cm} \text{ if }\, n \ge 3 \text{ odd}, \\
        0                   \hspace{1.5cm} \text{ if }\, n \text{ even},
    \end{cases}
\end{align}
with $C_n$ denoting the $n$-th Catalan number. It can be shown that these Feynman rules follow directly from the inverse KLT interaction Lagrangian \cite{Mizera:2017cqs}
\begin{align}\label{defKLTLag}
    \mathcal{L}_{\text{KLT}} = \frac{1}{2} \Big\langle\! \phi - \frac{1}{4}\arcsin(2\phi) - \frac{1}{2}\phi \sqrt{1-4\phi^2}  \Big\rangle,
\end{align}
when expanded into an infinite series in powers of $\phi$. As in the case of the Lagrangian \eqref{defLagTrPhi3} for the $\text{tr}\phi^3$ theory, $\langle \dots \rangle = \text{tr}(\dots)$ denotes the fundamental trace of the fields $\phi^{i}_{\, j}=\phi^{a}(t^a)^{i}_{\, j}$.

\subsubsection*{Some properties of $m_n^{\alpha'}$}
Using the Feynman rules \eqref{KLTFeynRules}, the lowest order diagonal matrix elements (up to $n=6$) in \eqref{invKLTex} can be computed from the diagram topologies
\begin{align}
\label{KLTfeynRules}
    m_3^{\alpha'} &= \begin{matrix}
        \hspace{-0.0cm}\vspace{-0.1cm}\includegraphics[width=0.9cm]{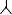}
    \end{matrix}\, , \hspace{1.0cm}
    m_4^{\alpha'} = \begin{matrix}
        \hspace{-0.0cm}\vspace{-0.1cm}\includegraphics[width=1.6cm]{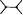}
    \end{matrix}\, , \hspace{1.0cm}
    m_5^{\alpha'} = \begin{matrix}
        \hspace{-0.0cm}\vspace{-0.1cm}\includegraphics[width=3.2cm]{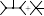}
    \end{matrix}\, ,\\
    \vspace{2.5cm} m_6^{\alpha'} &= \begin{matrix}
        \hspace{-0.0cm}\vspace{0.2cm}\includegraphics[width=13.0cm]{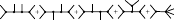}
    \end{matrix} \, .\nonumber
\end{align}
A distinctive feature of the stringy inverse KLT kernel, when compared to amplitudes in the BAS theory \eqref{defBASLag}, is the infinite tower of odd-point interaction vertices. Nevertheless, the two theories share the same lowest valency interaction, the three-point contact term $\sim C_0 \,{=}\,1$. As such, expressions for purely trivalent diagrams in \eqref{KLTfeynRules} are equivalent to those in the BAS theory except that the corresponding propagators have been stringified $1/X_{ij}\to 1/t_{ij}$ according to \eqref{KLTFeynRules}. In the low-energy limit $\alpha' \,{\to}\, 0$ each such propagator contributes a factor $1/\alpha'$ to the overall scaling of the diagram.

It is now easy to see that the additional diagrams in \eqref{KLTfeynRules}, involving vertices of valency $\ge 5$, only contribute at sub-leading orders in $\alpha'$ when compared to purely cubic diagrams. This is because at fixed $n$ trivalent graphs allow for the largest number of propagators ($n\,{-}\,3$). This ensures that at the leading order in the limit $\alpha' \,{\to}\, 0$ only cubic graphs contribute and we recover amplitudes of the tr$\phi^3$ theory,
\begin{align} \label{alpha0limBAS}
    m_n^{\alpha'} = (\pi\alpha')^{3-n}\big(m_n + \mathcal{O}(\alpha')\big).
\end{align}

Conversely, taking a bottom-up perspective, we can consider the inverse KLT kernel as a \textit{stringification} of BAS amplitudes. Due to propagators being of the form $t_{ij}^{-1}$, diagonal matrix elements $m_n^{\alpha'}\!(\mathds{1} | \mathds{1})$ exhibit a periodic pole structure with simple poles when some
\begin{align}\label{periodPoles}
    X_{ij} = k/\alpha', \hspace{0.5cm} k\in\mathds{Z},
\end{align}
on all of which they consistently factorize into a product of two lower-point matrix elements, 
\begin{align}\label{mfact}
    \mathop{\mathrm{Res}}_{X_{ij}=k/\alpha'} m_{n}^{\alpha'}\!(t_{ij}) = m_L^{\alpha'}\!(t_L)\,m_R^{\alpha'}\!(t_R).
\end{align}
From \eqref{periodPoles} and \eqref{mfact} we gather that the inverse KLT kernel exhibits an infinite spectrum of resonances corresponding to states of positive \textit{and negative} mass squared $m^2 = 0, \pm \frac{1}{\alpha'}, \pm\frac{2}{\alpha'},...$, all of which correspond to scalar excitations. This represents quite an unconventional stringification of BAS amplitudes, especially when compared to $Z$-theory \cite{Broedel:2013tta,Mafra:2016mcc,Carrasco:2016ygv,Carrasco:2016ldy} which possesses a more physical spectrum. Nevertheless the inverse KLT matrix elements exhibit many stringy properties, such as the monodromy relations,
\begin{align}\label{invKLTmono}
    m_n^{\alpha'}\! (\mathds{1}|\mathds{1}) + \sum_{k=2}^{n-1}e^{ix_k}\, m_n^{\alpha'}\!(2\dots k\,1\,k{+}1\dots n|\mathds{1}) = 0,
\end{align}
with phases given by $x_k\,{=}\,\pi\alpha'2p_1\!\cdot\!(p_2+\dots+p_k)$, and the associated (stringy) \textit{hidden zeros} studied extensively in the recent literature \cite{DAdda:1971wcy,Arkani-Hamed:2023swr,Li:2024qfp,Bartsch:2024amu,Cao:2024gln}.

Another distinct advantage of the inverse KLT kernel is that it can be efficiently computed in closed form. In fact, we will show that there is a concrete sense in which the matrix elements $m_n^{\alpha'}$ can be considered the simplest stringification of the tr$\phi^3$ theory. The situation is more complicated for other known stringifications of bi-adjoint scalars such as the (non-)abelian $Z$-functions for which explicit expressions are not known beyond $n\,{=}\, 5$. The inverse KLT kernel can therefore serve as a particularly simple toy model for the study of stringy amplitudes, not just for bi-adjoint scalars but also for NLSM pions as we will now show. 

\subsection{Pions and the $\alpha'$-shift}
Recently it was shown \cite{Bartsch:2025loa} that the inverse KLT kernel also directly encodes stringified pion amplitudes in the non-linear sigma model (NLSM) as well as stringy mixed amplitudes involving both bi-adjoint scalars $\phi$ and pions $\pi$.

Many amplitudes involving pions were shown to be \textit{equivalent} to the diagonal inverse KLT matrix elements $m_n^{\alpha'}$ via so-called $\alpha'$-shifts. The latter amount to simple $\alpha'$-dependent deformations of the kinematic $X$-variables. In the following, we will quickly review the $\alpha'$-shifts for the specific case of pure pion amplitudes.

To define the $\alpha'$-shift for pure pion amplitudes it will be necessary to make a distinction between even- and odd-particle channels. Concretely, for the channel corresponding to $X_{ij}$ we define its \textit{parity} as
\begin{align}\label{defChPar}
    P(X_{ij}) = 
    \begin{cases} 
        \text{even,} \hspace{0.5cm} \text{ if }   \hspace{0.2cm} p(i)=p(j), \\
        \text{odd,} \hspace{0.6cm} \text{ if }    \hspace{0.2cm} p(i)\neq p(j),
    \end{cases}
\end{align}
where $p(n)=\text{even/odd}$ refers to the even- or oddness (i.e. parity) of the integer $n$. The definition \eqref{defChPar} is such that it assigns even/odd parity to even/odd-particle channels. We now split up the set of $X$-variables according to their parity, defining
\begin{align}\label{defXoe}
    X_{\text{even/odd}} = \lbrace X_{ij},\,  P(X_{ij})  = \text{even/odd}\rbrace.
\end{align}
Of course the set of all $X\,{=}\,X_{\text{even}}{\cup} X_{\text{odd}}$. The $\alpha'$-shift consists of two parts. First, we rescale
\begin{align}\label{alphaResc}
    \alpha' \to \alpha'/2,
\end{align}
and then define $\alpha'$-shifted $X$-variables via
\begin{align}\label{alphaShft}
    \hat{X}_{\text{even}} = X_{\text{even}}\pm1/\alpha', \hspace{0.5cm}  \hat{X}_{\text{odd}} = X_{\text{odd}}.
\end{align}
Let us now see how this shift acts on the stringy $t_{ij}$ variables \eqref{tsVar} of the inverse KLT kernel. Thanks to simple trigonometric identities the $t$-variables transform under \eqref{alphaResc} and \eqref{alphaShft} as,
\begin{align} \label{alphaShftRep}
    \frac{1}{t_\text{even}}\to -\tau_\text{even}, \hspace{1.0cm} \frac{1}{t_\text{odd}}\to \frac{1}{\tau_\text{odd}}.
\end{align}
Here we use $\tau_{ij}=\tan(\frac{\pi}{2}\alpha' X_{ij})$ to denote the $\alpha'$-rescaled stringy variables $t_{ij}$.

\subsubsection*{Stringy Pions}
For $n\,{=}\,2k$ even multiplicity, we now consider the inverse KLT matrix elements $m_{2k}^{\alpha'}(X)$ directly as functions of $X$-variables. We can then define \textit{stringy pion functions} $A_{2k}^{\pi,\alpha'}$ by first applying the rescaling \eqref{alphaResc} to obtain $m_n^{\alpha'\!/2}$ and then evaluating the $\alpha'$-rescaled matrix element on $\alpha'$-shifted kinematics \eqref{alphaShft}, i.e.
\begin{align}\label{defPionStr}
    A_{2k}^{\pi,\alpha'}(X) = m_{2k}^{\alpha'\!/2}(\hat{X}).
\end{align}
The rescaling $\alpha'\to \alpha'/2$ is not essential but ensures that the stringy pion functions \eqref{defPionStr} satisfy the identically normalized monodromy relations \eqref{invKLTmono} as the inverse KLT matrix elements $m_n^{\alpha'}$ in their original tr$\phi^3$ form.
For reference, we reproduce here the simplest stringy pion functions at four and six points,
\begin{align}\label{Ashft46pt}
\begin{split}
    A_4^{\pi,\alpha'} \!\!&= -\tau_{13}-\tau_{24},\\
    A_6^{\pi,\alpha'} \!\!&= \frac{1}{2}\frac{(\tau_{13}+\tau_{24})(\tau_{46}+\tau_{15})}{\tau_{14}}-\tau_{13}-\frac{1}{3}\tau_{13}\tau_{35}\tau_{15} + \text{cyc.}
\end{split}
\end{align}
Using the transformation rules \eqref{alphaShftRep} and the explicit expressions \eqref{invKLTex} for the inverse KLT matrix elements the reader can easily convince themselves of the correctness of \eqref{Ashft46pt}.
The stringy pion functions reduce to ordinary $SU(N)$ non-linear sigma model (NLSM) amplitudes at leading order in $\alpha'$,
\begin{align}\label{piLowEng}
    A_{2k}^{\pi,\alpha'} \!= \frac{\pi\alpha'}{2} \Big(\!A_{2k}^{\text{NLSM}} + \mathcal{O}(\alpha'^{2})\!\Big).
\end{align}
Moreover, they obey the Adler zero \cite{Adler:1964um} at finite $\alpha'$,
\begin{align}\label{Aadler}
    \lim_{p_i \to 0} A_{2k}^{\pi,\alpha'} = 0,
\end{align}
retaining another highly characteristic feature of pion amplitudes.

It should be mentioned that the structure of the $\alpha'$-shift \eqref{alphaShft} for the stringy inverse KLT kernel takes significant inspiration from the known $\delta$-shift \cite{Arkani-Hamed:2024nhp,Arkani-Hamed:2024yvu} for tr$\phi^3$ theory amplitudes. The latter allows to extract NLSM field theory amplitudes directly from those of the tr$\phi^3$ theory by applying a kinematic shift $X_{ij}\to \hat{X}_{ij}(\delta)$ analogous to \eqref{alphaShft} upon identifying $\delta\,{\equiv}\, 1/\alpha'$, and expanding the amplitude around $\delta\,{=}\,\infty$,
\begin{align}\label{PiDeltaShft}
    m_{2k}(\hat{X}(\delta)) = \frac{1}{\delta^{2k-2}}\left( A_{2k}^{\text{NLSM}}(X) + \mathcal{O}(\delta^{-1}) \right).
\end{align}
We can obtain a very similar looking relation involving the stringy inverse KLT kernel by combining \eqref{defPionStr} with \eqref{piLowEng} which gives
\begin{align}\label{PiAlphaShft}
    m^{\alpha'\!/2}_{2k}(\hat{X}(\alpha')) = \frac{\pi\alpha'}{2}\left( A_{2k}^{\text{NLSM}}(X) + \mathcal{O}(\alpha'^{2}) \right).
\end{align}
The equivalence of the leading order contributions to the $\delta$-shifted tr$\phi^3$ amplitude \eqref{PiDeltaShft} and the $\alpha'$-shifted inverse KLT kernel \eqref{PiAlphaShft} is no coincidence. In fact, the connection between the $\delta$-shift and the $\alpha'$-shift can be given a geometric interpretation via the so-called associahedral grid \cite{Bartsch:2025mvy}, the positive geometry for the inverse KLT kernel.
However, beyond the leading order the $\delta$-shifted tr$\phi^3$ amplitude and the $\alpha'$-shifted inverse KLT kernel generate different higher-order corrections to pion amplitudes. For instance, comparing \eqref{PiDeltaShft} and \eqref{PiAlphaShft} at four points we find
\begin{align}\label{AlphaDeltaComp4}
    m_{4}(\hat{X}(\delta)) &= \frac{1}{\delta^2}\left(\! -(X_{13}+X_{24}) + \frac{1}{\delta} (X_{13}^2-X_{24}^2) - \frac{1}{\delta^2}(X_{13}^3 + X_{24}^3) + \dots \right)\!,\\
    m^{\alpha'\!/2}_{4}(\hat{X}(\alpha')) &= \frac{\pi\alpha'}{2}\left(\! -(X_{13}+X_{24}) - \frac{(\pi\alpha')^2}{12} (X_{13}^3+X_{24}^3) - \frac{(\pi\alpha')^4}{120} (X_{13}^5+X_{24}^5) + \dots  \right)\! . \nonumber
\end{align}
We observe that some of the higher-order corrections generated by the $\delta$-shift do not correspond to amplitudes, such as the \textit{anti-cyclic} $\mathcal{O}(\delta^{-3})$ contribution in \eqref{AlphaDeltaComp4}. This is not the case for the $\alpha'$-shifted inverse KLT kernel where higher-order corrections can be consistently interpreted in an effective field theory context. In fact, from a bottom-up perspective, the infinite tower of higher-order corrections provided by the $\alpha'$-shift is carefully chosen such that the $\alpha'$-complete four-point function \eqref{Ashft46pt} satisfies the string monodromy relations \eqref{invKLTmono}. This holds true also at higher multiplicities.

While both \eqref{PiDeltaShft} and \eqref{PiAlphaShft} accomplish the goal of extracting pions from a tr$\phi^3$-like function, the $\alpha'$-shift for the inverse KLT kernel does so in a way that makes the power counting of pion amplitudes and the structure of the NLSM contact terms manifest. To illustrate this, we study the stringy pion function in \eqref{Ashft46pt} for $n\,{=}\,6$ in detail. In particular, we can show at the level of individual Feynman diagrams how the factorization and contact terms of the stringy pion function emerge when the $\alpha'$-shift is applied. For instance, we can obtain the factorization term of $A_6^{\pi,\alpha'}$ in \eqref{Ashft46pt} by considering the following diagrams contributing to $m_6^{\alpha'}$,
\begin{align}
    \, \hspace{-1.9cm} \\
    \begin{matrix}
         \hspace{0.1cm}\vspace{-0.05cm}\includegraphics[width=15.2cm]{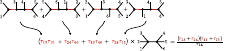}
    \end{matrix} \nonumber
\end{align}
Here the arrows represent the transformation \eqref{alphaShftRep} applied to each diagram. We see that for a given diagram topology the $\alpha'$-shift acts by collapsing all internal lines corresponding to even-particle channels. The resulting topology therefore only has odd-particle channels, as expected for NLSM amplitudes. The numerator of the resulting topology is given by the product of the $\tau_{ij}$-variables corresponding to the even-particle channels that were removed from the original topology.

Similarly, we can obtain the six-point NLSM contact term and its higher-order correction from the diagrams,
\begin{align}
    \begin{matrix}
         \hspace{0.1cm}\vspace{-0.05cm}\includegraphics[width=12.2cm]{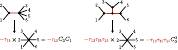}
    \end{matrix} 
\end{align}
where $C_k$ are the Catalan numbers. We observe that the diagram on the left, involving the five-point interaction for the inverse KLT kernel, is responsible for generating the leading-order six-point NLSM contact term $\sim\! \tau_{13} \,{+}\, \text{cyc.}$ (cf. \eqref{NptNLSMvertShft}). On the other hand, the ``snowflake'' diagram on the right generates a higher-order correction to the contact term not present in the NLSM Lagrangian. While this contribution is irrelevant for the field theory limit $\alpha'\,{\to}\, 0$ it ensures stringy properties of the pion function $A^{\pi,\alpha'}$ like the string monodromy \eqref{invKLTmono} at finite $\alpha'$.

Finally, let us consider the general case where $n\,{=}\,2k$ is even and show that the $\alpha'$-shift effectively interpolates between the infinite tower of odd-point interactions \eqref{KLTFeynRules} of the inverse KLT kernel and the infinite tower of even-point interactions derived from the NLSM Lagrangian in the so-called ``minimal'' parametrization,
\begin{align}\label{LagMinPar}
    \mathcal{L}_{\text{NLSM}} = \frac{1}{8}\langle \partial U\cdot\partial U^{\dagger} \rangle, \hspace{0.5cm} \text{ where } \hspace{0.5cm} U(\phi) = i2\phi + \sqrt{1-4\phi^2},
\end{align}
and $\phi = \phi^at^a$ now denotes the pion fields. Expanding the Lagrangian in powers of $\phi$ we can read off the flavor-ordered Feynman rules for the even-point interaction vertices \cite{Kampf:2013vha},
\begin{align}\label{NLSMvMin}
    V_{2k}(X) = -\frac{1}{2}\sum_{j=0}^{k-2} C_j C_{k-j-2} \, (X_{1,3+2j} + \text{cyc.}), \hspace{0.5cm} \text{ where } \hspace{0.5cm} k\ge 2.
\end{align}

With this in mind, we now look at the inverse KLT matrix element $m_{2k}^{\alpha'}$ and consider how the $\alpha'$-shift acts on the set of single propagator diagrams,
\begin{align}\label{NptNLSMvertShft}
    \frac{1}{2}\sum_{j=0}^{k-2} 
    \left(\begin{matrix}
         \hspace{0.1cm}\vspace{-0.05cm}\includegraphics[width=3.0cm]{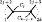}
    \end{matrix} 
    + \text{cyc.} \right)
    \xrightarrow{\eqref{alphaShftRep}} - \frac{1}{2}\sum_{j=0}^{k-2} C_jC_{k-j-2}(\xRed{\tau_{1,3+2j}} + \text{cyc.}),
\end{align}
where the factor $1/2$ is there to avoid double-counting of diagrams. It is easily verified that \eqref{NptNLSMvertShft} is in fact a sum over \textit{all} single-propagator Feynman diagrams consistent with the Feynman rules \eqref{KLTFeynRules} and contributing to $m_{2k}^{\alpha'}$. Under the $\alpha'$-shift we see that this class of diagrams is in one-to-one correspondence with individual terms in the minimal parametrization NLSM vertex \eqref{NLSMvMin}. This illustrates in a transparent way how the infinite tower of even-point interactions in the NLSM is encoded by (and equivalent to) the infinite tower of odd-point interactions of the inverse KLT kernel.

We will return to the connection between the inverse KLT kernel and pions later, when we discuss the generalization of the inverse KLT matrix elements $m_n^{\alpha'}$ to the inverse KLT integrand at loop level in Section \ref{KLTloop}.

\section{Cubic Recursion for the Stringy Inverse KLT Kernel (Trees)}\label{KLTtree}
%
Let us now turn to the first main result of this article. It lies in the realization that the infinite tower of interactions \eqref{KLTFeynRules} for the inverse KLT kernel is spurious and can be replaced by a single, suitably defined \textit{cubic} interaction vertex. Notably, this simplification does not occur in the conventional Feynman diagram representation of the inverse KLT kernel but only manifests itself when the matrix elements $m_n^{\alpha'}$ are recast in the form of a \textit{new kind of} Berends-Giele recursion, which we will simply call the \textit{cubic Berends-Giele recursion}. The latter allows for efficient computation of the stringy matrix elements $m_n^{\alpha'}$ and exposes their essential equivalence to amplitudes in the tr$\phi^3$ field theory.

Importantly, the cubic Berends-Giele recursion derived in this section for the tree-level inverse KLT kernel will serve as a natural foundation for the later definition of the inverse KLT loop integrand in Section \ref{KLTloop}.

\subsection{Diagonal Matrix Elements}
To start, let us briefly consider the structure of what we will call the \textit{naive} Berends-Giele recursion for the stringy matrix elements $m_n^{\alpha'}$ based on the inverse KLT Lagrangian \eqref{defKLTLag}.
Considering the generic form of the Berends-Giele recursion in Figure \ref{fig:bgdiaggengen}, and applying it to the stringy inverse KLT kernel, we expect that at $n$ points all root vertices with multiplicity ${\le}\, n$ have to enter the recursion. Schematically, the recursion of $m_n^{\alpha'}$ therefore involves Berends-Giele diagrams of the form
\vspace{-0.3cm}
\begin{align}\label{malphaBGnaive}
    \hspace{-0.5cm}m_n^{\alpha'} \simeq 
    \begin{matrix}
         \hspace{-0.0cm}\vspace{-0.3cm}\includegraphics[width=5.3cm]{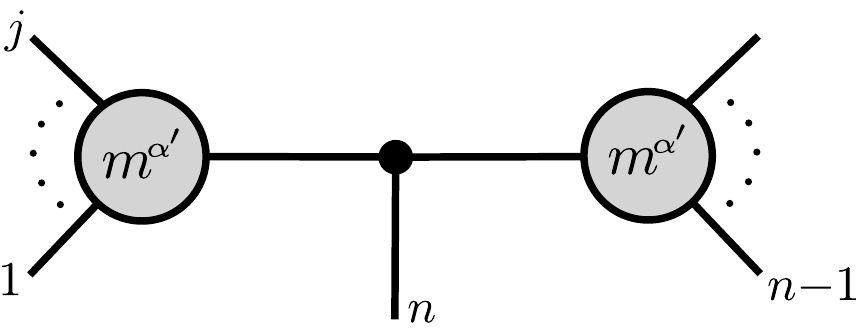}
    \end{matrix}
    \!\!\!\!+\,\,
    \sum_{\substack{v \ge 5 \\ \text{odd}} }^{n}
    \begin{matrix}
         \hspace{0.0cm}\vspace{1.3cm}\includegraphics[width=5.8cm]{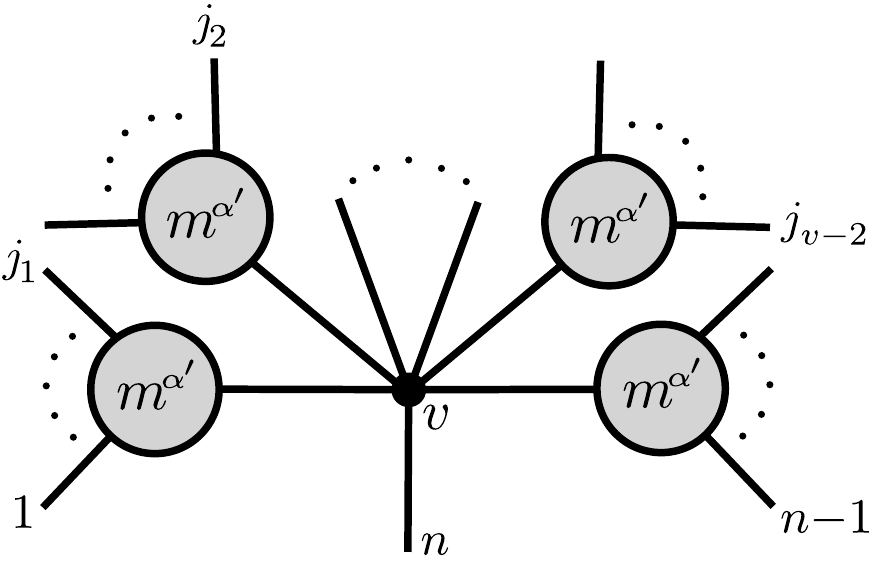}
    \end{matrix},
\end{align}
\vspace{-1.5cm}

\noindent The first term above is analogous to the cubic Berends-Giele diagrams \eqref{BASBGdiag} for the tr$\phi^3$ theory, except that internal lines now correspond to stringy propagators as defined in \eqref{KLTFeynRules}.
The remaining terms in \eqref{malphaBGnaive} combine contributions from Feynman diagrams where the root leg is connected to $5-,7-$ and higher-point root vertices of valency $v$. Indeed, the naive recursion \eqref{malphaBGnaive} is valid in principle and can be used to compute $m_n^{\alpha'}$. However, its structure is not particularly compelling as it canonically follows from the Lagrangian \eqref{defKLTLag}.

\subsubsection*{A cubic simplification}
We will now show that the naive recursion \eqref{malphaBGnaive} based on the infinite tower of interactions \eqref{KLTFeynRules} fails to take advantage of a remarkable hidden simplicity of the inverse KLT kernel. In fact, the recursive structure of the stringy matrix elements $m_n^{\alpha'}$ turns out to be \textit{just as simple} as that of their field theory counterparts $m_n$ \eqref{BASBGdiag} in the tr$\phi^3$ theory. In particular, \textit{all} higher valency diagrams ($v \,{\ge}\, 5$) in \eqref{malphaBGnaive} can be absorbed into a simple $\alpha'$-dependent redefinition of the cubic root vertex
\begin{align}\label{rootValphaExp}
    \begin{matrix}
         \hspace{-0.0cm}\vspace{-0.3cm}\includegraphics[width=2.4cm]{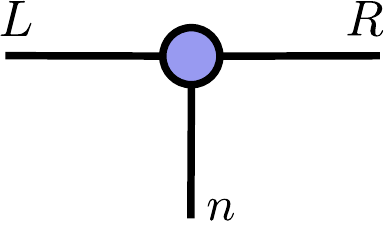}
    \end{matrix}
    = 1 + \tan(\pi\alpha' X_L)\tan(\pi\alpha' X_R),
\end{align}
where, as before, $X_{L/R}$ are the invariants associated to the momenta flowing into the vertex from the left/right. In particular, if either of the left/right-hand momenta is on-shell, i.e. if $X_{L/R}\,{=}\,0$, the stringy root vertex becomes independent of $\alpha'$ and reduces to the tr$\phi^3$ field theory vertex \eqref{rootVBAS}. We encourage the reader to pause for a moment and familiarize themselves with the remarkably simple structure of the \textit{effective root vertex} \eqref{rootValphaExp}. In a sense, the remainder of this article will be concerned just with working out the consequences of this vertex for the inverse KLT kernel at tree- and loop-level.

The first surprising consequence of the structure of the effective root vertex \eqref{rootValphaExp} is that it allows us to formulate a \textit{purely cubic} Berends-Giele recursion for the matrix elements of the inverse KLT kernel,
\begin{align}\label{malphaBGschem}
\begin{split}
    m_n^{\alpha'} &= 
    \begin{matrix}
         \hspace{-0.0cm}\vspace{-0.15cm}\includegraphics[width=3.8cm]{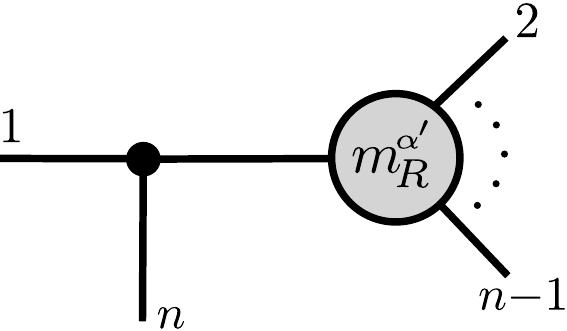}
    \end{matrix}
    +\hspace{-0.2cm}
    \begin{matrix}
         \hspace{-0.0cm}\vspace{-0.15cm}\includegraphics[width=4.0cm]{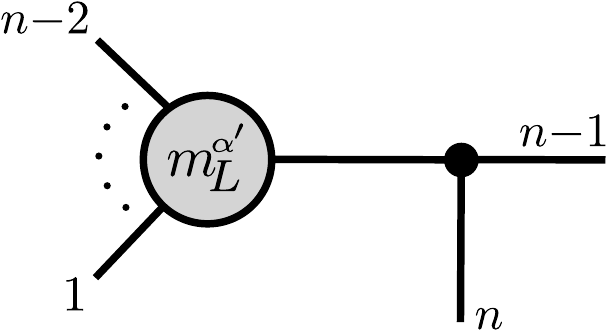}
    \end{matrix}
    \\
    &\hspace{0.55cm} +\sum_{j=3}^{n-2}\!\! 
    \begin{matrix}
         \hspace{0.2cm}\vspace{-0.25cm}\includegraphics[width=5.8cm]{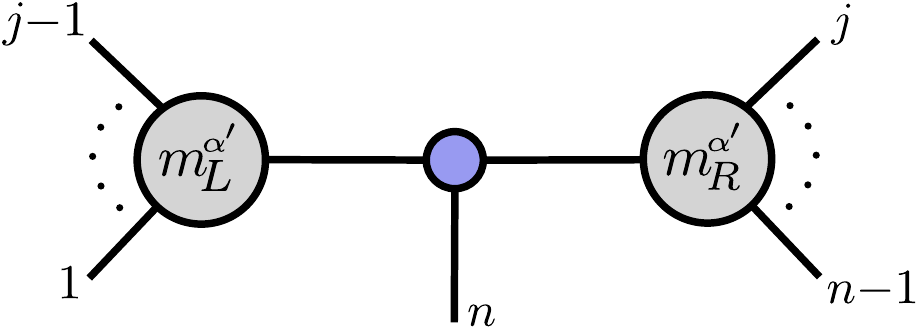}
    \end{matrix}.
\end{split}
\end{align}
Comparing the Berends-Giele diagrams above to the ones for the tr$\phi^3$ amplitude in \eqref{BASBGdiag}, we see that they are structurally identical, with the exception of the appearance of the effective root vertex \eqref{rootValphaExp} for the inverse KLT kernel. Notably, the first two terms in \eqref{malphaBGschem} involve the regular tr$\phi^3$ root vertex, as either the left- or right-hand momentum entering the effective vertex is on-shell.

Evaluating the Berends-Giele diagrams in \eqref{malphaBGschem} we arrive at a strikingly concise recursive formula for the diagonal matrix elements of the inverse KLT kernel, 
\begin{align}\label{BGKLT}
\begin{split}
    m_n^{\alpha'}\!(1\dots n) &= \frac{1}{t_{2n}}m_{n-1}^{\alpha'}(2\dots n) + \frac{1}{t_{1,n-1}}m_{n-1}^{\alpha'}(1\dots n{-}1)\\
    &+\! \sum_{j=3}^{n-2} \frac{1+t_{1j}t_{jn}}{t_{1j}t_{jn}}\, m_j^{\alpha'}\!(1\dots j)m_{n-j+1}^{\alpha'}(j\dots n).
\end{split}
\end{align}
where all matrix elements are considered to be in $X$-label representation.

Of course, the validity of the cubic Berends-Giele recursion \eqref{BGKLT} is not at all obvious from the way we have stated the result, and we will prove its consistency shortly. First, however, we want to show how the recursion works for a simple, but non-trivial example and discuss some of its properties.

\subsubsection*{A simple example ($n\,{=}\,5$)}

Let us demonstrate the cubic recursion \eqref{BGKLT} for a simple example at $n\,{=}\,5$. We can easily compute the matrix element $m_5^{\alpha'}$ from the following Berends-Giele diagrams,
\begin{align}\label{KLTBGex5diag}
    \hspace{-0.6cm} m_5^{\alpha'} = \begin{matrix}
         \hspace{-0.0cm}\vspace{-0.3cm}\includegraphics[width=3.2cm]{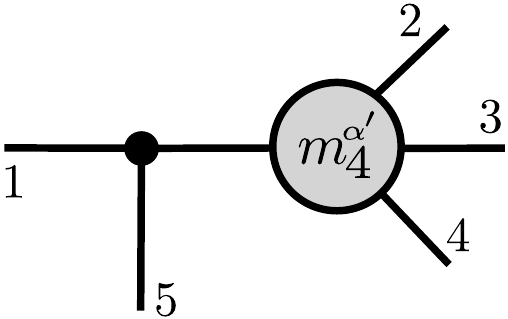}
    \end{matrix}
    \,+\,
    \begin{matrix}
         \hspace{-0.0cm}\vspace{-0.46cm}\includegraphics[width=3.2cm]{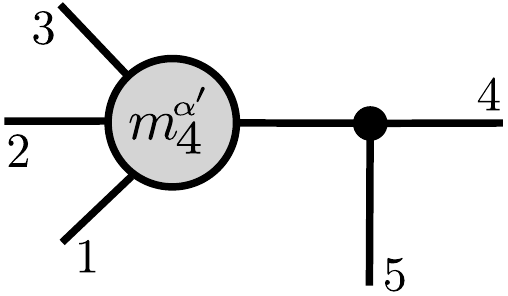}
    \end{matrix}\,
    \,+\!
    \begin{matrix}
         \hspace{-0.0cm}\vspace{-0.3cm}\includegraphics[width=4.2cm]{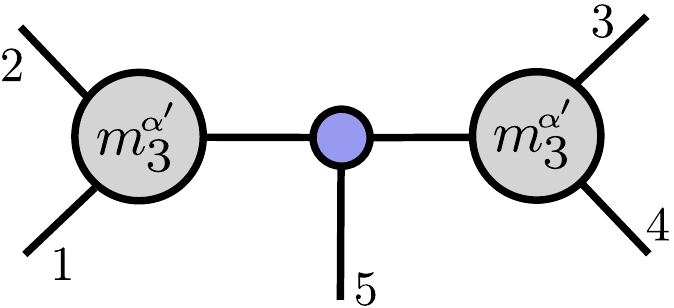}
    \end{matrix}.
\end{align}
Noting the similarity to the tr$\phi^3$ recursion \eqref{BASBGex5diag} we evaluate the diagrams and obtain
\begin{align}\label{KLTBGex5expr}
\begin{split}
    m_5^{\alpha'} &= \frac{1}{t_{25}}m_4^{\alpha'}\!(2345) + \frac{1}{t_{14}}m_4^{\alpha'}\!(1234) + \frac{1+t_{13}t_{35}}{t_{13}t_{35}}m^{\alpha'}_3\!(123)m^{\alpha'}_3\!(345)\\
    &= \frac{1}{t_{25}}\!\left(\!\frac{1}{t_{24}} \,{+}\, \frac{1}{t_{35}}\!\right) + \frac{1}{t_{14}}\!\left(\!\frac{1}{t_{13}} \,{+}\, \frac{1}{t_{24}}\! \right) + \frac{1+t_{13}t_{35}}{t_{13}t_{35}},
\end{split}
\end{align}
which agrees with the matrix element shown in \eqref{invKLTex}. 
In particular, we observe that the diagram involving the effective root vertex \eqref{rootValphaExp} generates the correct five-point contact term $\sim\! 1$. This simple mechanism by which the effective root vertex cancels propagators in the Berends-Giele diagram to generate contact terms turns out to be the origin of the entire infinite tower of interactions \eqref{KLTFeynRules} for the inverse KLT kernel.
We will study how the recursion generates these contact interactions more generally once we prove the validity of the cubic Berends-Giele recursion at arbitrary multiplicity in Section \ref{BGKLTtreeProof}.

\subsubsection*{Discussion}
By collapsing the infinite tower of interactions \eqref{KLTFeynRules} to just a single, stringy cubic vertex \eqref{rootValphaExp}, the recursion \eqref{BGKLT} achieves a drastic reduction in the number of Berends-Giele diagrams required to compute a given matrix element $m_n^{\alpha'}$. When compared to the naive Berends-Giele recursion \eqref{malphaBGnaive}, for which the number of diagrams grows exponentially as $2^{n-3}$, the cubic recursion \eqref{malphaBGschem} requires only $n{-}2$ contributions. This shows that from a computational point of view the string theory inverse KLT kernel is just as simple as amplitudes in the tr$\phi^3$ theory. The apparent complexity of the stringy matrix elements associated with the infinite tower of odd-point interactions \eqref{KLTFeynRules} is spurious.

It is no surprise then, that owing to the structure of the effective root vertex, the field theory limit \eqref{alpha0limBAS} of the stringy matrix elements $m_n^{\alpha'}$ is manifest when expressed in terms of the cubic Berends-Giele recursion. Since
\begin{align}
    \frac{1+t_{1j}t_{jn}}{t_{1j}t_{jn}} \xrightarrow{\,\alpha'\to 0\,}\frac{1}{\alpha'^2 X_{1j}X_{jn}}+ \mathcal{O}(1),
\end{align}
each Berends-Giele diagram in \eqref{BGKLT} precisely reduces to the corresponding diagram in the recursion \eqref{BASBGdiag} for bi-adjoint scalars.

Its similarity to the recursion for the tr$\phi^3$ theory notwithstanding, it should be emphasized that the Berends-Giele recursion \eqref{BGKLT} is highly non-trivial and exposes a previously unknown simplicity of the stringy matrix elements $m_n^{\alpha'}$. In particular, the cubic recursion \textit{cannot} be understood as a naive regrouping of appropriately redefined cubic Feynman diagrams. 
To illustrate this, let us suppose that the recursion \eqref{KLTBGex5expr} \textit{is} indeed just a regrouping of properly defined cubic Feynman diagrams. The structure of the root vertex \eqref{rootValphaExp} suggests that there should be a Lagrangian that generates a single cubic off-shell vertex
\begin{align}\label{rootVertGuess}
V_3(p_1,p_2,p_3) =
\begin{matrix}
         \hspace{-0.0cm}\vspace{-0.25cm}\includegraphics[width=2.5cm]{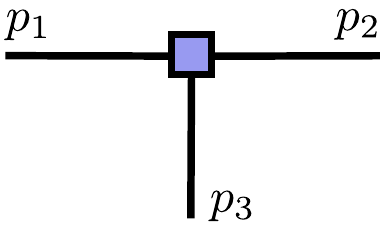}
\end{matrix} = 1+t_1t_2+t_2t_3+t_3t_1,
\end{align}
where we use the notation $t_i\,{=}\,\tan(\pi\alpha' p_i^2)$. Note that if any two momenta entering the vertex are on-shell, $p_i^2\,{=}\,p_j^2\,{=}\,0$, it reduces to that of the BAS theory, $V_3(p_i,p_j,P)=1$ . If instead only one momentum is on-shell, $p^2\,{=}\,0$, and two momenta $p_L,p_R$ are off-shell we recover the effective root vertex \eqref{rootValphaExp}, i.e. $V_3(p,p_L,p_R)=1+t_Lt_R$. 
To compute matrix elements we just sum over all cubic Feynman diagrams involving the vertex \eqref{rootVertGuess}.

Returning to the five-point example, only a single cubic diagram topology contributes,
\vspace{-0.1cm}
\begin{align}
\begin{split}
    \tilde{m}_5^{\alpha'} = 
    \begin{matrix}
         \hspace{-0.0cm}\vspace{-0.5cm}\includegraphics[width=3.5cm]{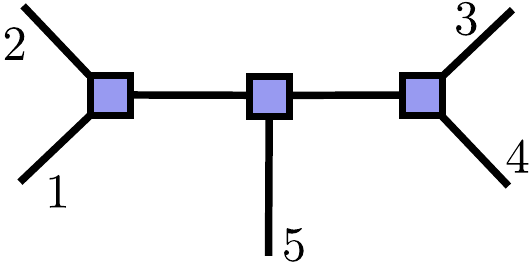}
    \end{matrix}
    + \text{cyc.} = \frac{1+t_{13}t_{35}}{t_{13}t_{35}} + \text{cyc.} \,\overset{!}{\neq}\, m_5^{\alpha'} .
\end{split}
\end{align}
Comparing this to the inverse KLT matrix element in \eqref{KLTBGex5expr} we see that the above prescription fails to give the correct contact term as $\tilde{m}_5^{\alpha'}\,{-}\,m_5^{\alpha'} =4$.

This suggests that the cubic recursion \eqref{BGKLT} represents a \textit{new kind of} Berends-Giele recursion, one where the root vertex does not follow from a Lagrangian. In other words, the inverse string theory KLT kernel is \textit{not} cubic when expressed in terms of Feynman diagrams, but \textit{is} cubic only once it is expressed in terms of Berends-Giele diagrams involving the effective root vertex \eqref{rootValphaExp}.

\subsection{Proof}\label{BGKLTtreeProof}
We prove that the cubic Berends-Giele recursion proposed in \eqref{BGKLT} indeed reproduces the diagonal matrix elements $m_n^{\alpha'}$ of the inverse string theory KLT kernel. This will be accomplished in two steps. First, we will show that the functions generated via \eqref{BGKLT} have contact terms corresponding to those expected from the Feynman rules \eqref{KLTFeynRules}. Then we show that the functions obtained through the recursion consistently factorize according to \eqref{mfact}. Since we are dealing with rational functions of $t_{ij}$, satisfying both these criteria is sufficient to establish equality with the matrix elements of the inverse KLT kernel \eqref{defInvKLT}. Readers not interested in the technical details of the proof are free to skip this section and move on to the introduction of the inverse KLT integrand in Section \ref{KLTloop}.

An alternative proof of the recursion at the level of the generating functional of Berends-Giele currents can be found in Appendix \ref{app:KLTBGprf2}.
\subsubsection*{Contact Terms}
We start by showing how the contact interactions \eqref{KLTFeynRules} of the inverse KLT kernel arise from the cubic recursion \eqref{BGKLT}. This can be understood using a two-step inductive argument. First, we show that there are no contact interactions at even points $n=2K$, for some fixed $K\ge 2$. For the induction we assume that for all $k<K$ contact terms are in fact absent,
\begin{align}\label{KLTctEvenAss}
    m_{2k}^{\alpha'}(1\dots 2k)\Big\rvert_{\text{ct.}} \!\!\!\!\!= 0, \hspace{0.5cm}\text{for } k<K.
\end{align}
The notation $\dots\rvert_{\text{ct.}}$ indicates the restriction to contact terms of the matrix element, i.e. to those contributions that cannot be fixed from factorization \eqref{mfact} on poles. Since the matrix elements $m_n^{\alpha'}$ have a unique representation in terms of $t_{ij}$-variables we can find their contact terms simply by discarding all terms that involve at least one propagator $t_{ij}^{-1}$. In the following discussion, the same logic is applied also to products of matrix elements.
With this in mind we can use the explicit form of the recursion \eqref{BGKLT} to compute
\begin{align}\label{KLTctEven}
    m_{2K}^{\alpha'}\Big\rvert_{\text{ct.}} \!\!&= \frac{1}{t_{1,2K-1}}m^{\alpha'}_{2K-1}\Big\rvert_{\text{ct.}} \!\!+ \sum_{j=3}^{2K-2} \!\left(\!1+\frac{1}{t_{1,j}t_{j,2K}}\!\right) m_j^{\alpha'}m_{2K-j+1}^{\alpha'}\Big\lvert_{\text{ct.}} \!\!+ \frac{1}{t_{2,2K}}m^{\alpha'}_{2K-1}\Big\rvert_{\text{ct.}} \nonumber \\
    &= \sum_{j=3}^{2K-2} m_j^{\alpha'}m_{2K-j+1}^{\alpha'}\Big\lvert_{\text{ct.}},
\end{align}
where for visual clarity we have suppressed the $X$-label dependence of the individual matrix elements. To arrive at the second line we discard all terms in the recursion that explicitly involve a propagator $t_{ij}^{-1}$. Since there are no cancellations between numerators and denominators in the Berends-Giele recursion, these terms cannot possibly contribute to the contact term of $m_{2K}^{\alpha'}$. Now we observe that the remaining terms in \eqref{KLTctEven} for any $j=3,\dots,{K{-}2}$ involve a product of an odd- and an even-point matrix element. Therefore, assumption \eqref{KLTctEvenAss} implies that one factor always vanishes. This concludes the inductive step and shows that, as desired,
\begin{align}\label{KLTctEvenRes}
    \kappa_{2K} \equiv m_{2K}^{\alpha'}\Big\rvert_{\text{ct.}} \!\!= 0,
\end{align}
where, for later reference we have introduced the notation $\kappa_{2k}$ to denote the contact term.

Next we want to show that at $n=2K{+}1$ odd, the recursion generates matrix elements with non-zero contact terms as given by \eqref{KLTFeynRules}. Again, we start the induction by assuming that for $k<K$ the matrix elements already satisfy
\begin{align}\label{KLTctOddAss}
    m_{2k+1}^{\alpha'}\Big\rvert_{\text{ct.}} \!\!= C_{k-1}.
\end{align}
In analogy to \eqref{KLTctEven} we use the recursion \eqref{BGKLT} to compute
\begin{align}\label{KLTctOdd}
\begin{split}
    m_{2K+1}^{\alpha'}\Big\rvert_{\text{ct.}} \!\!&= \!\sum_{j=3}^{2K-1} m_j^{\alpha'}m_{2K-j+2}^{\alpha'}\Big\lvert_{\text{ct.}}\\
    &= \sum_{j=2}^{K-1} m_{2j}^{\alpha'}m_{2(K-j+1)}^{\alpha'}\Big\lvert_{\text{ct.}}\!\! +\sum_{j=1}^{K-1} m_{2j+1}^{\alpha'}m_{2(K-j)+1}^{\alpha'}\Big\lvert_{\text{ct.}},
\end{split}
\end{align}
where we split the sum into terms involving products of odd- and even-point matrix elements. The first term on the second line vanishes due to the previous result \eqref{KLTctEvenRes}. What remains is to use the assumption \eqref{KLTctOddAss} and evaluate
\begin{align}\label{KLTctOddRes}
    \kappa_{2K+1} \equiv m_{2K+1}^{\alpha'}\Big\rvert_{\text{ct.}} \!\!= \sum_{j=1}^{K-1} C_{j-1}C_{K-j-1} = C_{K-1},
\end{align}
which, owing to the recursive nature of Catalan numbers, yields the correct contact term. 

This further illustrates that the Berends-Giele recursion \eqref{BGKLT} does not correspond to a straightforward regrouping of Feynman diagrams generated from the Feynman rules \eqref{KLTFeynRules}. In fact, reading \eqref{KLTctOddRes} from right to left suggests that the contact term $C_{K-1}$ is split up into a sum of $K{-}1$ terms which are distributed among a set of Berends-Giele diagrams,
\begin{align}\label{BGKLTcont}
    \begin{matrix}
        \hspace{-0.0cm}\vspace{-0.1cm}\includegraphics[width=1.9cm]{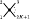}
    \end{matrix} \!\!\!\!=\sum_{j=1}^{K-1} C_{j-1}C_{K-j-1} \,\,\,\xhookrightarrow{\,\,\,\,\,\,\,\,}\,\,\, \sum_{j=1}^{K-1} \begin{matrix}
         \hspace{0.1cm}\vspace{-0.25cm}\includegraphics[width=5.7cm]{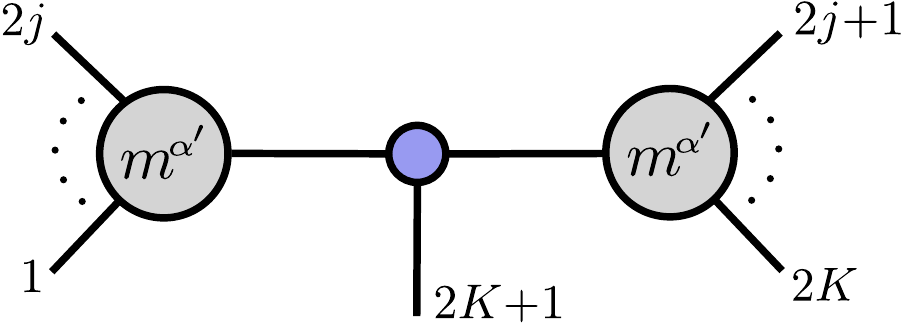}
    \end{matrix}\,.
\end{align}

Therefore, unlike for the bi-adjoint scalar amplitudes, it is in general not possible to uniquely assign a given Feynman diagram to a single Berends-Giele diagram when using the cubic recursion for the inverse string theory KLT kernel.

\subsubsection*{Factorization}
To complete the proof we still need to show that the functions obtained from the recursion \eqref{BGKLT} correctly factorize on poles. To facilitate this, we will first adapt the factorization condition \eqref{mfact} to the $X$-label representation of the matrix elements $m_n^{\alpha'}$. Since we are dealing  with rational functions of the variables $t_{ij}\,{=}\,\tan (\pi\alpha' X_{ij})$, instead of treating the periodic poles at $X_{ij} \,{=}\, k/\alpha'$ with $k\,{\in}\, \mathds{Z}$ separately (cf. \eqref{periodPoles}), we will equivalently just consider poles when $t_{ij}\,{=}\,0$. This allows us to write the factorization condition \eqref{mfact} in $X$-label representation as
\begin{align}\label{defFact}
\begin{split}
    \frac{\partial}{\partial t^{-1}_{1i}}\, 
    \begin{matrix}
        \hspace{-0.0cm}\vspace{-0.1cm}\includegraphics[width=1.9cm]{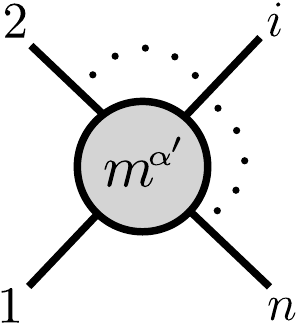}
    \end{matrix} &=  
    \begin{matrix}
        \hspace{-0.1cm}\vspace{-0.1cm}\includegraphics[width=4.8cm]{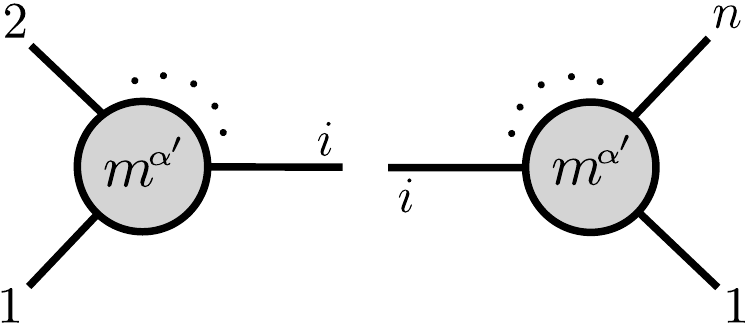}
    \end{matrix}
    \\
    \Leftrightarrow \frac{\partial}{\partial t^{-1}_{1i}}\, m^{\alpha'}_n\!(1\dots i\dots n) &= 
    m_i^{\alpha'}\!(1\dots i)m_{n-i+2}^{\alpha'}(i\dots n\,1),
\end{split}
\end{align}
for $i\,{=}\,3,\dots,n{-}1$, where we formally replace taking a residue at $t_{ij}\,{=}\,0$ by a derivative with respect to the corresponding propagator $t_{ij}^{-1}$. Due to the cyclicity of the matrix elements we can restrict, without loss of generality, to studying poles of the form $t_{1i}^{-1}$.

As we did for the contact terms before, we will prove consistent factorization of the cubic Berends-Giele recursion \eqref{BGKLT} inductively for a matrix element $m^{\alpha'}_K$ at fixed multiplicity $K$, assuming that the factorization property \eqref{defFact} holds for matrix elements $m_k^{\alpha'}$ with $k\,{<}\,K$. We start with the special case of factorization on the pole at $t_{1,K-1}\,{=}\,0$ as it is particularly simple. Indeed, from the explicit form of the recursion \eqref{BGKLT} we find that only a single term contributes, giving
\begin{align}
    \frac{\partial}{\partial t^{-1}_{1K-1}}\, m^{\alpha'}_K(1\dots K) &= \frac{\partial}{\partial t^{-1}_{1K-1}} \frac{1}{t_{1,K-1}}  m^{\alpha'}_{K-1}(1\dots K-1) = m^{\alpha'}_{K-1}(1\dots K-1).
\end{align}
Comparing this to \eqref{defFact} for the case $i\,{=}\,n\,{-}\,1$, we see that this is already the desired result, as the other matrix element expected to appear on the pole is $m_{3}(K{-1}\,K\,1)=1$.

Moving on to the generic case for poles of the form $t_{1i}\,{=}\,0$ with $i=3,\dots,K{-}2$ the derivation is slightly more involved. Using the recursion we find
\begin{align}
\begin{split}
    \frac{\partial}{\partial t^{-1}_{1i}}\, m^{\alpha'}_K(1\dots K) &= \frac{1}{t_{1,K-1}} \frac{\partial}{\partial t^{-1}_{1i}} m^{\alpha'}_{K-1}(1\dots K-1) + \frac{1}{t_{2,K}} \frac{\partial}{\partial t^{-1}_{1i}} m^{\alpha'}_{K-1}(2\dots K)\\ 
    &+ \sum_{j=3}^{K-2}\! \bigg\lbrace\! \frac{\partial}{\partial t^{-1}_{1i}}\left(\!1+ \frac{1}{t_{1j}t_{jK}}\!\right) \!\!\bigg\rbrace  m_j^{\alpha'}(1\dots j)m_{K-j+1}^{\alpha'}(j\dots K)\\
    &+ \sum_{j=3}^{K-2}\!\left(\!1+ \frac{1}{t_{1j}t_{jK}}\!\right) \!\frac{\partial}{\partial t^{-1}_{1i}} m_j^{\alpha'}(1\dots j)m_{K-j+1}^{\alpha'}(j\dots K).
\end{split}
\end{align}
Using consistent factorization \eqref{defFact} of the lower point matrix elements $m_k^{\alpha'}$ with $k\,{<}\, K$ the above reduces to
\begin{align}
    \frac{\partial}{\partial t^{-1}_{1i}}\, m^{\alpha'}_K(1\dots K) &= m_i^{\alpha'}\!(1\dots i)\bigg\lbrace \!\frac{1}{t_{1,K-1}}m_{K-i+1}^{\alpha'}(i\dots K{-}1\,1) + \frac{1}{t_{i,K}}m_{K-i+1}^{\alpha'}(i\dots K) \nonumber \\
    & \hspace{2.6cm} +\!\!\sum_{j=i+1}^{K-2}\!\left(\!1+ \frac{1}{t_{1j}t_{jK}}\!\right)  m_{j-i+2}^{\alpha'}(i\dots j\,1)m_{K-j+1}^{\alpha'}(j\dots K)\!\bigg\rbrace \nonumber \\
    &\equiv m^{\alpha'}_i\!(1\dots i)m_{K-i+2}^{\alpha'}(i\dots K\, 1),
\end{align}
where we recognize the terms in curly brackets as the Berends-Giele recursion for the matrix element $m_{K-i+2}^{\alpha'}(i\dots K\, 1)$ appearing on the factorization channel. 

This concludes the proof of consistent factorization for the matrix element $m_K^{\alpha'}$. Together with the correct structure of contact terms shown in \eqref{KLTctEvenRes} and \eqref{KLTctOddRes} this proves that the cubic Berends-Giele recursion \eqref{BGKLT} correctly reproduces diagonal matrix elements of the stringy inverse KLT kernel.

\section{Cubic Recursion for the Stringy Inverse KLT Kernel (Loops)}\label{KLTloop}

We will now take advantage of the previously defined cubic recursion relation \eqref{BGKLT} for the inverse KLT kernel $m_n^{\alpha'}$ at tree-level, and generalize it in order to \textit{define the inverse KLT loop integrand}. This will be accomplished by taking seriously the cubic nature of inverse KLT matrix elements when expressed in terms of the effective root vertex \eqref{rootValphaExp} and appropriately adapting the structure of the recursion to loop integrands. This will allow us to constructively define planar loop integrands $m^{\alpha'}_{\xRed{L},n}$ for any multiplicity $n$ and number of loops $\xRed{L}$, while also providing a highly efficient means of calculating them.

In lifting the cubic Berends-Giele recursion to the loop-level, we will make the astounding observation that the recursion does not just encode the infinite tower of tree-level contact interactions \eqref{KLTFeynRules} but in fact encapsulates an \textit{infinite number of infinite towers} of contact terms. These additional interactions contribute at higher loop orders and only modify the loop \textit{integrand}, not the integrated matrix element. Nevertheless, the existence of these contact terms for the inverse KLT integrand will turn out to be crucial when we want to make the connection to pion integrands via $\alpha'$-shifts, ensuring key properties such as an integrand-level Adler zero. We will also show that this infinity of infinite towers of interactions can be neatly summed into a remarkably simple all-loop Lagrangian for the inverse KLT integrand that generalizes the tree-level Lagrangian \eqref{defKLTLag}.

\subsection{Surface kinematics for the inverse KLT integrand}
Before delving into the definition of the inverse KLT integrand, we first have to discuss the appropriate choice of kinematic variables for loop integrands. As we saw earlier, the stringy $t$-variables \eqref{tsVar} were adequate to capture the dependence of inverse KLT matrix elements on external kinematics at tree-level. For an $L$-loop integrand we have to additionally take into account its dependence on the loop momenta $\ell_i$ where $i\,{=}\,1,\dots,L$.

\begin{figure}[t]
    \centering
    \includegraphics[width=3.8cm]{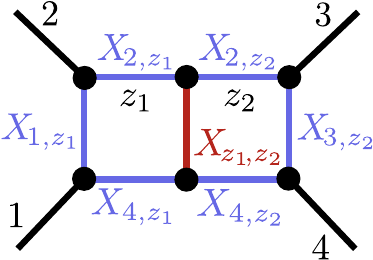}
    \caption{Example for the canonical labeling of propagators in a planar Feynman diagram by the planar loop variables \eqref{defXloop1}.}
    \label{fig:plLoopVars1}
\end{figure}

Since we will be studying only planar loop integrands in this work, there is a natural choice of invariants corresponding to inverse propagators of planar Feynman diagrams \cite{Arkani-Hamed:2010pyv},
\begin{align}\label{defXloop1}
    X_{i,z_{j}} = (\ell_j + p_1+\dots+p_i)^2, \hspace{1.0cm} X_{z_i,z_j} = (\ell_i - \ell_j)^2,
\end{align}
where at $n$ points and $L$ loops the labels run over $i\,{=}\,1,\dots,n$ and $z_i\,{=}\,z_1,\dots z_L$. To give an example, we can canonically label the propagators of a two-loop, four-point Feynman diagram in the $\text{tr}\phi^3$ theory using the variables \eqref{defXloop1} as shown in Figure \ref{fig:plLoopVars1}.

\begin{figure}[H]
    \centering
    \includegraphics[width=2.3cm]{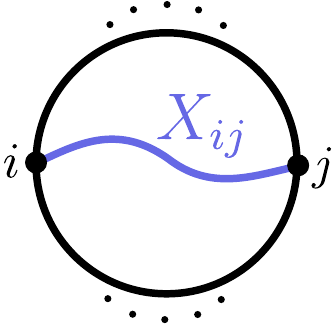}
    \hspace{1.2cm}
    \includegraphics[width=2.3cm]{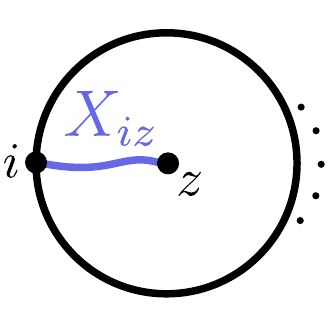}
    \hspace{1.2cm}
    \includegraphics[width=2.3cm]{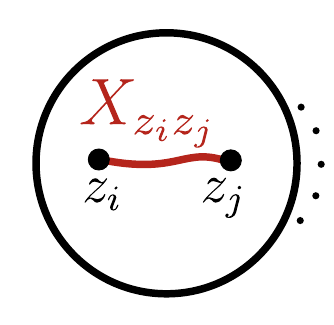}
    \caption{Kinematic $X$-variables can be associated with curves on the \textit{kinematic surface}. From left to right: examples for the tree-level $X$-variables \eqref{defX} and the loop-level $X$-variables \eqref{defXloop1}.}
    \label{fig:SurfVars1}
\end{figure}

In \eqref{defXloop1} we chose to denote by $z_i$ the $X$-label corresponding to loop momentum $\ell_i$, adopting the notational convention established in the context of the ``surfaceology'' framework \cite{Arkani-Hamed:2023lbd,Arkani-Hamed:2023mvg,Arkani-Hamed:2024tzl}. There, loop variables $z_i$ are identified with punctures on the interior of a disk, the \textit{kinematic surface}, whereas external momenta are associated with punctures on its boundary. Variables $X_{i,j}$ can then be interpreted as curves between punctures $i$ on the boundary and/or interior punctures $z_i$ of the kinematic surface as shown in Figure \ref{fig:SurfVars1}. 

\begin{figure}[H]
    \centering
    \includegraphics[width=2.3cm]{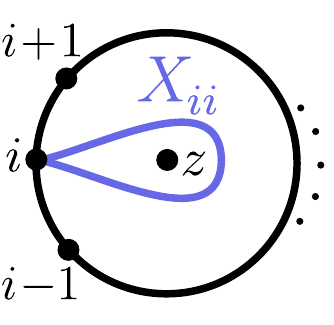}
    \hspace{1.2cm}
    \includegraphics[width=2.3cm]{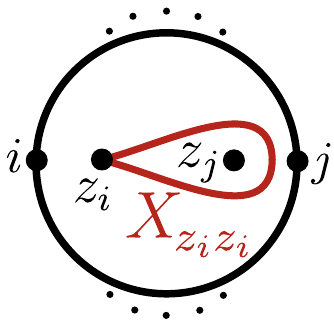}
    \hspace{1.2cm}
    \includegraphics[width=2.7cm]{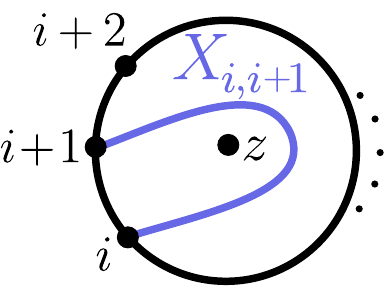}
    \caption{Kinematic tadpole variables $X_{ii}$, $X_{z_iz_i}$ and off-shellness variables $X_{i,i+1}$ have a natural interpretation as curves on the kinematic surface involving one or more loop punctures $z_i$.}
    \label{fig:SurfVars2}
\end{figure}

The surfaceology framework offers a natural extension of kinematic variables beyond those introduced so far which will be indispensable to define the inverse KLT loop integrand. These additional \textit{tadpole} and \textit{off-shellness} variables have a natural interpretation as curves on the kinematic surface as shown in Figure \ref{fig:SurfVars2}.

\begin{figure}[t]
    \centering
    \includegraphics[width=2.8cm]{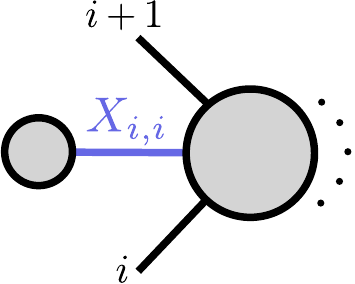}
    \hspace{1.2cm}
 $   \begin{matrix}
        \vspace{1.7cm}
        \includegraphics[width=3.4cm]{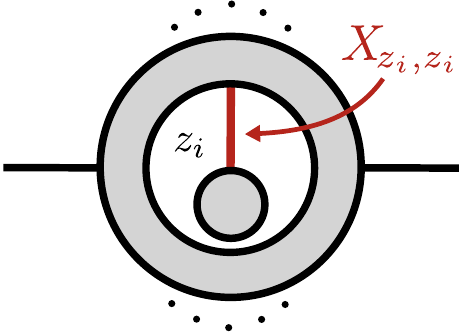}    
   \end{matrix}$
    \hspace{1.2cm}
    \includegraphics[width=3.9cm]{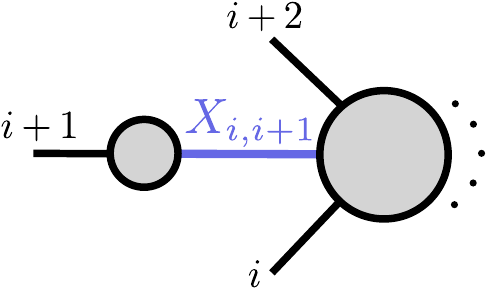}
    \vspace{-1.7cm}
    \caption{Diagrammatic interpretation of kinematic tadpole variables $X_{ii}$, $X_{z_iz_i}$ and off-shellness variables $X_{i,i+1}$ as tadpole and bubble corrections on external and internal lines.}
    \label{fig:DiagSurfVars2}
\end{figure}

They are associated with the appearance of 1-point and 2-point functions as corrections to external and internal lines at the loop level as shown diagrammatically in Figure \ref{fig:DiagSurfVars2}.
However, while the variables $X_{i,i+1}\,{=}\,p_i^2$ can still be meaningfully associated with the off-shellness of an external particle via \eqref{defX}, the tadpole variables $X_{i,i}$ and $X_{z_i,z_i}$ cannot be interpreted in terms of momenta $p_i$ and loop momenta $\ell_i$ at all. Instead, they should be considered as formal variables that regularize the divergences associated with diagrams involving the propagators shown in Figure \ref{fig:DiagSurfVars2}. Nevertheless, the inverse KLT integrand we will construct here crucially depends on all of these variables for its consistency, making it naturally a function defined on the kinematic surface.

\begin{figure}[H]
    \centering
    \includegraphics[width=2.4cm]{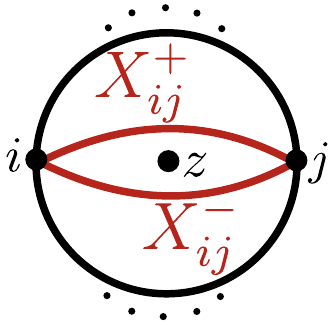}
    \caption{Homologically distinct curves $X_{ij}^{\pm}$ on the kinematic surface cannot be continuously deformed into each other in the presence of a loop puncture $z_i$. For most of our construction we identify both $X_{ij}^{\pm}$ as corresponding to the same physical channel $X_{ij}$.}
    \label{fig:curveHom}
\end{figure}

To define the inverse KLT integrand, we need to highlight one further peculiarity of the surface picture at the loop level. Since, by definition, kinematic variables correspond to \textit{homologically inequivalent} curves on the surface \cite{Arkani-Hamed:2024nhp}, there is a proliferation of $X$-variables associated to a given physical channel. For instance, at one loop, the tree-level variables $X_{ij}$ can be distinguished into $X_{ij}^{+}$ and $X_{ij}^{-}$ according to whether they correspond to curves that pass the loop puncture $z$ on the left/right as shown in Figure \ref{fig:curveHom}.
In the construction of the KLT integrand provided subsequently, we will (mostly) choose not to make this distinction, mainly for the sake of readability. Instead, we identify all $X$-variables associated to the same channel. However, the integrands we consider can be readily supplemented with homology information if needed.

\subsubsection*{Surface structure of the effective root vertex}
Before we get into the explicit construction of the inverse KLT integrand, we first need to understand how the effective root vertex \eqref{rootValphaExp} of the cubic Berends-Giele recursion is sensitive to particular properties of the kinematic surface.

We start by noting the simple fact that a given Berends-Giele diagram can be canonically associated with a tiling of the kinematic surface into regions. For instance, taking an $n$-point Berends-Giele diagram from the tree-level recursion \eqref{malphaBGschem}, the corresponding tiling of the kinematic surface can be obtained simply by drawing the curves associated to the left- and right-hand propagators $X_{L/R}$ of the diagram as follows,
\begin{align}\label{surfDualTree}
\begin{split}
    \begin{matrix}
         \hspace{0.1cm}\vspace{-0.05cm}\includegraphics[width=6.2cm]{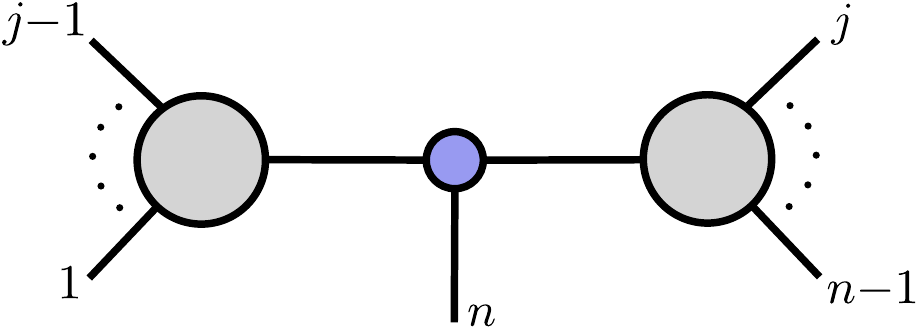}
    \end{matrix}
    &\simeq\hspace{0.2cm}
    \begin{matrix}
         \hspace{0.1cm}\vspace{-0.05cm}\includegraphics[width=2.9cm]{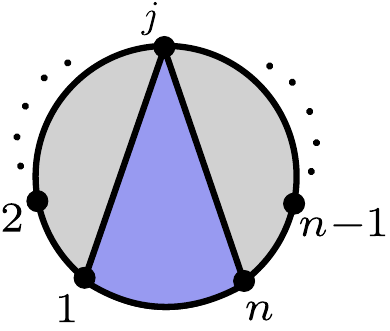}
    \end{matrix}\\
    &\simeq
    \frac{1+t_{1j}t_{jn}}{t_{1j}t_{jn}} m_L^{\alpha'}\!(1\dots j) m_R^{\alpha'}\!(j\dots n).
\end{split}
\end{align}
In this way we construct what is called the \textit{dual} of the diagram on the surface, which, in principle, contains the same information as the original Berends-Giele diagram. The three regions of the surface dual canonically correspond to the effective root vertex as well as the left/right-hand sub-amplitudes of the original diagram.
For the generic example in \eqref{surfDualTree} we have chosen the case where there are at least two external lines connecting to each left/right sub-amplitude (i.e. $j\,{=}\,3,\dots, n\,{-}\,2$) such that the momentum invariants $X_{L/R}$ appearing in the propagators of the Berends-Giele diagram can be considered off-shell. 

Now, a consistent definition of Berends-Giele diagrams involving ``on-shell'' momenta, i.e. diagrams where only one external leg connects to a given side of the effective vertex, requires the particular structure of the kinematic surface. For concreteness, let us consider the tree-level diagram in \eqref{surfDualTree} where the right-hand propagator is on-shell (i.e. $j\,{=}\,n\,{-}\,1$). Drawing the dual diagram on the kinematic surface, we observe that the curve corresponding to the right-hand propagator can be continuously deformed to the boundary segment between punctures $n$ and $n\,{-}\,1$, making the two curves homologically equivalent,
\begin{align}\label{surfDualTreeOS}
    \hspace{-0.2cm}
    \begin{matrix}
         \hspace{0.1cm}\vspace{-0.25cm}\includegraphics[width=5.6cm]{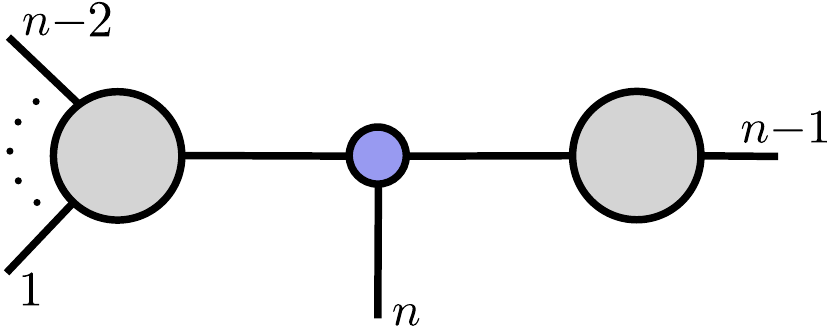}
    \end{matrix}
    \hspace{0.2cm}\simeq\hspace{0.2cm}
    \begin{matrix}
         \hspace{0.1cm}\vspace{-0.05cm}\includegraphics[width=2.9cm]{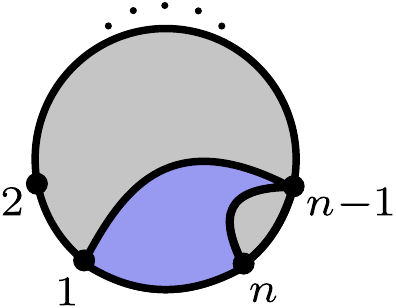}
    \end{matrix}
    \hspace{0.2cm}\simeq\hspace{0.2cm}
    \begin{matrix}
         \hspace{0.1cm}\vspace{-0.05cm}\includegraphics[width=2.9cm]{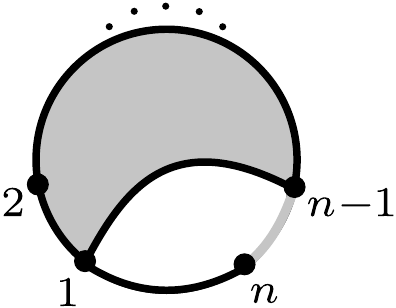}
    \end{matrix}.
\end{align}
Identifying the homologous curves as shown above can be interpreted effectively as collapsing the surface region of the right-hand sub-amplitude into a curve by deforming the curve corresponding to the right-hand propagator to the boundary curve ($n\,{-}\,1\, \!,\! n$). We will now instate the rule that, \textit{whenever a region on the surface can be collapsed in such a way, the effective root vertex similarly collapses} into that of the tr$\phi^3$ theory, 
\begin{align}\label{surfDualTreeOS2}
    \begin{matrix}
         \hspace{0.1cm}\vspace{-0.05cm}\includegraphics[width=2.9cm]{figures/surfTilingTreeOS2.pdf}
    \end{matrix}
    \hspace{0.2cm}\simeq\hspace{0.2cm}
    \begin{matrix}
         \hspace{0.1cm}\vspace{-0.25cm}\includegraphics[width=4.1cm]{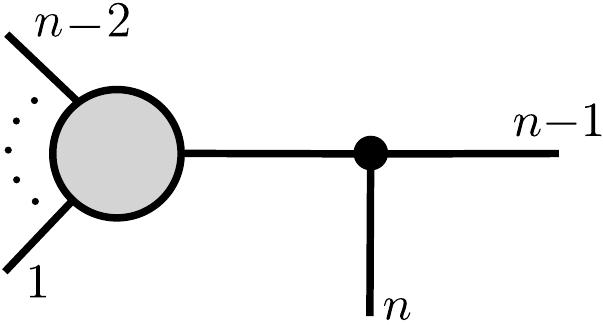}
    \end{matrix}\,  = \frac{1}{t_{1,n-1}}m_L^{\alpha'}\!(1 \dots n\,{-}\,1).
\end{align}
This provides a surface interpretation for the appearance of the tr$\phi^3$ vertex in the Berends-Giele diagrams involving an on-shell leg in the tree-level recursion \eqref{malphaBGschem}.

Let us now anticipate what the homology-sensitive definition of the effective root vertex implies for loop integrands. The loop-level recursion will include Berends-Giele diagram topologies of the type \eqref{surfDualTree} which, however, now involve lower-point sub-integrands on either side of the root vertex. In particular, let us focus on the one-loop analogue of the diagram \eqref{surfDualTreeOS}, except that the right-hand sub-integrand is now itself of one loop order as shown in \eqref{surfDualLoopOS}. In the dual picture this situation corresponds to a tiling of the kinematic surface into regions identical to \eqref{surfDualTreeOS} except that the right-hand region now contains a puncture representing the loop-dependence of the sub-integrand,
\begin{align}\label{surfDualLoopOS}
    \begin{matrix}
         \hspace{0.1cm}\vspace{-0.25cm}\includegraphics[width=5.8cm]{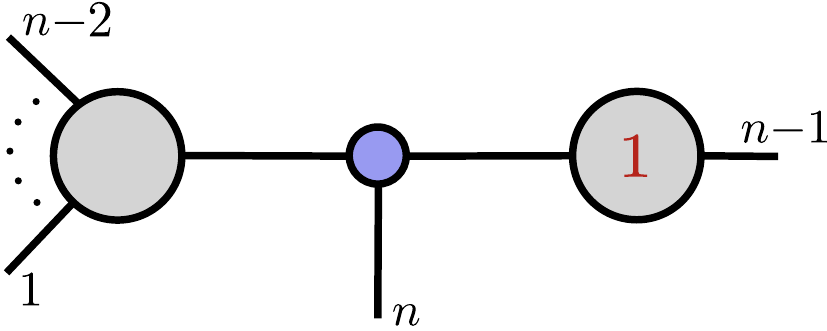}
    \end{matrix}
    \hspace{0.2cm}&\simeq\hspace{0.2cm}
    \begin{matrix}
         \hspace{0.1cm}\vspace{-0.05cm}\includegraphics[width=2.9cm]{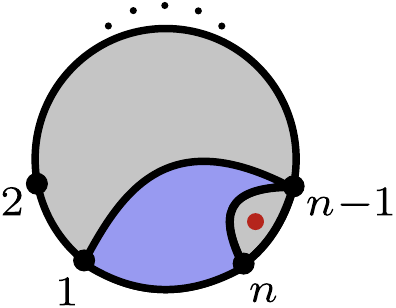}
    \end{matrix}\\
    &\simeq \frac{1+t_{1,n-1}t_{n-1,n}}{t_{1,n-1}t_{n-1,n}} m_L^{\alpha'}(1\dots n\,{-}\,1) m^{\alpha'}_{\xRed{1},R}(n\,{-}\,1\, n).  \nonumber  
\end{align}
The appearance of the puncture in the dual picture is crucial as the curve corresponding to the right-hand propagator is \textit{no longer homologous} to the boundary. As such, we are not entitled to collapse the right-hand region the same way we did in \eqref{surfDualTreeOS} and the full, $\alpha'$-dependent structure of the effective root vertex \eqref{rootValphaExp} remains intact.

Having laid out the subtleties related to the sensitivity of the effective root vertex to homology on the kinematic surface, we are now fully equipped to define the inverse KLT integrand at one loop and beyond.

\subsection{One-Loop Recursion}
We will first consider the case of the one-loop ($\xRed{L}\,{=}\,1$) integrands $m_{\xRed{1},n}^{\alpha'}$ for the inverse string theory KLT kernel. This gives us an opportunity to discuss the basic idea behind generalizing the cubic Berends-Giele recursion to the loop-level, and to understand some of the new features encountered beyond tree-level in as simple a setting as possible.

The generalization to loop integrands is most easily understood at the level of Berends-Giele diagrams. Just as for the tree-level recursion \eqref{malphaBGschem} before, there will be certain contributions to the loop-level recursion where the effective root vertex \eqref{rootValphaExp} is in a tree-level configuration, for example,
\begin{align}\label{BG1loopTreeTop}
    m_{\xRed{1},n}^{\alpha'} \,\supset \,
    \begin{matrix}
         \hspace{0.1cm}\vspace{-0.05cm}\includegraphics[width=5.0cm]{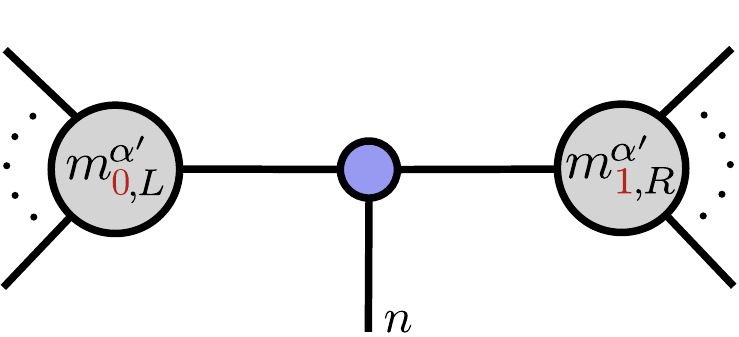}
    \end{matrix}
    \,\,\,+\,
    \begin{matrix}
         \hspace{0.1cm}\vspace{-0.05cm}\includegraphics[width=5.0cm]{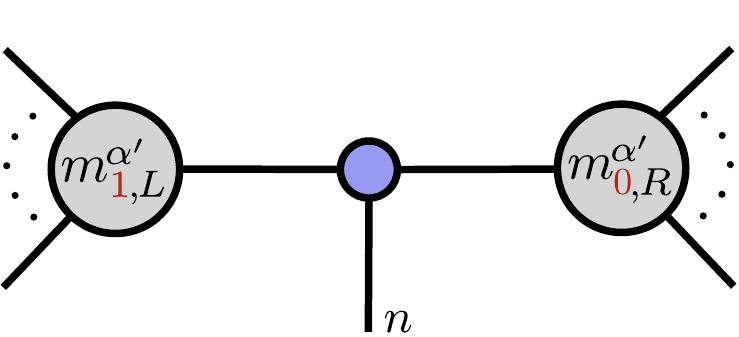}
    \end{matrix}
\end{align}
This is a straightforward generalization of the tree-level diagram topology except that now one of the left- or right-hand sub-amplitudes is a lower-point one-loop integrand $m_{\xRed{1},L/R}^{\alpha'}$. Going forward, we will also use the notation $m_{\xRed{0},n}^{\alpha'} \,{\equiv}\, m_{n}^{\alpha'}$ for the tree-level inverse KLT matrix elements.

However, there is also an entirely new Berends-Giele diagram topology where the effective root vertex itself is fused into a loop,
\begin{align}\label{BG1loopLoopTop}
    m_{\xRed{1},n}^{\alpha'} \,\supset \!
    \begin{matrix}
         \hspace{0.1cm}\vspace{-0.05cm}\includegraphics[width=2.1cm]{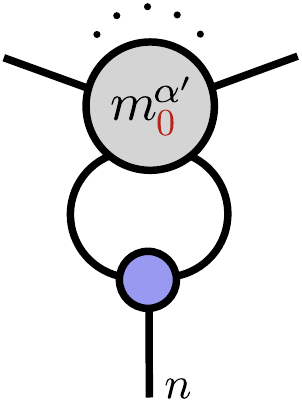}
    \end{matrix}.
\end{align}
At $n$ points, this topology involves a tree-level matrix element $m_{\xRed{0},n{+}1}^{\alpha'}$ at \textit{higher} multiplicity ($n\,{+}\,1$) but at \textit{lower} loop-order ($\xRed{L}\,{=}\,0$) which therefore does not spoil the recursion.

In setting up the recursion for the matrix elements $m_{\xRed{1},n}^{\alpha'}$ we will discuss the special cases of integrands with $n\,{=}\,1,2,3$ separately as they are slightly degenerate owing to their low multiplicity. Only then will we proceed to the general case where $n\,{\ge}\, 4$ and a discussion of the novel structures encountered for the inverse KLT integrand at one loop.

\subsubsection*{Tadpole (n=1)}
Due to the existence of the loop topology \eqref{BG1loopLoopTop} we encounter the first non-zero inverse KLT integrand already for $n\,{=}\,1$. The one-loop tadpole is given by just a single Berends-Giele diagram,
\begin{align}\label{BG1loop1}
    m_{\xRed{1},1}^{\alpha'}(1) = 
    \begin{matrix}
         \hspace{0.1cm}\vspace{-0.05cm}\includegraphics[width=1.05cm]{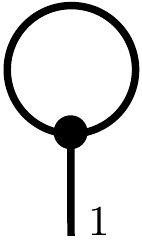}
    \end{matrix}
    \,= \frac{1}{t_{1z}}\,.
\end{align}
Equivalently, this is the only Feynman diagram following from the Feynman rules \eqref{KLTFeynRules}.

More importantly, however, we see that even for the simplest example of the one-loop tadpole the dual description of the effective root vertex is essential. This is because in the diagram \eqref{BG1loop1} the effective root vertex appears collapsed. Indeed, this is consistent with the vertex rule based on homology of curves on the surface, which can be understood by considering the following Berends-Giele diagram (and its dual),
\begin{align}\label{surfDualTadpOS}
    \hspace{-0.2cm}
    \begin{matrix}
         \hspace{0.1cm}\vspace{-0.05cm}\includegraphics[width=1.2cm]{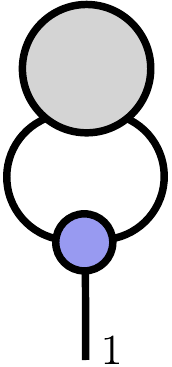}
    \end{matrix}
    \hspace{0.2cm}\simeq\hspace{0.2cm}
    \begin{matrix}
         \hspace{0.1cm}\vspace{-0.25cm}\includegraphics[width=1.9cm]{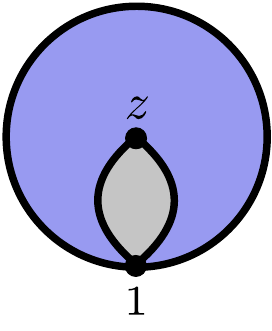}
    \end{matrix}
    \hspace{0.2cm}\simeq\hspace{0.2cm}
    \begin{matrix}
         \hspace{0.1cm}\vspace{-0.25cm}\includegraphics[width=1.9cm]{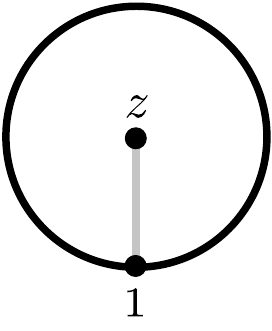}
    \end{matrix}\, .
\end{align}
As is apparent from the surface dual of the diagram, the curves corresponding to left- and right-hand propagators are homologous. Thus, according to the rule \eqref{surfDualTreeOS}, the region on the kinematic surface corresponding to the two-point sub-amplitude as well as the effective root vertex can be collapsed and we arrive at the contribution in \eqref{BG1loop1}.

Having shown how to analyze the behavior of the effective root vertex under homology for various examples, we will not continue belaboring this point in detail for each Berends-Giele diagram. Instead, in the remainder of this article, we simply state the Berends-Giele diagrams contributing to a given integrand with the homology rules for the effective vertex already implicitly implemented.

\subsubsection*{Bubble (n=2)}
We continue with the two-point (bubble) integrand which consists of three contributions,
\begin{align}\label{BG1loop2}
    m_{\xRed{1},2}^{\alpha'}(12) &=
    \begin{matrix}
         \hspace{-0.0cm}\vspace{-0.8cm}\includegraphics[width=3.0cm]{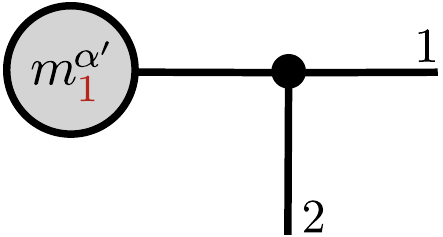}
    \end{matrix}
    \,\,+\,\,
    \begin{matrix}
         \hspace{-0.0cm}\vspace{-0.8cm}\includegraphics[width=3.0cm]{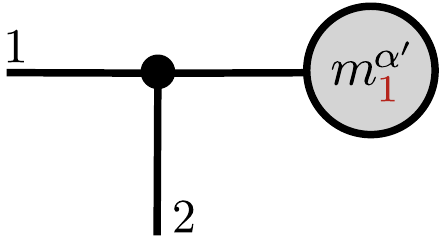}
    \end{matrix}
    \,\,+\,\,
    \begin{matrix}
         \hspace{-0.0cm}\vspace{-0.10cm}\includegraphics[width=1.2cm]{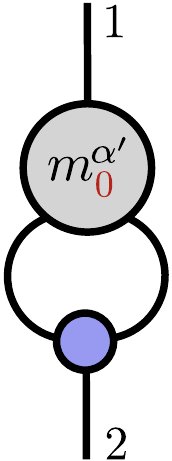}
    \end{matrix}
    \\
    &= \frac{1}{t_{11}}m_{\xRed{1},1}^{\alpha'}(1) +\frac{1}{t_{22}}m_{\xRed{1},1}^{\alpha'}(2) + \frac{1+t_{1z}t_{2z}}{t_{1z}t_{2z}}m_{\xRed{0},3}^{\alpha'}(12z) = \frac{1}{t_{11}t_{1z}} +\frac{1}{t_{22}t_{2z}} + \frac{1}{t_{1z}t_{2z}} + 1. \nonumber
\end{align}
Here we observe an interesting phenomenon. Similar to what we saw at tree-level, the effective root vertex in the loop topology now generates a \textit{loop-level} contact interaction $\sim\! 1$. However, this contact term \textit{does not} originate from the inverse KLT Lagrangian \eqref{defKLTLag}. Indeed, computing the integrand from Feynman diagrams we get contributions only from cubic diagrams
\begin{align}
    m_{\xRed{1},2,\text{Feyn}}^{\alpha'} = 
    \begin{matrix}
         \hspace{-0.0cm}\vspace{-0.0cm}\includegraphics[width=2.1cm]{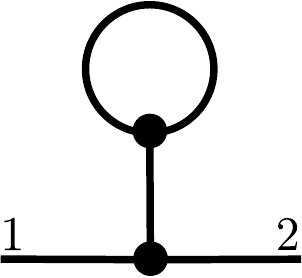}
    \end{matrix}
    \,\,+\,\,
    \begin{matrix}
         \hspace{-0.0cm}\vspace{-0.2cm}\includegraphics[width=2.1cm]{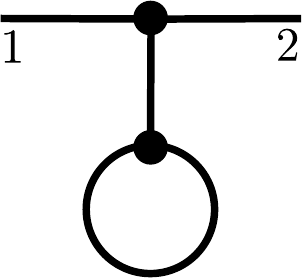}
    \end{matrix}
    \,\,+\,\,
    \begin{matrix}
         \hspace{-0.0cm}\vspace{-0.10cm}\includegraphics[width=2.7cm]{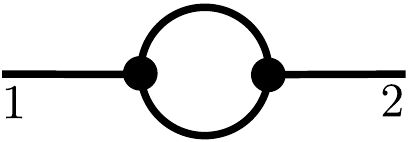}
    \end{matrix}\, .
\end{align}
While these diagrams look to be in one-to-one correspondence with the Berends-Giele diagrams in \eqref{BG1loop2} they only involve the three-point vertex of the inverse KLT Lagrangian. In particular, they cannot make use of the effective root vertex \eqref{rootValphaExp}. For this reason we find a discrepancy
\begin{align}\label{Feyn1loop2}
    m_{\xRed{1},2}^{\alpha'} - m_{\xRed{1},2,\text{Feyn}}^{\alpha'} = 1 = \,
    \begin{matrix}
         \hspace{-0.0cm}\vspace{0.2cm}\includegraphics[width=1.8cm]{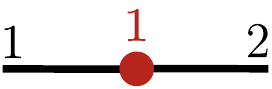}\,,
    \end{matrix}
\end{align}
which we denote diagrammatically as an inherently one-loop, two-point contact interaction. This interaction is generated naturally by the cubic Berends-Giele recursion \eqref{BG1loop2} but is not present when computing the integrand from Feynman diagrams. Compared to the cubic diagrams in \eqref{Feyn1loop2} it is sub-leading in $\alpha'$ and therefore does not contribute in the field theory limit $\alpha' \to 0$. Curiously, the contact term also does not contribute to the integrated matrix element as it is scaleless,
\begin{align}
    \int\! \text{d}^d\ell \,\, \begin{matrix}
         \hspace{-0.0cm}\vspace{0.2cm}\includegraphics[width=1.8cm]{figures/2pt1loopCt.pdf}
    \end{matrix} \,=\! \int\! \text{d}^d\ell \,\, 1  \simeq 0,
\end{align}
and should therefore be considered a feature of the \textit{integrand} per se.

Nonetheless, the occurrence of inherently loop-level contact terms will turn out to be a highly characteristic feature of the inverse KLT integrand, also at higher multiplicities. We will return to study their structure and significance once we have defined the cubic one-loop recursion for the general $n$-point case.

\subsubsection*{Three Points}
Moving on to the three-point inverse KLT integrand, there are now five contributions to the cubic Berends-Giele recursion,
\begin{align}\label{BG1loop3}
\begin{split}
    m_{\xRed{1},3}^{\alpha'}(123) 
    &= \frac{1+t_{11}t_{13}}{t_{11}t_{13}}m^{\alpha'}_{\xRed{1},1}(1)m^{\alpha'}_{\xRed{0},3}(123) + \frac{1+t_{13}t_{33}}{t_{13}t_{33}}m^{\alpha'}_{\xRed{0},3}(123)m^{\alpha'}_{\xRed{1},1}(3) \\
    &+ \frac{1}{t_{12}}m_{\xRed{1},2}^{\alpha'}(12) +\frac{1}{t_{23}}m_{\xRed{1},2}^{\alpha'}(23)  + \frac{1+t_{1z}t_{3z}}{t_{1z}t_{3z}}m^{\alpha'}_{\xRed{0},4}(123z)\\
    &= \frac{1}{3}\frac{1}{t_{1z}t_{2z}t_{3z}} + \frac{1}{t_{12}t_{1z}t_{2z}} + \frac{1}{t_{11}t_{12}t_{1z}} + \frac{1}{t_{11}t_{13}t_{1z}}  + \frac{1}{t_{1z}} + \frac{1}{t_{12}} + \text{cyc.}
\end{split}
\end{align}
As before, we can compare the inverse KLT integrand above to the one obtained from Feynman diagrams to find
\begin{align}\label{Feyn1loop3}
    m_{\xRed{1},3}^{\alpha'} - m_{\xRed{1},3,\text{Feyn}}^{\alpha'} = \frac{1}{t_{12}} + \text{cyc.} = \,
    \begin{matrix}
         \hspace{-0.0cm}\vspace{0.2cm}\includegraphics[width=1.8cm]{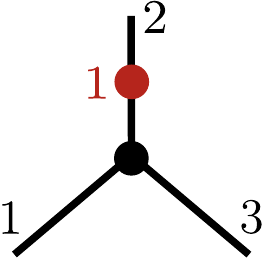}
    \end{matrix}
    \!+ \text{cyc.}
\end{align}
We see that the inverse KLT integrand computed from the Berends-Giele recursion consistently incorporates the two-point interaction \eqref{Feyn1loop2} as a 1-loop correction to the external propagators of the tree-level topology. The insertion of the two-point interaction causes the tree-level diagram in \eqref{BG1loop3} to be formally of one-loop order.

\subsubsection*{General case (n \!$\ge$\! 4)}
Having gathered some intuition for the structure of the loop-level Berends-Giele recursion, we move on to describe the all-multiplicity case where $n\,{\ge}\, 4$. To make sense of the various kinds of contributions in the recursion, it will be convenient to split the Berends-Giele diagrams up into four categories. Firstly, we distinguish diagrams for which the root vertex is in a tree configuration as in \eqref{BG1loopTreeTop}, or fused into a loop as in \eqref{BG1loopLoopTop}. The tree-topologies are further split up according to whether they involve tadpoles ($1$-point integrands), bubbles ($2$-point integrands) or whether both left- and right-hand integrands have multiplicity ${\ge}\, 3$. The cubic Berends-Giele recursion for the $n$-point one-loop inverse KLT integrand can therefore be decomposed as
\begin{align}\label{BG1loopContrib}
    m_{\xRed{1},n}^{\alpha'} = m_{\xRed{1},n,(1)}^{\alpha'} + m_{\xRed{1},n,(2)}^{\alpha'} + m_{\xRed{1},n,(\ge 3)}^{\alpha'} + m_{\xRed{1},n,(\text{loop})}^{\alpha'}.
\end{align}

Let us take the individual contributions \eqref{BG1loopContrib} in turn and consider which Berends-Giele diagrams contribute. Starting with the tadpole contributions, there are only two diagrams,
\begin{align}\label{BG1loopGen1}
\begin{split}
    m_{\xRed{1},n,(1)}^{\alpha'} &=
    \begin{matrix}
         \hspace{-0.0cm}\vspace{-0.3cm}\includegraphics[width=4.6cm]{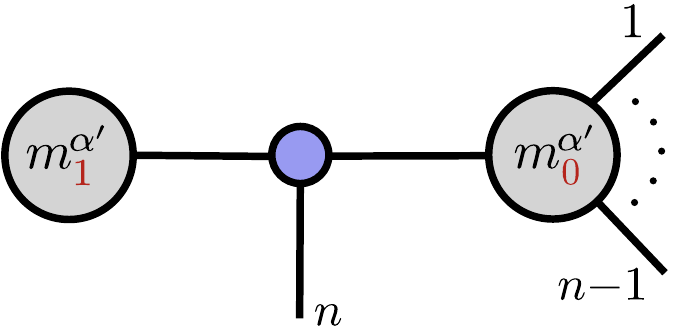}
    \end{matrix} \,+\,
    \begin{matrix}
         \hspace{-0.0cm}\vspace{-0.3cm}\includegraphics[width=4.6cm]{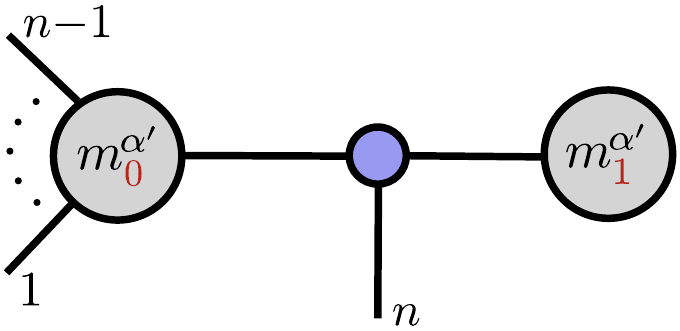}
    \end{matrix}\\
    &= \frac{1+t_{11}t_{1n}}{t_{11}t_{1n}} m_{\xRed{1},1}^{\alpha'}(1)m_{\xRed{0},n}^{\alpha'}(1\dots n) + \frac{1+t_{1n}t_{nn}}{t_{1n}t_{nn}} m_{\xRed{0},n}^{\alpha'}(1\dots n)m_{\xRed{1},1}^{\alpha'}(n).
\end{split}
\end{align}
Moving on to bubble contributions we have four Berends-Giele diagrams,
\begin{align}\label{BG1loopGen2}
    m_{\xRed{1},n,(2)}^{\alpha'} &= 
    \begin{matrix}
         \hspace{-0.0cm}\vspace{-0.35cm}\includegraphics[width=4.0cm]{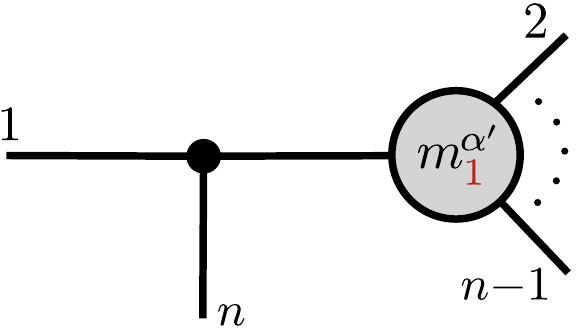}
    \end{matrix} \,+\,
    \begin{matrix}
         \hspace{-0.0cm}\vspace{-0.35cm}\includegraphics[width=4.0cm]{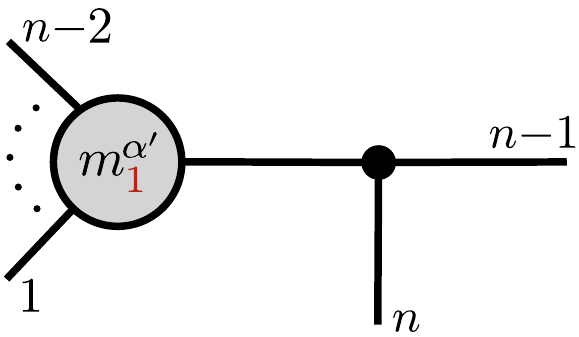}
    \end{matrix} \nonumber
    \\
    &+
    \begin{matrix}
         \hspace{-0.0cm}\vspace{-0.3cm}\includegraphics[width=4.6cm]{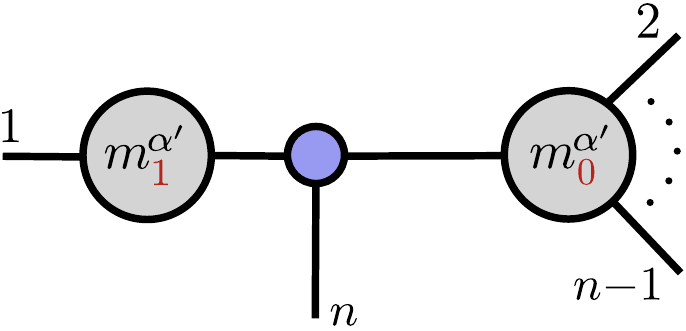}
    \end{matrix} \,+\,
    \begin{matrix}
         \hspace{-0.0cm}\vspace{-0.3cm}\includegraphics[width=4.6cm]{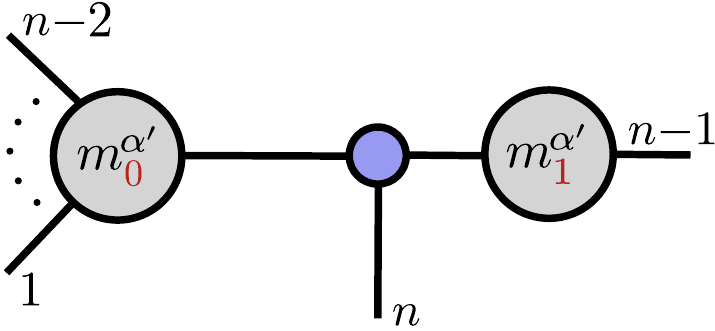}
    \end{matrix}\\
    &= \frac{1}{t_{2n}} m_{\xRed{1},n-1}^{\alpha'}(2\dots n) + \frac{1}{t_{1n-1}} m_{\xRed{1},n-1}^{\alpha'}(1\dots n{-}1) \nonumber \\
    &+ \frac{1+t_{12}t_{2n}}{t_{12}t_{2n}} m_{\xRed{1},2}^{\alpha'}(12)m_{\xRed{0},n-1}^{\alpha'}(2\dots n) + \frac{1+t_{1n-1}t_{n-1n}}{t_{1n-1}t_{n-1n}} m_{\xRed{0},n-1}^{\alpha'}(1\dots n{-}1)m_{\xRed{1},2}^{\alpha'}(n{-}1\,n). \nonumber
\end{align}
The first two diagrams are a straightforward generalization of the contributions in the three-level recursion \eqref{BGKLT} where one of the legs in the effective vertex is on-shell. However, at one loop, there are additional contributions from diagrams with a one-loop bubble integrand $m_{\xRed{1},2}^{\alpha'}$ sitting on the external line. In this case, the full $\alpha'$-dependence of the effective root comes to bear due to the homology argument made in \eqref{surfDualLoopOS}.

Next, we obtain contributions from diagrams without tadpoles and bubbles where the root vertex is in a tree-level configuration \eqref{BG1loopTreeTop},
\begin{align}\label{BG1loopGen3}
    \hspace{-0.25cm}m_{\xRed{1},n,(\ge 3)}^{\alpha'} &= 
    \sum_{j=3}^{n-2} \left(\,
    \begin{matrix}
         \hspace{-0.0cm}\vspace{-0.0cm}\includegraphics[width=5.0cm]{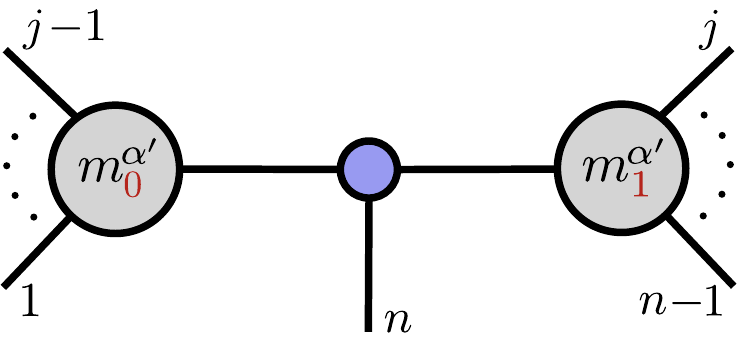}
    \end{matrix} \,+\,
    \begin{matrix}
         \hspace{-0.0cm}\vspace{-0.0cm}\includegraphics[width=5.0cm]{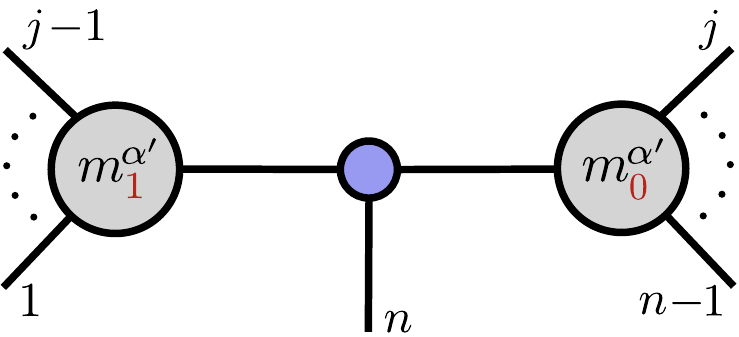}
    \end{matrix}
    \,\right) \\
    &= \sum_{j=3}^{n-2}\frac{1+t_{1j}t_{jn}}{t_{1j}t_{jn}} \!\left\lbrace\!  m_{\xRed{0},j}^{\alpha'}(1\dots j)m_{\xRed{1},n-j+1}^{\alpha'}(j\dots n)\! + m_{\xRed{1},j}^{\alpha'}(1\dots j)m_{\xRed{0},n-j+1}^{\alpha'}(j\dots n) \!\right\rbrace\!. \nonumber
\end{align}
Note that these contributions are non-trivial only when $n\,{\ge}\, 5$. The subscript ${}_{({\ge}\, 3)}$ in $m_{\xRed{1},n,(\ge 3)}^{\alpha'}$ just refers to the fact that the corresponding Berends-Giele diagrams have left- and right-hand integrands with multiplicity $\ge 3$, i.e. they do not involve tadpoles or bubbles.

We can recognize the diagrams in \eqref{BG1loopGen3} as a direct generalization of the tree-level contributions in \eqref{BGKLT}. A novel feature at one loop is that we have to take into account the two ways of distributing the loop-dependence (for fixed $j$) between the left- and the right-hand lower-point integrand.

Finally we have a single contribution with the root vertex fused into a loop as in \eqref{BG1loopLoopTop},
\begin{align}\label{BG1loopGenLoop}
\begin{split}
    m_{\xRed{1},n,(\text{loop})}^{\alpha'} &= 
    \begin{matrix}
         \hspace{-0.0cm}\vspace{-0.4cm}\includegraphics[width=2.15cm]{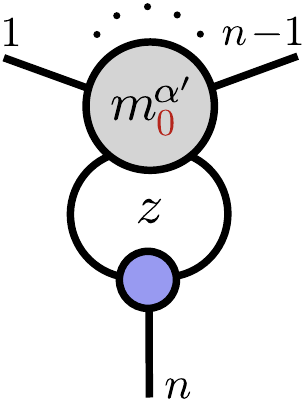}
    \end{matrix}
    = \frac{1+t_{1z}t_{nz}}{t_{1z}t_{nz}}  m_{\xRed{0},n{+}1}^{\alpha'}(1\dots n z).
\end{split}
\end{align}
Combining all contributions to the recursion as in \eqref{BG1loopContrib} we obtain a total of $2n\,{-}\,1$ terms. This shows that, compared to $n\,{-}\,2$ terms at tree-level, the computational difficulty for the one-loop KLT integrand increases only marginally and is still $\mathcal{O}(n)$. Again, this is exactly the same number of contributions as in the Berends-Giele recursion for the field theory integrand in the $\text{tr}\phi^3$ theory. This is of course no coincidence as the recursion relation \eqref{BG1loopContrib} for the KLT integrand manifestly reduces, term by term, to the Berends-Giele recursion relation for the field theory $\text{tr}\phi^3$ integrand
\begin{align}\label{KLT1loopAlphaZero}
    m_{\xRed{1},n}^{\alpha'} = (\pi\alpha')^{-n} \big( m_{\xRed{1},n} + \mathcal{O}(\alpha'^{2})\big)
\end{align}
The integrand $m_{\xRed{1},n}$ obtained in this limit agrees exactly with the one obtained from the field theory limit of the surfacehedron \cite{Arkani-Hamed:2023lbd,Arkani-Hamed:2023mvg}.

\subsubsection*{Example (n=4)}

Using the recursion \eqref{BG1loopContrib} we can compute the four-point inverse KLT integrand,
\begin{align}\label{BG1loop4}
\begin{split}
    m_{\xRed{1},4}^{\alpha'} &= \frac{1}{4}\frac{1}{t_{1z}t_{2z}t_{3z}t_{4z}} +\frac{1}{t_{1z}t_{2z}t_{3z}}\frac{1}{t_{13}} + \frac{1}{t_{1z}t_{2z}}\! \left\lbrace  \!\frac{1}{t_{12}}\!\left(\! \frac{1}{t_{13}} \,{+}\,\frac{1}{t_{24}} \!\right) + 1\! \right\rbrace + \frac{1}{2}\frac{1}{t_{1z}t_{3z}} \frac{1}{t_{13}^2} \\
    &+ \frac{1}{t_{1z}}\left\lbrace\! \frac{1}{t_{11}} \!\left(\! \frac{1}{t_{12}t_{13}} + \frac{1}{t_{13}t_{14}} + \frac{1}{t_{14}t_{24}} + \frac{1}{t_{24}t_{12}} + \frac{1}{t_{13}^2} + 1  \!\right)\! + \frac{2}{t_{13}} + \frac{1}{t_{24}} \!\right\rbrace \\
    &+ \frac{1}{t_{12}}\!\left(\! \frac{1}{t_{13}} \,{+}\,\frac{1}{t_{24}} \!\right) + \frac{1}{2}\frac{1}{t_{13}^2} + \frac{3}{4} + \text{cyc.}
\end{split}
\end{align}
Here we observe that, similarly to the case of the two-point integrand \eqref{BG1loop2}, the cubic Berends-Giele recursion generates another, inherently loop-level, four-point contact term $\sim\! 3$. Furthermore, the discrepancy between the four-point Feynman integrand computed from the inverse KLT Lagrangian \eqref{defKLTLag} and the inverse KLT integrand computed from the Berends-Giele recursion is given exactly by the terms in the last line of \eqref{BG1loop4}. They can be interpreted diagrammatically as follows,
\begin{align}\label{Feyn1loop4}
    m_{\xRed{1},4}^{\alpha'}-m_{\xRed{1},4,\text{Feyn}}^{\alpha'} = 
    \left(
    \begin{matrix}
        \hspace{-0.0cm}\vspace{-0.1cm}\includegraphics[width=2.2cm]{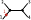}
    \end{matrix}
    +
    \begin{matrix}
        \hspace{-0.0cm}\vspace{-0.1cm}\includegraphics[width=2.2cm]{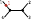}
    \end{matrix}
    +
    \frac{1}{2}\,\begin{matrix}
        \hspace{-0.0cm}\vspace{-0.1cm}\includegraphics[width=2.2cm]{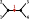}
    \end{matrix}
    + \text{cyc.} \right)
    +
    \begin{matrix}
        \hspace{-0.0cm}\vspace{-0.1cm}\includegraphics[width=1.4cm]{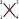}
    \end{matrix}\, ,
\end{align}
where, in addition to the four-point one-loop contact term, the two-point interaction \eqref{Feyn1loop2} appears as an insertion on external and internal propagators of tree-level diagrams.

This is indicative of a more general structure where the inverse KLT integrands computed from the Berends-Giele recursion can be identified with consistent integrands in a theory described by supplementing the tree-level inverse KLT Lagrangian \eqref{defKLTLag} with an effective one-loop Lagrangian. We will return to this observation presently when we study the general structure of the one-loop contact terms for $n \,{\ge}\, 4$. First, however, we wish to argue for the plausibility and consistency of the inverse KLT integrand generated by the recursion \eqref{BG1loopContrib} by considering its single cuts for general multiplicity $n$.

\subsubsection*{Single Cuts}
The one-loop inverse KLT integrand \eqref{BG1loopContrib} is a novel kinematic function \textit{defined} via the cubic Berends-Giele recursion relation. As such, we cannot prove its validity by comparing to any previously known results in the literature. However, we can check certain consistency conditions that the integrand must satisfy to ensure that our definitions are meaningful.

One such crucial property are the single cuts of the integrand which are known to fix the integrand up to contact terms. The explicit form of the cubic Berends-Giele recursion relations allows us to study the behavior of the inverse KLT integrand on single-cuts in a straightforward fashion.

Let us start by investigating single-cuts involving propagator with loop-dependence. That is, we want to study the residue of the $n$-point integrand $m_{\xRed{1},n}^{\alpha'}$ when e.g. $t_{1z}\,{=}\,0$. The corresponding pole is expected to involve the forward limit of the $(n\,{+}\,2)$-point tree-level function $m_{\xRed{0},n+2}^{\alpha'}$. 
As we are working with the $X$-label representation of the integrand, we can replace the residue by a derivative with respect to the corresponding propagator $t_{1z}^{-1}$, as we did in the proof of consistent factorization in \eqref{defFact}. In this case we expect the integrand to satisfy
\begin{align}\label{1loopFwdLim}
\begin{split}
    \frac{\partial}{\partial t_{1z}^{-1}} 
    \begin{matrix}
        \hspace{-0.0cm}\vspace{-0.1cm}\includegraphics[width=2.2cm]{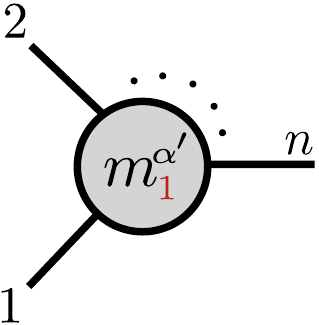}
    \end{matrix}\,\,\, &=
    \begin{matrix}
        \hspace{-0.0cm}\vspace{-0.1cm}\includegraphics[width=2.2cm]{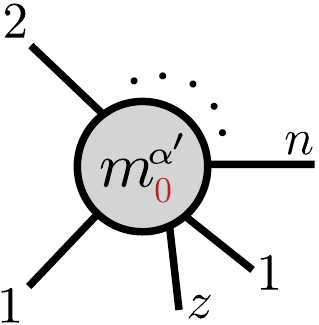}
    \end{matrix}
    \\
    \Leftrightarrow \frac{\partial}{\partial t_{1z}^{-1}} m_{\xRed{1},n}^{\alpha'}(1\dots n) &= m_{\xRed{0},n+2}^{\alpha'}(1z12\dots n), 
\end{split}
\end{align}
where both sides should be interpreted as functions of $X$-labels. The above coincides with the expected behavior of integrands defined on the kinematic surface \cite{Arkani-Hamed:2024nhp}. We can easily verify \eqref{1loopFwdLim} for the inverse KLT integrands at $n=1,2,3$ by inspection,
\begin{align}\label{1loopFwdLim123}
\begin{split}
    \frac{\partial}{\partial t_{1z}^{-1}} m_{\xRed{1},1}^{\alpha'}(1) &= 1 = m_{\xRed{0},3}^{\alpha'}(1z1), \hspace{0.7cm} \frac{\partial}{\partial t_{1z}^{-1}} m_{\xRed{1},2}^{\alpha'}(12) = \frac{1}{t_{11}} + \frac{1}{t_{2z}} = m_{\xRed{0},4}^{\alpha'}(1z12), \\
    \frac{\partial}{\partial t_{1z}^{-1}} m_{\xRed{1},3}^{\alpha'}(123) &= \frac{1}{t_{11}t_{12}} +\frac{1}{t_{11}t_{13}} + \frac{1}{t_{12}t_{2z}} + \frac{1}{t_{13}t_{3z}} + \frac{1}{t_{2z}t_{3z}} +1 = m_{\xRed{0},5}^{\alpha'}(1z123).
\end{split}
\end{align}
Note here the importance of the tadpole variables $t_{i,i}$ and off-shellness variables $t_{i,i+1}$ in regularizing the divergent behavior of certain propagators in the forward limit.
With the results from \eqref{1loopFwdLim123} we can move on to inductively prove the general case where $n\,{\ge}\, 4$. Computing the single-cut for each sub-integrand in the recursion \eqref{BG1loopContrib} according to \eqref{1loopFwdLim} (i.e. using the inductive assumption) we find after some rearranging
\begin{align}
\begin{split}
    \frac{\partial}{\partial t_{1z}^{-1}} m_{\xRed{1},n}^{\alpha'}(1\dots n) &= \frac{1}{t_{nz}} m_{\xRed{0},n+1}^{\alpha}(z12\dots n) + \frac{1}{t_{1n-1}}m_{\xRed{0},n+1}^{\alpha}(1z1\dots n\,{-}\,1)\\
    &+ \sum_{j=1}^{n-2} \frac{1+t_{1j}t_{jn}}{t_{1j}t_{jn}} m_{\xRed{0}j+2}^{\alpha'}(1z1\dots j)m_{\xRed{0},n-j+1}^{\alpha'}(j\dots n)\\
    &\hspace{-0.15cm}\overset{\eqref{BGKLT}}{\equiv} m_{\xRed{0},n+2}^{\alpha'}(1z1\dots n),
\end{split}
\end{align}
which, upon comparing to the tree-level recursion \eqref{BGKLT} evaluated on the appropriate labels, yields the forward limit as desired. This proves the consistency of the integrand with respect to the single cut \eqref{1loopFwdLim}.

At one loop, there is another type of single-cut involving the tree-level propagators $t_{1i}\,{=}\,0$. These single-cuts require a bit more care in their definition. Considering, for instance, the four-point integrand in \eqref{BG1loop4} we see that it has single- and double-poles in $t_{13}$. For the corresponding single-cut to be well-behaved these higher-order poles have to be resolved into simple poles. This can be achieved by taking into account the homology of curves on the kinematic surface as shown in Figure \ref{fig:curveHom} such that double poles $1/t_{1i}^2$ are effectively replaced with $1/t_{1i}^{+}t_{1i}^{-}$ which has only simple poles in $t_{1i}^{\pm}$. Once this distinction is made, the single cuts are well defined and the inverse KLT integrands satisfy
\begin{align}\label{1loopCutX1i}
\begin{split}
    \frac{\partial}{\partial (t_{1i}^{+})^{-1}} m_{\xRed{1},n}^{\alpha'}(1\dots n) &= m_{\xRed{0},i}^{\alpha'}(1\dots i) m^{\alpha'}_{\xRed{1},n-i+2}(i\dots n1),\\
    \frac{\partial}{\partial (t_{1i}^{-})^{-1}} m_{\xRed{1},n}^{\alpha'}(1\dots n) &= m_{\xRed{1},i}^{\alpha'}(1\dots i) m^{\alpha'}_{\xRed{0},n-i+2}(i\dots n1),
\end{split}
\end{align}
which corresponds to the canonical behavior of an integrand defined on the kinematic surface.
With the single-cuts given by \eqref{1loopFwdLim} and \eqref{1loopCutX1i} the 1-loop KLT integrand $m_{\xRed{1},n}^{\alpha'}$ is fixed up to contact terms which we will study now.

\subsubsection*{Contact terms}
The efficiency of the cubic Berends-Giele recursion is all the more remarkable as the KLT integrand turns out to receive contributions from an additional infinite tower of inherently \textit{one-loop contact terms} which, in analogy to \eqref{KLTctEvenRes} and \eqref{KLTctOddRes}, we denote by
\begin{align}\label{1loopCnt}
    \kappa_{\xRed{1},n} \equiv m_{\xRed{1},n}^{\alpha'}\Big\lvert_{\text{ct.}}.
\end{align}
We already encountered two examples of these contact terms at two and four points, \eqref{BG1loop2} and \eqref{BG1loop4} respectively, where we found $\kappa_{\xRed{1},2}\,{=}\,1$ and $\kappa_{\xRed{1},4}\,{=}\,3$.
In contrast, for the odd-point integrands at $n\,{=}\,1,3$ in \eqref{BG1loop1} and \eqref{BG1loop3} contact terms were apparently absent, $\kappa_{\xRed{1},1}\,{=}\,\kappa_{\xRed{1},3}\,{=}\,0$. It can be shown that this holds more generally and contact terms vanish at odd multiplicity, i.e. $\kappa_{\xRed{1},2k+1}\,{=}\,0$.

Thanks to the recursive nature of the inverse KLT integrand, the non-vanishing contact terms \eqref{1loopCnt} at even multiplicity ($n\,{=}\,2k$) can be determined recursively in terms of the tree-level interactions $\kappa_{\xRed{0},n}$ in \eqref{KLTFeynRules},
\begin{align}\label{BG1loopContRec}
    \kappa_{\xRed{1},2k} = \sum_{j=1}^{k-1}\left( \kappa_{\xRed{0},2j+1}\,\kappa_{\xRed{1},2(k-j)} + \kappa_{\xRed{1},2j} \,\kappa_{\xRed{0},2(k-j)+1} \right) + \kappa_{\xRed{0},2k+1}.
\end{align}
This relation can be obtained directly from the recursion \eqref{BG1loopContrib} for loop integrands $m_{\xRed{1},2k}^{\alpha'}$ by an analogous procedure to the one used to find the explicit form of the tree-level contact terms in \eqref{KLTctEven} and \eqref{KLTctOdd}.
Since the tree-level contact terms are given by $\kappa_{\xRed{0},2j+1}=C_{j-1}$, this recursion can be solved for $\kappa_{\xRed{1},2k}$ as a (rather complicated) polynomial in the Catalan numbers $C_i$. However, explicitly calculating the first few non-vanishing contact terms up to $n\,{=}\,10$ we observe that
\begin{align}\label{KLT1loopCt}
    \hspace{-0.35cm}
    \begin{matrix}
        \hspace{-0.0cm}\vspace{-0.1cm}\includegraphics[width=1.4cm]{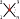}
    \end{matrix}
    =\kappa_{\xRed{1},n} =
    \begin{cases}
    1,\,3\,,\,10\,,\,35\,,\,126\,,\dots  = (n-1)\, C_{\frac{n-2}{2}}, \hspace{0.5cm}\text{if }\,n\,{\ge}\, 2\, \text{ is even,}\\
    0 \hspace{6.68cm}\text{if }\,n\, \text{ is odd,}\\
    \end{cases}
\end{align}
and therefore the non-vanishing one-loop contact terms turn out to be proportional to Catalan numbers, just as at tree-level. In fact, the structure of the contact terms \eqref{KLT1loopCt} as a polynomial in $n$ times Catalan numbers allows to derive a simple form for a corresponding one-loop Lagrangian,
\begin{align}\label{KLTLag1Loop}
    \mathcal{L}_{\xRed{1},\text{KLT}} = -\frac{1}{2} \bigg\langle\! \log\!\bigg(\! \frac{1+\sqrt{1-4\phi^2}}{2}\! \bigg) \!\bigg\rangle.
\end{align}
The interactions generated by the above Lagrangian should be treated formally as being of one-loop order, similarly to a counter-term Lagrangian, and serve to supplement the tree-level KLT Lagrangian $\mathcal{L}_{\xRed{0},\text{KLT}}$ in \eqref{defKLTLag}. 
The inverse KLT integrand $m_{\xRed{1},n}^{\alpha'}$ can then be computed from the \textit{effective one-loop Lagrangian},
\begin{align}\label{LagKLTL1}
\mathcal{L}^{\xRed{(1)}}_{\text{KLT}} = \mathcal{L}_{\xRed{0},\text{KLT}} + \hbar \mathcal{L}_{\xRed{1},\text{KLT}},
\end{align}
using Feynman diagrams by considering all one-loop diagram topologies with vertices coming from $\mathcal{L}_{\xRed{0},\text{KLT}}$ and tree-level topologies with a single vertex insertion from $\mathcal{L}_{\xRed{1},\text{KLT}}$. For example, the one-loop four-point integrand \eqref{BG1loop4} can be computed from contributions of the form
\begin{align}\label{KLT1loopEffDiags}
\begin{split}
    \mathcal{L}^{\xRed{(1)}}_{\text{KLT}}&\bigg\lvert_{1-\text{loop}}   \subset 
    \begin{matrix}
        \hspace{-0.0cm}\vspace{-0.1cm}\includegraphics[width=1.8cm]{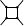}
    \end{matrix}
    + \dots + 
    \begin{matrix}
        \hspace{-0.0cm}\vspace{-0.1cm}\includegraphics[width=1.4cm]{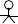}
    \end{matrix} +
    \begin{matrix}
        \hspace{-0.0cm}\vspace{-0.1cm}\includegraphics[width=1.9cm]{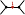}
    \end{matrix}
    + \dots +
    \begin{matrix}
        \hspace{-0.0cm}\vspace{-0.1cm}\includegraphics[width=1.0cm]{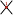}
    \end{matrix},
\end{split}
\end{align}
where black vertices derive from the tree-level Lagrangian $\mathcal{L}_{\xRed{0},\text{KLT}}$ and red vertices correspond to insertions of $\mathcal{L}_{\xRed{1},\text{KLT}}$ in \eqref{KLTLag1Loop}. The diagrams in \eqref{KLT1loopEffDiags} involving one-loop vertices are of course exactly the contributions in \eqref{Feyn1loop3} which where missing when comparing the Feynman integrand computed just with $\mathcal{L}_{\xRed{0},\text{KLT}}$ to the inverse KLT integrand computed from the cubic Berends-Giele recursion \eqref{BG1loopContrib}.

Let us briefly remark on the significance of the additional one-loop interactions for the inverse KLT integrand, whose existence essentially follows from the requirement that the cubic simplicity observed for the tree-level matrix elements $m_{n}^{\alpha'}$  when expressed in terms of a Berends-Giele recursion is preserved at loop level.

Since tree-level diagrams with insertions of \eqref{KLTLag1Loop} always involve fewer propagators than the maximum number possible at the given order, they do not contribute to the leading order in the limit $\alpha' \to 0$ so that the connection \eqref{KLT1loopAlphaZero} to the $\text{tr}\phi^3$ surfacehedron integrand is not spoiled.  However, even at finite values of $\alpha'$, the nature of these contact terms might seem perplexing. This is because each diagram topology involving some one-loop interaction \eqref{KLT1loopCt} does not depend on the loop momentum $\ell$. As such, the given diagram is ``scaleless'' with respect to the loop integration over $\ell$ and does not contribute to the integrated inverse KLT matrix element since
\begin{align}
    \int\! \text{d}^d\ell \,\,
    \begin{matrix}
        \hspace{-0.0cm}\vspace{-0.1cm}\includegraphics[width=1.6cm]{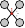}
    \end{matrix} \,\,=\, \begin{matrix}
        \hspace{-0.0cm}\vspace{-0.1cm}\includegraphics[width=1.6cm]{figures/1loopCntDiagTop.pdf}
    \end{matrix} \times \left( \int\! \text{d}^d\ell\,\, 1 \right) \simeq 0.
\end{align}
Therefore, as far as the physical, integrated, matrix element is concerned, the effective one-loop Lagrangian \eqref{LagKLTL1} is fully equivalent to the tree-level Lagrangian \eqref{defKLTLag}. Indeed, the one-loop Lagrangian \eqref{KLTLag1Loop} contributes non-trivially only to the inverse KLT \textit{integrand}. 

However, once we connect the inverse KLT integrand to integrands for NLSM pions in Section \ref{piAlphaLoop}, we will see that the loop-level contact interactions are vital to make sure that certain desirable properties of pion integrands, such as the Adler zero, are satisfied. Furthermore, given their emergence from the effective root vertex, it is not unreasonable to suspect that these contact terms might reflect certain combinatorial properties of the kinematic surface.

\subsection{All-Loop Recursion}
Before we can define the recursion for the general $L$-loop integrand, a short remark on conventions and notation is in order.
At $L$ loops an integrand $m_{\xRed{L},n}^{\alpha'}$ will depend on $L$ loop variables $Z\,{=}\,\lbrace z_1,\dots, z_L\rbrace$. By convention, we will construct integrands using the Berends-Giele recursion that are completely symmetric with respect to their dependence on these loop variables, that is
\begin{align}\label{mLnZSym}
    m_{\xRed{L},n}^{\alpha'}(z_1,z_2,\dots,z_L) = m_{\xRed{L},n}^{\alpha'}(z_{\sigma(1)},z_{\sigma(2)},\dots,z_{\sigma(L)}),
\end{align}
for any permutation $\sigma\,{\in}\, S_L$. In the above and for the remainder of this discussion we suppress the dependence of integrands on a fixed set of external labels in the notation.

To generate symmetric integrands like \eqref{mLnZSym} using the recursion, it will be useful to introduce the symmetric product of integrands. Suppose we have a left- and a right-hand integrand $m_{\xRed{L_L},n_L}^{\alpha'}$ and $m_{\xRed{L_R},n_R}^{\alpha'}$. Taking $L\,{=}\,L_L\,{+}\,L_R$ we define the symmetric product of these integrands with respect to the punctures $Z\,{=}\,\lbrace z_1,\dots, z_L\rbrace$ as
\begin{align}\label{defSymPr}
    m_{\xRed{L_L},n_L}^{\alpha'} \cdot m_{\xRed{L_R},n_R}^{\alpha'} \equiv \frac{L_L! L_R!}{L!} \!\sum_{Z_L \subset Z} m_{\xRed{L_L},n_L}^{\alpha'}\!(Z_L)\, m_{\xRed{L_R},n_R}^{\alpha'}\!(Z_R),
\end{align}
where the sum runs over all subsets $Z_L \subset Z$ of size $\lvert Z_L \rvert \,{=}\, L_L$. Consequently the complement has size $\lvert Z_R \rvert \,{=}\, \lvert Z\setminus\! Z_L \rvert \,{=}\, L_R$. Since each term on the right-hand side of \eqref{defSymPr} is expected to produce the same contribution once the loop integrations over all $z_i$ are carried out, we normalize by a binomial factor $\frac{L_L! L_R!}{L!}$ to avoid over-counting.

To give an example, let us consider the symmetric product of a one-loop integrand $m_{\xRed{1},n_L}^{\alpha'}$ and a two-loop integrand $m_{\xRed{2},n_R}^{\alpha'}$ as it would naturally appear in the recursion of an $L\,{=}\,1\,{+}\,2\,{=}\,3$ loop integrand. In this case the symmetric product has three terms,
\begin{align}
    m_{\xRed{1},n_L}^{\alpha'} \cdot m_{\xRed{2},n_R}^{\alpha'} = \frac{1}{3}\!\left( \! m_{\xRed{1},n_L}^{\alpha'}\!(z_1) m_{\xRed{2},n_R}^{\alpha'}\!(z_2z_3) + m_{\xRed{1},n_L}^{\alpha'}\!(z_2)  m_{\xRed{2},n_R}^{\alpha'}\!(z_1z_3) + m_{\xRed{1},n_L}^{\alpha'}\!(z_3) m_{\xRed{2},n_R}^{\alpha'}\!(z_1z_2)\! \right).
\end{align}
Applying this procedure to each Berends-Giele diagram, we make sure that the full integrand is constructed in a $z$-symmetric fashion term by term in the recursion.

\subsubsection*{Recursion}
\begin{figure}[t]
    \centering
    \includegraphics[width=4.5cm]{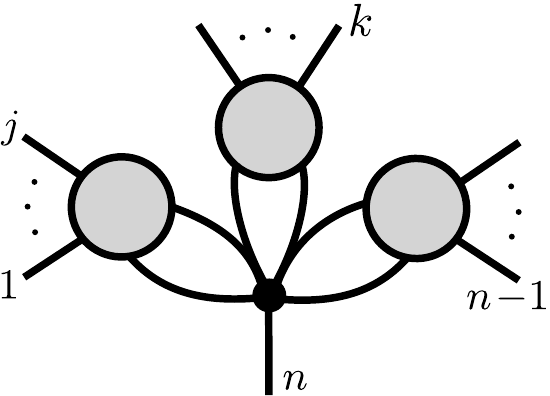}
    \hspace{2.2cm}
        \includegraphics[width=1.7cm]{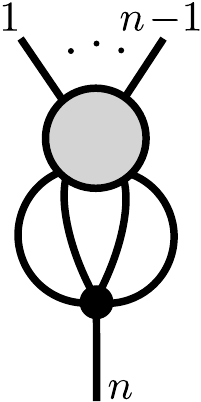}    
    \caption{Non-trivial contributions to the three-loop matrix element $m^{\alpha'}_{\xRed{3},n}$ when computed using the naive Berends-Giele recursion based on the Lagrangian \eqref{defKLTLag}. The cubic structure of the inverse KLT integrand allows to capture these diagrams using simpler diagram topologies of the form \eqref{BGloopTreeTop} and \eqref{BGloopGenLoop} involving the effective root vertex. }
    \label{fig:naiveBG3loopDiag}
\end{figure}

With notational conventions fixed, let us now define the $L$-loop inverse KLT integrand $m_{\xRed{L},n}^{\alpha'}$ via the cubic Berends-Giele recursion. The generalization from the one-loop recursion is straightforward, as there are no new diagram topologies to include in the recursion beyond those already considered in \eqref{BG1loopTreeTop} and \eqref{BG1loopLoopTop}. It is worth pointing out that this is a highly non-trivial feature of the inverse KLT integrand and the cubic Berends-Giele recursion based on the effective root vertex \eqref{rootValphaExp}. In fact, at loop orders $L\,{\ge}\, 2$, the analogue of the naive Berends-Giele recursion \eqref{defKLTLag} following from tree-level inverse KLT Lagrangian \eqref{defKLTLag} includes contributions from increasingly intricate Berends-Giele diagram topologies. Some representative diagrams contributing to the naive recursion at $L=3$ are shown in Figure \ref{fig:naiveBG3loopDiag}. However, these diagrams can in fact be subsumed into the cubic Berends-Giele diagram topologies in \eqref{BGloopTreeTop} and \eqref{BGloopGenLoop}, involving just the effective root vertex, illustrating the inherently cubic structure of the inverse KLT integrand.

A novel feature of the recursion at higher loop orders is that each Berends-Giele diagram needs to be summed over \textit{all} ways of distributing the $L$ loops between the left- and right-hand integrands. For Berends-Giele diagrams where the root vertex is in a tree-level configuration as in \eqref{BG1loopTreeTop} this will require a sum over the \textit{number of loops} $L_{L/R}$ of the right- and left-hand integrands such that $L\,{=}\,L_R\,{+}\,L_L$,
\begin{align}\label{BGloopTreeTop}
    m_{\xRed{L},n}^{\alpha'} \,\supset \, \hspace{-0.3cm}\sum_{L=L_R+L_L}\hspace{-0.1cm}
    \begin{matrix}
         \hspace{0.1cm}\vspace{-0.05cm}\includegraphics[width=5.0cm]{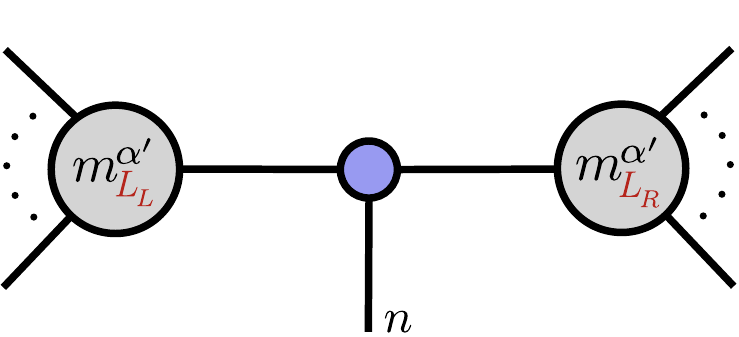}
    \end{matrix}
    \,\simeq\, \frac{1+t_Lt_R}{t_Lt_R} \hspace{-0.3cm}\sum_{L=L_R+L_L}\hspace{-0.3cm} m_{\xRed{L_L},n_L}^{\alpha'}\cdot m_{\xRed{L_R},n_R}^{\alpha'},
\end{align}
as well as a symmetrization over loop variables $z_i$ in each sub-integrand as in \eqref{defSymPr}.

For loop topology Berends-Giele diagrams of the type \eqref{BG1loopLoopTop}, one loop variable always appears explicitly in the root vertex while the remaining $L\,{-}\,1$ loops are contained in the ($n{+}1$)-point sub-integrand. At $L$ loops we therefore only have to include a symmetrization over $z$-variables for this diagram topology, which we will denote
\begin{align}\label{BGloopGenLoop}
\begin{split}
    m_{\xRed{L},n}^{\alpha'} \,\supset 
    \begin{matrix}
         \hspace{-0.0cm}\vspace{-0.4cm}\includegraphics[width=2.15cm]{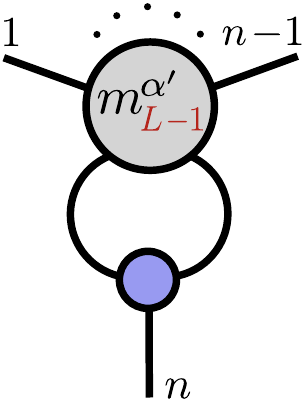}
    \end{matrix}
    &= \frac{1+t_{1Z}t_{nZ}}{t_{1Z}t_{nZ}}\cdot m_{\xRed{L-1},n+1}^{\alpha'}(1\dots n Z)\\[-0.5cm]
    & \equiv \frac{1}{L}\sum_{i=1}^{L} \frac{1+t_{1z_i}t_{nz_i}}{t_{1z_i}t_{nz_i}}\, m_{\xRed{L-1},n+1}^{\alpha'}(1\dots n z_i).
\end{split}
\end{align}

With these conventions in mind, we can now proceed with the explicit construction of the all-loop inverse KLT integrand. Like before, we discuss the special cases $n\,{=}\,1,2,3$ first, and then move on to the general case for $n\,{\ge}\, 4$. We provide the full, detailed expressions for all integrands to make it as easy as possible for the interested reader to implement the recursion in a computer program.

\subsubsection*{Special cases (n=1,2,3)}
For $L {\ge} 2$ the tadpole integrand receives some additional contributions compared to its one-loop counterpart \eqref{BG1loop1}. In particular, we have
\begin{align}
\begin{split}\label{BGLloop1}
    m_{\xRed{L},1}^{\alpha'} &= \frac{1+t_{11}^2}{t_{11}^2} \sum_{l=1}^{L-1} m_{\xRed{l},1}^{\alpha'}(1) \cdot m_{\xRed{L-l},1}^{\alpha'}(1) + \frac{1+t_{1Z}^2}{t_{1Z}^2}\cdot m_{\xRed{L-1},2}^{\alpha'}(1Z).
\end{split}
\end{align}
The first set of Berends-Giele diagrams, where two lower-loop tadpoles are joined into a tree topology,
\begin{align}\label{BGdiagLloopTadp}
    m_{\xRed{L},1}^{\alpha'} \,\supset \, \hspace{-0.1cm}\sum_{l=1}^{L-1}\hspace{-0.0cm}
    \begin{matrix}
         \hspace{0.1cm}\vspace{-0.05cm}\includegraphics[width=4.4cm]{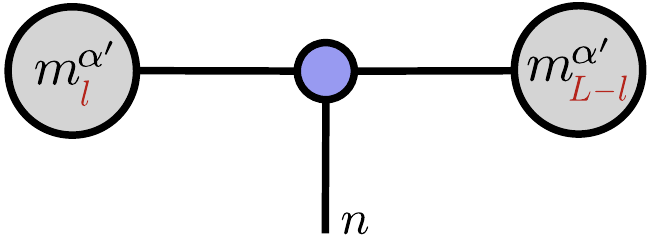}
    \end{matrix}
    \,=\, \frac{1+t_{11}^2}{t_{11}^2} \sum_{l=1}^{L-1} m_{\xRed{l},1}^{\alpha'}(1) \cdot m_{\xRed{L-l},1}^{\alpha'}(1),
\end{align}
is entirely new for $L\,{\ge}\,2$, while the loop-topology diagram, i.e. the second term in \eqref{BGLloop1}, is a natural generalization of the original contribution to the one-loop tadpole \eqref{BG1loop1}. 

The two-point inverse KLT integrand at $L$ loops is given by
\begin{align}
\begin{split}
    m_{\xRed{L},2}^{\alpha'} &= \frac{1+t_{11}t_{12}}{t_{11}t_{12}} \sum_{l=1}^{L-1} m_{\xRed{l},1}^{\alpha'}(1) \cdot m_{\xRed{L-l},2}^{\alpha'}(12) + \frac{1+t_{12}t_{22}}{t_{12}t_{22}} \sum_{l=1}^{L-1} m_{\xRed{l},2}^{\alpha'}(12)\cdot m_{\xRed{L-l},1}^{\alpha'}(2) \\
    &+ \frac{1}{t_{11}}m_{\xRed{L},1}^{\alpha'}(1) + \frac{1}{t_{22}}m_{\xRed{L},1}^{\alpha'}(2) + \frac{1+t_{1Z}t_{2Z}}{t_{1Z}t_{2Z}}\cdot m^{\alpha'}_{\xRed{L-1},3}(12Z).
\end{split}
\end{align}

Finally, at three points, the general $L$-loop recursion reads
\begin{align}
\begin{split}
    m_{\xRed{L},3}^{\alpha'} &= \frac{1+t_{11}t_{13}}{t_{11}t_{13}} \sum_{l=1}^{L} m_{\xRed{l},1}^{\alpha'}(1)\cdot m_{\xRed{L-l},3}^{\alpha'}(123) + \frac{1+t_{13}t_{33}}{t_{13}t_{33}} \sum_{l=0}^{L-1} m_{\xRed{l},3}^{\alpha'}(123)\cdot m_{\xRed{L-l},1}^{\alpha'}(3)\\
    &+ \frac{1}{t_{12}} m_{\xRed{L},2}(12) + \frac{1}{t_{23}} m_{\xRed{L},2}(23) + \frac{1+t_{12}t_{23}}{t_{12}t_{23}} \sum_{l=1}^{L-1} m_{\xRed{l},2}^{\alpha'}(12)\cdot m_{\xRed{L-l},2}^{\alpha'}(23)\\
    &+ \frac{1+t_{1Z}t_{3Z}}{t_{1Z}t_{3Z}}\cdot m^{\alpha'}_{\xRed{L-1},4}(123Z).
\end{split}
\end{align}

Here, and in the following, we do not explicitly draw the Berends-Giele diagrams corresponding to individual contributions in the recursion. The diagram topologies are, with the exception of the tadpole diagram \eqref{BGdiagLloopTadp}, identical to those encountered at one loop.

\subsubsection*{General case (n \!$\ge$\! 4)}
Let us now finally turn to the most general case of the $L$-loop integrand for $n\,{\ge}\, 4$. As in the case of the one-loop integrand \eqref{BG1loopContrib} we split the integrand up into tadpole-, bubble- and higher-valency contributions as well as the contribution from the loop-topology diagram,
\begin{align}\label{BGLloopContrib}
    m_{\xRed{L},n}^{\alpha'} = m_{\xRed{L},n,(1)}^{\alpha'} + m_{\xRed{L},n,(2)}^{\alpha'} + m_{\xRed{L},n,(\ge 3)}^{\alpha'} + m_{\xRed{L},n,(\text{loop})}^{\alpha'}.
\end{align}
Firstly, the Berends-Giele diagrams involving tadpoles are given by
\begin{align}\label{BGLloopGen1}
\begin{split}
    m_{\xRed{L},n,(1)}^{\alpha'} &= \sum_{l=1}^{L} \!\bigg( \!\frac{1+t_{11}t_{1n}}{t_{11}t_{1n}}\, m_{\xRed{l},1}^{\alpha'}(1)\cdot m_{\xRed{L-l},n}^{\alpha'}(1\dots n) + \frac{1+t_{1n}t_{nn}}{t_{1n}t_{nn}}\, m_{\xRed{L-l},n}^{\alpha'}(1\dots n)\cdot m_{\xRed{l},1}^{\alpha'}(n) \! \bigg),
\end{split}
\end{align}
which we recognize as a straightforward generalization of \eqref{BG1loopGen1} to $L$ loops.
Likewise, the general $L$-loop form of the bubble contributions \eqref{BG1loopGen2} is simply
\begin{align}\label{BGLloopGen2}
\begin{split}
    m_{\xRed{L},n,(2)}^{\alpha'} &= \frac{1}{t_{2n}} m_{\xRed{L},n-1}^{\alpha'}(2\dots n) + \frac{1}{t_{1n-1}} m_{\xRed{L},n-1}^{\alpha'}(1\dots n{-}1) \\
    &+ \sum_{l=1}^{L} \!\bigg(\! \frac{1+t_{12}t_{2n}}{t_{12}t_{2n}} m_{\xRed{l},2}^{\alpha'}(12) \cdot m_{\xRed{L-l},n-1}^{\alpha'}(2\dots n) \\
    &\hspace{0.8cm} + \frac{1+t_{1n-1}t_{n-1n}}{t_{1n-1}t_{n-1n}} m_{\xRed{L-l},n-1}^{\alpha'}(1\dots n{-}1) \cdot m_{\xRed{l},2}^{\alpha'}(n{-}1\,n) \!\bigg).
\end{split}
\end{align}
Continuing with the higher-valency contributions we get
\begin{align}\label{BGLloopGen3}
    m_{\xRed{L},n,(\ge 3)}^{\alpha'} = \sum_{j=3}^{n-2}\frac{1+t_{1j}t_{jn}}{t_{1j}t_{jn}} \sum_{l=0}^{L} m_{\xRed{l},j}^{\alpha'}(1\dots j) \cdot m_{\xRed{L-l},n-j+1}^{\alpha'}(j\dots n).
\end{align}
Finally we have the contribution from the loop-topology Berends-Giele diagram,
\begin{align}\label{BGLloopGenLoop}
    m_{\xRed{L},n,(\text{loop})}^{\alpha'} = \frac{1+t_{1Z}t_{nZ}}{t_{1Z}t_{nZ}} \cdot m_{\xRed{L-1},n{+}1}^{\alpha'}(1\dots n Z),
\end{align}
where we employ the notation \eqref{BGloopGenLoop} for symmetrizing the $z$-dependence between the root vertex and the ($n\,{+}\,1$)-point integrand $m_{\xRed{L-1},n{+}1}^{\alpha'}$.

This concludes the definition of the $L$-loop inverse KLT integrand. Counting up the number of Berends-Giele diagrams in each of the contributions to \eqref{BGLloopContrib} we arrive at a total of just $n(L\,{+}\,1)\,{-}\,1$ diagrams (up to $z$-symmetrization). This shows that the recursion scales exceptionally well in both the number of loops and legs. For instance, the 4-loop, 8-point inverse KLT integrand $m_{\xRed{4},8}^{\alpha'}$ can be computed in less than a second on a standard desktop computer using the cubic Berends-Giele recursion.

Let us now consider a simple explicit example and discuss some general properties of the matrix elements $m_{\xRed{L},n}^{\alpha'}$.

\subsubsection*{A simple example at two loops}
To illustrate the recursion at work, we compute the simplest two-loop example of the inverse KLT integrand at one point. There are only two Berends-Giele diagram topologies corresponding to the two contributions in \eqref{BGLloop1}, yielding
\begin{align}\label{BG2loop1}
\begin{split}
    m_{\xRed{2},1}^{\alpha'} &= \begin{matrix}
         \hspace{-0.0cm}\vspace{-0.8cm}\includegraphics[width=3.8cm]{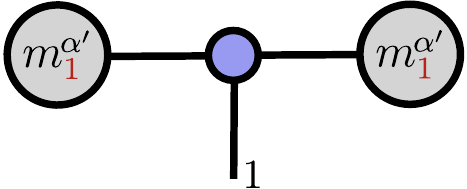}
    \end{matrix}
    \,\,\,+\,\,\,
    \begin{matrix}
         \hspace{-0.0cm}\vspace{-0.3cm}\includegraphics[width=1.2cm]{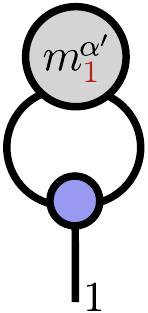}
    \end{matrix}\\
    &= \frac{1}{2!} \bigg( \frac{1}{t_{1z_1}^2t_{z_1z_2}t_{1z_2}} + \frac{1}{t_{1z_1}^2t_{z_1z_1}t_{z_1z_2}} + \frac{1}{t_{1z_1}^2t_{11}t_{1z_2}} + \frac{1}{t_{1z_1}t_{11}^2t_{1z_2}}\\
    &\hspace{0.5cm} + \frac{1}{t_{1z_1}t_{1z_2}} + \frac{1}{t_{1z_1}t_{z_1z_2}} + \frac{1}{t_{11}t_{1z_1}} + \frac{1}{t_{z_1z_1}t_{z_1z_2}} + \frac{1}{t_{1z_1}^2} + (z_1 \leftrightarrow z_2) \!\bigg) +1.
\end{split}
\end{align}
While the Berends-Giele recursion is certainly the most efficient way to compute the matrix element $m^{\alpha'}_{\xRed{2},1}$, we want to emphasize that the contributions involving poles can again be interpreted consistently in terms of Feynman diagrams arising from the effective one-loop Lagrangian $\mathcal{L}_{\text{KLT}}^{\xRed{(1)}}$ defined below \eqref{KLTLag1Loop}. The corresponding Feynman diagrams are shown in Appendix \ref{app:m21Feyn}.

However, and more importantly, we observe a recurring phenomenon; that of the emergence of loop-level contact terms from the cubic Berends-Giele recursion. From the example above we can read off a one-point contact interaction,
\begin{align}
    \begin{matrix}
         \hspace{-0.0cm}\vspace{0.2cm}\includegraphics[width=0.4cm]{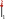} 
    \end{matrix}
    =\,1,
\end{align}
which does \textit{not} derive from the one-loop inverse KLT Lagrangian. In fact, this contact term is nothing but the first representative of yet another infinite tower of inherently \textit{two-loop} contact interactions on top of the already known tree-level and one-loop interactions \eqref{KLTFeynRules} and \eqref{KLT1loopCt} which is mandated by the cubic structure of the inverse KLT integrand.

With this apparent proliferation of contact terms at hand, it is time we study their structure more generally.

\subsubsection*{General structure of contact terms}
We can analyze the general $L$-loop Berends-Giele recursion \eqref{BGLloopContrib} for contact terms in much the same way as we did at $L\,{=}\,1$ in \eqref{KLT1loopCt}. Curiously, we find that there exists a separate infinite tower of interactions contributing to the inverse KLT integrand \textit{at each loop order}, giving rise to an \textit{infinity} of infinite towers of contact terms. The structure of the recursion implies that the contact terms for inverse KLT matrix elements $m_{\xRed{L},n}^{\alpha'}$ satisfy
\begin{align}\label{mLContPatt}
    m_{\xRed{L},n}^{\alpha'}\Big\lvert_{\text{ct.}} \equiv \kappa_{\xRed{L},n} \!= 
    \begin{cases}
        \neq 0, \hspace{1.0cm} \text{ if $\,L\,{+}\,n\,$ is odd,}    \\
        0,   \hspace{1.55cm} \text{if $\,L\,{+}\,n\,$ is even},
    \end{cases}
\end{align}
i.e. they are only non-zero when (I): $L$ is even and $n$ is odd, or (II): $L$ is odd and $n$ is even. This pattern can be explained most easily by considering the leading power counting of the $L$-loop, $n$-point inverse KLT integrand in the stringy $t$-variables,
\begin{align}\label{mLpw}
    m_{\xRed{L},n}^{\alpha'} \simeq \mathcal{O}\!\left( \!\frac{1}{t^{3(L-1)+n}}\!\! \right),
\end{align}
and recognizing that the effective root vertex \eqref{rootValphaExp} in the cubic Berends-Giele recursion can only increase the $t$-counting of individual terms contributing to $m_{\xRed{L},n}^{\alpha'}$ in multiples of two by canceling select pairs of propagators in leading order contributions. Therefore, in order for the recursion to generate a contact term of $\mathcal{O}(t^0)$, the leading power counting \eqref{mLpw} of the matrix element must be $3(L\,{-}\,1)\,{+}\,n=\text{even}$, which is the case if and only if $L\,{+}\,n=\,$odd. 

\begin{figure}[H]
    \centering
    \includegraphics[width=8.0cm]{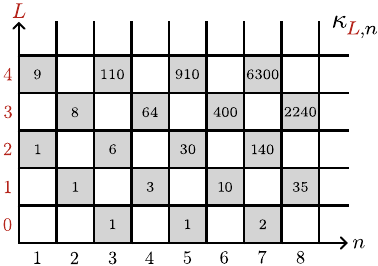}
    \caption{Outline of the structure and explicit values of non-trivial contact terms $\kappa_{\xRed{L},n}$ (gray squares) up to $\xRed{L}\,{=}\,4$ and $n\,{=}\,8$. Vanishing contact terms are indicated by white squares.}
    \label{fig:conTab}
\end{figure}

More concretely, the cubic Berends-Giele recursion for the matrix elements $m_{\xRed{L},n}^{\alpha'}$ implies a corresponding recursion relation for the non-zero contact terms in \eqref{mLContPatt} which reads
\begin{align}\label{contRec}
    \kappa_{\xRed{L},n} = \sum_{l=0}^L\sum_{k=1}^n \kappa_{\xRed{l},k}\kappa_{\xRed{L-l},n-k+1} + \kappa_{\xRed{L-1},n+1}.
\end{align}
Fortunately, it is possible to solve this recursion and derive a generating function for the contact terms $\kappa_{\xRed{L},n}$. Of course, the generating function for contact terms has a natural interpretation as (the derivative of) the corresponding interaction Lagrangian. For the contact interactions  \eqref{contRec} the derivative of the \textit{all-loop Lagrangian} for the inverse KLT integrand takes the remarkably simple form (cf.  Appendix \ref{app:LKLTderiv})
\begin{align}\label{KLTLagL}
    \frac{\partial \mathcal{L}_{\text{KLT}}}{\partial\phi} \equiv \mathcal{L}'_{\text{KLT}} = \frac{1}{2}\Bigg\langle \!\!\!  \left(\! 1-\frac{\hbar}{\phi}\!\right)\!\! \Bigg[\!1-\! \sqrt{1+4\frac{\tfrac{\hbar}{\phi}K_1(\hbar)-\phi^2}{(1-\tfrac{\hbar}{\phi})^2}} \,\Bigg] \!\Bigg\rangle.
\end{align}
Surprisingly, while the derivative of the Lagrangian is simple, it is not straightforward to perform the elliptic integral over $\phi$ in order to obtain the Lagrangian proper. However, for practical purposes this is not an issue as the closed form \eqref{KLTLagL} is sufficient to extract contact terms at any desired order by evaluating
\begin{align}
    \kappa_{\xRed{L},n} = \frac{1}{L!(n-1)!} \frac{\partial^{L}}{\partial\hbar^{L}} \frac{\partial^{n-1}}{\partial\phi^{n-1}} \mathcal{L}'_{\text{KLT}}\Bigg\rvert_{\substack{\hbar=0\\ \phi=0}} \,.
\end{align}
We list the explicit values for the first few non-zero contact terms at lowest order in Figure \ref{fig:conTab} together with an illustration of the overall structure \eqref{mLContPatt} of non-trivial contact interactions appearing for the inverse KLT integrand.

Let us briefly discuss the function $K_1$ appearing in the Lagrangian \eqref{KLTLagL}. It records information about tadpole ($n\,{=}\,1$) contact terms at all loop orders,
\begin{align}\label{K1}
    K_1(\hbar) = \sum_{l=1}^\infty \kappa_{\xRed{2l},1} \hbar^{2l} = \sum_{l=1}^\infty \,\,\,
    \begin{matrix}
         \hspace{-0.0cm}\vspace{0.2cm}\includegraphics[width=0.35cm]{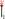} 
    \end{matrix} \,\,\,  \hbar^{2l}.
\end{align}
Using $K_1$ as initial data, the Lagrangian \eqref{KLTLagL} effectively codifies the recursive nature of contact terms and allows to extract all higher-valency interaction vertices. We should mention that further applying the recursion \eqref{contRec}, the tadpole contact terms $\kappa_{\xRed{2l},1}$ themselves can, in principle, be solved entirely in terms of tree-level contact terms \eqref{KLTFeynRules}. However, finding this relation in closed form turns out to be non-trivial. Therefore, we instead \textit{conjecture} the explicit form of the tadpole contact terms and the function \eqref{K1} to be
\begin{align}\label{K1conj}
    \kappa_{\xRed{2l},1} = \frac{1}{2}\frac{3^l}{4l-1} C_{2l}  \hspace{0.5cm} \Rightarrow\hspace{0.5cm} K_1(\hbar) = \sum_{l=1}^{\infty} \kappa_{\xRed{2l},1} \hbar^{2l} =  \frac{1}{2}+\frac{1}{6}\frac{(1-x)^{3/2}-(1+x)^{3/2}}{x},
\end{align}
where $x=4\sqrt{3}\hbar$. The conjectured form \eqref{K1conj} is obtained by fitting to the data generated by explicitly running the recursion for $\kappa_{\xRed{L},1}$ up to $L\,{=}\,22$. A comprehensive derivation of the Lagrangian \eqref{KLTLagL} is given in Appendix \ref{app:LKLTderiv}.

Let us also briefly remark on the structure of higher valency vertices. Explicitly evaluating the recursion up to eight loops we observe that the contact terms for the KLT integrands seem to exhibit a simple structure that generalizes the pattern already observed at the tree- and one-loop level. \textit{Conjecturally}, the non-zero $L$-loop ($L\,{\ge}\, 1$), $n$-point ($n\,{\ge}\, 1$) contact terms take the form
\begin{align}\label{KLTLloopCt}
    \kappa_{\xRed{L},n} = \frac{1}{L!} 4^{\lfloor\! \frac{L-1}{2} \!\rfloor} P_L(n)\, C_{\lfloor\! \frac{n-1}{2} \!\rfloor},
\end{align}
whenever $L\,{+}\, n$ is odd. The functions $P_L(n)$ appearing in \eqref{KLTLloopCt} are degree-$L$ polynomials in $n$ with integer coefficients. For instance, comparing with the one-loop contact term \eqref{KLT1loopCt}, we see that $P_1(n)= n-1 $. Fitting polynomials to the explicit solution (\ref{L_L_solution}) of the  recursion we obtain, up to six loops,
\begin{align}\label{contPoly}
   P_2(n) &= n(n+1), \hspace{4.7cm} P_3(n) = n(n-1)(n+4), \nonumber \\
   P_4(n) &= n(n+1)(n+2)(n+8), \hspace{2.3cm} P_5(n) = n(n-1)(n^3+20n^2+112n+228), \nonumber\\
   P_6(n) &= n(n+1)(n+2)(n^3 +30n^2 + 272n + 912), \hspace{2.0cm} \,\,\,\dots \,\, .
\end{align}
At loop orders $L\,{\ge}\, 5$ the polynomials become irreducible over the integers such that they cannot fully cancel with factorials appearing in the Catalan numbers $C_{\lfloor\! \frac{n-1}{2} \!\rfloor}$. This strengthens the assumption that the factorization of the contact term \eqref{KLTLloopCt} into the polynomials $P_L(n)$ and Catalan numbers is meaningful. It would be interesting to further understand the general pattern behind these polynomials and look for a combinatorial interpretation.

To conclude, we want to emphasize that although the explicit formulas for the contact terms \eqref{K1conj} and \eqref{KLTLloopCt} have been presented as conjectures, there is nothing ambiguous about the cubic recursion \eqref{BGLloopContrib} itself, which, in principle, allows to compute the contact terms $\kappa_{\xRed{L},n}$ to any desired order. The difficulty lies merely in extracting the explicit functional dependence of these contact terms on $n$ and $L$ (in closed form) from the recursion.
%

\section{Pion Integrands from $\alpha'$-shift}\label{piAlphaLoop}

Having discussed in detail how to efficiently construct the inverse KLT integrand for any number of loops and legs, let us briefly return to the connection between the inverse KLT kernel and stringy NLSM pions and generalize it to the loop level. At tree-level, the $\alpha'$-shift allowed to deform matrix elements $m_{2k}^{\alpha'}$ at even multiplicity into stringy pions $A_{2k}^{\alpha'}$ through a simple shift of the kinematic variables $X_{ij}$ (cf. \eqref{alphaShft}).
For loop integrands, we have to extend the shift prescription to \textit{all} kinematic $X$-variables \eqref{defXloop1} as well as the tadpole variables defined formally on the surface as in Figure \ref{fig:SurfVars2}. To do this, we take inspiration from the $\delta$-shift prescription that is known to connect $\text{tr}\phi^3$ field theory integrands to those of the NLSM as well as mixed amplitudes \cite{Arkani-Hamed:2024nhp}. For simplicity, we will focus here only on the shift prescription for pure pion integrands.

We recall that at tree-level the $\alpha'$-shift required an assignment of even or odd parity to each channel $X_{ij}$ according to \eqref{defChPar}. The most straightforward way to extend the notion of parity to channels involving loop labels $Z\,{=}\,\lbrace z_1, \dots, z_L\rbrace$ is to simply \textit{prescribe} some parity for each loop label,
\begin{align}\label{defZpar}
    P_Z^{(i)} = \lbrace p^{(i)}(z_1),\dots ,p^{(i)}(z_L) \rbrace, \hspace{1.0cm} p^{(i)}(z_j)\equiv \text{even/odd}.
\end{align}
At $L$ loops, the index $i\,{=}\,1,\dots, 2^L$ runs over all possible binary assignments of parities to loop labels $z_i$. Having defined the parity for loop labels $z_i$, we can immediately assign parities to channels such as $P(X_{iz_j})$ or $P(X_{z_iz_j})$ by employing the same definition \eqref{defChPar} as at tree-level where the parities of $z_i$ are determined according to \eqref{defZpar}.

Now to obtain the stringy pion integrand at $L$ loops we have to perform $2^L$ distinct $\alpha'$-shifts. For each parity assignment of $z_i$ in \eqref{defZpar} we split the set of $X$-variables into channels of even and odd parity,
\begin{align}
    X = X^{(i)}_{\text{even}}\cup X^{(i)}_{\text{odd}}, \hspace{1.5cm} i=1,\dots, 2^L,
\end{align}
where $X^{(i)}_{\text{even/odd}}$ are defined in analogy to \eqref{defXoe} at tree-level. For fixed $i$ we then define the corresponding $i$-th $\alpha'$-shift via
\begin{align}\label{XshftLoop}
    \alpha' \to \alpha'/2, \hspace{1.0cm} \text{and}\hspace{1.0cm}  \hat{X}^{(i)}_{\text{even}} = X^{(i)}_{\text{even}}\pm1/\alpha', \hspace{1.0cm}  \hat{X}^{(i)}_{\text{odd}} = X^{(i)}_{\text{odd}},
\end{align}
again in complete analogy to the tree-level shift \eqref{alphaResc} and \eqref{alphaShft}. Finally, the stringy $L$-loop pion integrand is then given by
\begin{align}\label{ApionLloop}
    A^{\pi,\alpha'}_{\xRed{L},2k}(X) = \sum_{i=1}^{2^L} m_{\xRed{L},2k}^{\alpha'\!/2}\big(\hat{X}^{(i)}_{\text{even}},\hat{X}^{(i)}_{\text{odd}}\big).
\end{align}

As we mentioned before, the above shift procedure is highly reminiscent of the so-called $\delta$-shift for field theory amplitudes and integrands involving cubic scalars and pions. There, integrands in the $\text{tr}\phi^3$ theory are deformed using a similar kinematic shift to \eqref{XshftLoop} with a parameter $\delta\equiv 1/\alpha'$. The field theory pion integrand can then be extracted from the shifted $\text{tr}\phi^3$ (surface) integrand by expanding around $\delta \,{=}\, \infty$,
\begin{align}\label{ApionDeltaShft}
    A^{\pi}_{\xRed{L},2k} = \sum_{i=1}^{2^L} \lim_{\delta\to \infty} \delta^{2k+2L-2} m_{\xRed{L},2k}\big(\hat{X}^{(i)}_{\text{even}}(\delta),\hat{X}^{(i)}_{\text{odd}}(\delta)\big).
\end{align}
The specific pion integrands generated through this procedure have been studied in great detail \cite{Arkani-Hamed:2024nhp,Arkani-Hamed:2024yvu,Arkani-Hamed:2024fyd,Bartsch:2024ofb} and exhibit many nice properties such as an exact Adler zero in appropriately defined ``surface soft'' limits and non-trivial split factorizations in multi-soft limits. They also match all single-cuts when defined in terms of surface variables. For this reason, the integrand \eqref{ApionDeltaShft} is sometimes dubbed the \textit{perfect} NLSM pion integrand.

The stringy pion integrand \eqref{ApionLloop} derived here from the inverse KLT integrand represents a simple stringification of precisely this perfect NLSM integrand in the sense that
\begin{align}
    A^{\pi,\alpha'}_{\xRed{L},2k} = \left(\!\frac{\pi\alpha'}{2}\!\right)^{\!\!1-L} \!\!\left(A^{\pi}_{\xRed{L},2k} + \mathcal{O}(\alpha'^2)\right).
\end{align}
The above equality is exact, meaning that at leading order in $\alpha'$, the integrands match as rational functions of kinematic surface variables, even including unphysical contributions involving tadpole and off-shellness variables that would vanish upon integration. We have checked this correspondence explicitly up to four loops and four points, as well as higher multiplicities at lower loop orders.

\subsubsection*{Example (n\,${=}$\,2)}
To get a more concrete understanding of the general shift prescription discussed above, we consider the simplest example of the stringy pion integrand at loop.

We start with the simplest non-trivial one-loop integrand at $n\,{=}\,2$. At one loop, there are only two possible assignments for the parity of the single loop label $z$ as described in \eqref{defZpar}. That is, either $p^{(1)}(z)=\text{even}$ or $p^{(2)}(z)=\text{odd}$. In each case we split the set of $X$-variables according to their parity,
\begin{align}
\begin{split}
    X^{(1)}_{\text{even}} &= \lbrace X_{11},X_{22},X_{2z}\rbrace, \hspace{1.0cm} X^{(1)}_{\text{odd}} = \lbrace X_{12},X_{1z}\rbrace, \\
    X^{(2)}_{\text{even}} &= \lbrace X_{11},X_{22},X_{1z}\rbrace, \hspace{1.0cm} X^{(2)}_{\text{odd}} = \lbrace X_{12},X_{2z}\rbrace,
\end{split}
\end{align}
Evaluating the two-point KLT integrand $m_{\xRed{1},2}^{\alpha'\!/2}$ in \eqref{BG1loop2} on shifted kinematics according to \eqref{XshftLoop} we obtain
\begin{align}\label{Pi1Loop2}
    m_{\xRed{1},2}^{\alpha'\!/2}\big(\hat{X}^{(1)}_{\text{even}},\hat{X}^{(1)}_{\text{odd}}\big) &= 1-\frac{\tau_{11}+\tau_{2z}}{\tau_{1z}}+\tau_{22}\tau_{2z}, \hspace{0.3cm} m_{\xRed{1},2}^{\alpha'\!/2}\big(\hat{X}^{(2)}_{\text{even}},\hat{X}^{(2)}_{\text{odd}}\big) = 1-\frac{\tau_{22}+\tau_{1z}}{\tau_{2z}}+\tau_{11}\tau_{1z}, \nonumber \\
    \eqref{ApionLloop} & \Rightarrow A_{\xRed{1},2}^{\pi,\alpha'} = 2-\frac{\tau_{11}+\tau_{2z}}{\tau_{1z}}-\frac{\tau_{22}+\tau_{1z}}{\tau_{2z}}+\tau_{11}\tau_{1z}+\tau_{22}\tau_{2z},
\end{align}
where we recall that the $\alpha'$-rescaled stringy variables are given by $\tau_{ij}=\tan\!\big( \frac{\pi}{2}\alpha' X_{ij} \big)$. We can now apply the \textit{surface soft limit} \cite{Arkani-Hamed:2024yvu} to the integrand \eqref{Pi1Loop2}. Taking $t_{2z}\mapsto t_{1z}$ and $t_{11}\,{=}\,t_{22} \mapsto 0$ we find $A_{\xRed{1},2}^{\pi,\alpha'} \mapsto 0$. This is the \textit{surface Adler zero} for pion integrands. Note that the contact term $\sim 2$ in the pion integrand \eqref{Pi1Loop2} was crucial to make the Adler zero work exactly. Of course, this contact term ultimately derives, via the $\alpha'$-shift and  \eqref{ApionLloop}, from the loop-level interactions \eqref{KLTLloopCt} of the inverse KLT integrand.

The same phenomenon persists at higher multiplicities and loop orders, where the loop-level interactions \eqref{KLTLagL} ensure that the surface soft limit for the stringy pion integrands
\begin{align}
    A_{\xRed{L},2k}^{\alpha',\pi} \xmapsto{\text{surf. soft}} 0,
\end{align}
is preserved exactly when carefully taking into account the curve homology associated to the kinematic $X$-variables.

\section{Conclusion and Outlook}
In this article we have introduced the \textit{inverse KLT integrand} $m_{\xRed{L},n}^{\alpha'}$, a generalization of the inverse string theory KLT kernel \cite{Mizera:2016jhj} to all loop orders. It is a simple, stringy function defined on the kinematic surface \cite{Arkani-Hamed:2023lbd,Arkani-Hamed:2023mvg} and encodes the scattering of cubic scalars and pions.
The cubic Berends-Giele recursion relations formulated in this work, in particular their striking similarity to the known recursion for the tr$\phi^3$ field theory, reveal an underlying simplicity of the stringy matrix elements and suggest that the inverse KLT integrand can be considered as a toy model for the ``simplest stringy amplitude''.

The primary focus of this work was on the construction of diagonal matrix elements for the inverse KLT integrand. An immediate and interesting opportunity for generalization lies in finding an appropriate definition for \textit{off-diagonal} matrix elements at loop-level. At tree-level, Mizera showed that off-diagonal inverse KLT matrix elements can be given an interpretation as intersection numbers of twisted cycles tiling the moduli space $\mathcal{M}_{0,n}$ \cite{Mizera:2017cqs}. It is reasonable to suspect that the integrands presented here allow for a similar interpretation as intersection numbers on appropriately defined loop-level moduli spaces $\mathcal{M}_{\xRed{L},n}$. 

Presumably, this conjecture is intimately connected with another extension of the inverse KLT integrand not touched upon in this work. It concerns the consistent treatment of \textit{closed curves} that exist on the kinematic surface at loop-level. At one loop, this problem was solved by the so-called halohedron integrand \cite{Salvatori:2018aha} for the tr$\phi^3$ theory which includes an additional kinematic variable $X_0$,
\begin{align}
    \begin{matrix}
             \hspace{-0.0cm}\vspace{-0.0cm}\includegraphics[width=2.2cm]{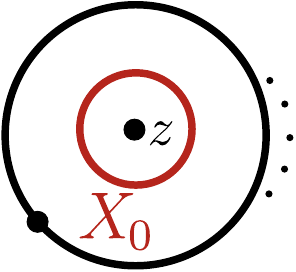}
    \end{matrix}\, ,
\end{align}
corresponding to the (only) closed curve on the kinematic surface with a single puncture.
We can similarly define stringy versions of the halohedron integrand that correspond to simple extensions of the inverse KLT integrand including the dependence on $X_0$. For instance, for the one-loop tadpole and bubble we can write
\begin{align}
    m_{\xRed{1},1}^{(X_0)} = \frac{1}{t_{1z}} + \frac{1}{t_{0}}, \hspace{0.8cm} m_{\xRed{1},2}^{(X_0)} = \frac{1}{t_{11}t_{1z}} +\frac{1}{t_{11}t_{0}} + \frac{1}{t_{22}t_{2z}}+ \frac{1}{t_{22}t_{0}} + \frac{1}{t_{1z}t_{2z}} + 1, 
\end{align}
where $t_0 \,{=}\,\tan(\pi\alpha' X_0)$ as usual. It is no coincidence that these integrands are functionally \textit{identical} to the four- and five-point tree-level matrix elements \eqref{invKLTex}. It merely reflects the fact that the moduli spaces $\mathcal{M}_{\xRed{1},1} \,{\simeq}\, \mathcal{M}_{\xRed{0},4}$ and $\mathcal{M}_{\xRed{1},2} \,{\simeq}\, \mathcal{M}_{\xRed{0},5}$ are known to be combinatorially identical \cite{devadoss2011}. This strongly suggests that the one-loop inverse KLT integrand, even at higher multiplicities, correctly captures the underlying boundary structure of moduli spaces of the string worldsheet \textit{directly in kinematic space}.

Of course, a particularly compelling question is whether this correspondence extends to the two-loop level and beyond. There, the structure of the corresponding moduli spaces of Riemann surfaces with more than two boundaries ($\xRed{L}\ge 2$) is known to be non-polytopal \cite{devadoss2011}. This will presumably require much more non-trivial closed-curve extensions of the inverse KLT integrand compared the one shown at one loop above. Still, the stringy nature of the inverse KLT integrand allows for a much richer set of functions to accommodate the boundary structure of the moduli space directly in kinematic space, especially when contrasted with its field theory analogue, the tr$\phi^3$ surface integrand, which is constrained to be a rational function of $X$-variables.

In a similar vein, a further open question regards the existence of (positive) geometries for the inverse KLT integrand. At tree-level, the inverse KLT kernel was recently found to be the canonical function of the so-called associahedral grid \cite{Bartsch:2025mvy}. It would be interesting to see if a similar geometric picture extends to the integrands studied in this work.

On a less technical note, there remains a more fundamental question related to the inverse KLT integrand: what is its purpose in life?
At tree-level, the \textit{inverse} KLT kernel is exactly what it proclaims to be. It is the matrix inverse of the KLT kernel appearing in the double copy of tree-level open and closed string amplitudes.
This raises the question of whether the inverse KLT integrand presented here, once inverted, can be utilized in a similar way to facilitate a KLT-type double copy of stringy loop integrands directly in kinematic space, potentially providing another perspective on the one-loop double copy relations \cite{Stieberger:2023nol,Mazloumi:2024wys} recently observed for full string integrands.

Finally, it would be interesting to explore how far the simple stringification provided by the inverse KLT kernel for the BAS theory and NLSM pions can be extended to other known effective field theories with special soft limit behavior and those that satisfy double copy relations \cite{Cheung:2014dqa,Cheung:2015ota,Cheung:2016drk,Cheung:2018oki,Elvang:2018dco,Low:2019ynd,Kampf:2019mcd,Rodina:2021isd,Kampf:2021bet,Chi:2021mio,Armstrong:2022vgl,Cheung:2022vnd,Bartsch:2022pyi,Kampf:2023elx,Brown:2023srz,Brauner:2024juy,Kampf:2024xjy}. This includes unordered theories, raising the question of a canonical definition of the corresponding integrands due to the lack of (global) planar variables. The inverse KLT integrand might aid in finding such a canonical description.

\section*{Acknowledgments}

We thank Shruti Paranjape and Jonah Stalknecht
for useful discussions. This work is supported by GA CR 24-11722S, MEYS LUAUS23126, OPJAK CZ.02.01.01/00/22 008/0004632, DOE grant No.
SC0009999 and the funds of the University of California.

\appendix

\section{Alternative proof of the cubic Berends-Giele recursion (Tree-Level)}\label{app:KLTBGprf2}

\subsection{Notation and general formulas}

The diagonal amplitudes $m_{diag}^{\alpha ^{\prime }}\left( \sigma \right)
\equiv m^{\alpha ^{\prime }}(\sigma |\sigma )$ corresponding to the inverse
stringy KLT kernel can be formally obtained from the Lagrangian
\begin{equation}
\mathcal{L}_{\alpha ^{\prime }}=-\frac{1}{2}\left\langle \phi \Delta
^{-1}\phi \right\rangle +V\left( \phi \right)  \label{lagrangian}
\end{equation}
where the inverse propagator reads%
\[
\Delta ^{-1}=\tan \left( \pi \alpha ^{\prime }\square \right)
\]
and the potential is
\begin{equation}
V\left( \phi \right) =\left\langle \frac{1}{2}\phi -\frac{1}{8}\arcsin 2\phi
-\frac{1}{4}\phi \sqrt{1-4\phi ^{2}}\right\rangle =\sum_{k=1}^{\infty }%
\frac{C_{k-1}}{2k+1}\left\langle \phi ^{2k+1}\right\rangle ,
\label{potential}
\end{equation}%
Here%
\[
C_{k}=\frac{1}{k+1}\left( 
\begin{array}{c}
2k \\ 
k%
\end{array}%
\right)
\]%
are the Catalan numbers and we use the notation $\phi =\phi ^{a}T^{a}$ and $%
\left\langle \cdot \right\rangle \equiv \mathrm{Tr}\left( \cdot \right) $.

Let us define the generator $\mathcal{T}[\phi ]$ of the connected amplitudes
associated with the Lagrangian $\mathcal{L}_{\alpha ^{\prime }}$ as%
\[
\exp \left( \frac{\mathrm{i}}{\hbar }\mathcal{T}[\phi ]\right) \equiv \exp
\left( -\mathrm{i}\frac{\hbar }{2}\frac{\delta }{\delta \phi }\cdot \Delta
\cdot \frac{\delta }{\delta \phi }\right) \exp \left( \frac{\mathrm{i}}{%
\hbar }S_{int}\left[ \phi \right] \right) ,
\]%
where we abbreviated%
\[
\frac{\delta }{\delta \phi }\cdot \Delta \cdot \frac{\delta }{\delta \phi }%
\equiv \int \mathrm{d}x\mathrm{d}y\frac{\delta }{\delta \phi ^{a}\left(
x\right) }\Delta \left( x,y\right) \frac{\delta }{\delta \phi ^{a}\left(
y\right) },
\]%
The above formula is a compact expression of the Wick expansion of the $S-$%
matrix, which can be obtained formally as%
\begin{equation}
S=\colon \exp \left( \mathrm{i}\mathcal{T}[\widehat{\phi }]\right) \colon
\label{S-matrix}
\end{equation}%
where $\widehat{\phi }$ is the operator of the free field in the interaction
picture satisfying 
\[
\square \widehat{\phi }=0
\]%
and $\colon \cdot \colon $ means normal ordering. The amplitudes are then
obtained from $\mathcal{T}[\phi ]$ according to the prescription (all the
momenta $p_{j}$ are treated as outgoing and on-shell)%
\begin{eqnarray}
A\left( 1^{a_{1}},2^{a_{2}},\ldots ,n^{a_{n}}\right) &=&\sum_{\sigma \in
S_{n}/%
\mathbb{Z}
_{n}}\left\langle T^{a_{\sigma (1)}}T^{a_{\sigma (2)}}\ldots T^{a_{\sigma
(n)}}\right\rangle m_{diag}^{\alpha ^{\prime }}\left( \sigma \right)
\label{amplitudes} \\
&=&\left( \prod\limits_{j=1}^{n}\int \mathrm{d}x_{j}e^{\mathrm{i}p_{j}\cdot
x_{j}}\frac{\delta }{\delta \phi ^{a_{j}}\left( x_{j}\right) }\right) 
\mathcal{T}[\phi ]|_{\phi \rightarrow 0}.  \label{amplitudes 1}
\end{eqnarray}%
This formula corresponds algebraically to the contraction of the $S-$ matrix
(\ref{S-matrix}) with the annihilation operators defining the out state.

Using the above definition of $\mathcal{T}[\phi ]$, after some algebra, we
can derive a compact form of the Berends-Giele relations as follows. Let us write
\begin{eqnarray*}
\frac{\delta \mathcal{T}\left[ \phi \right] }{\delta \phi _{a}\left(
x\right) } &=&\mathrm{e}^{-\frac{\mathrm{i}}{\hbar }\mathcal{T}\left[ \phi %
\right] }\exp \left( -\mathrm{i}\frac{\hbar }{2}\frac{\delta }{\delta \phi }%
\cdot \Delta \cdot \frac{\delta }{\delta \phi }\right) \frac{\delta S_{int}%
\left[ \phi \right] }{\delta \phi _{a}\left( x\right) }\exp \left( \frac{%
\mathrm{i}}{\hbar }S_{int}\left[ \phi \right] \right) \\
&=&\mathrm{e}^{-\frac{\mathrm{i}}{\hbar }\mathcal{T}\left[ \phi \right]
}\exp \left( -\mathrm{i}\frac{\hbar }{2}\frac{\delta }{\delta \phi }\cdot
\Delta \cdot \frac{\delta }{\delta \phi }-\mathrm{i}\frac{\hbar }{2}\frac{%
\delta }{\delta \chi }\cdot \Delta \cdot \frac{\delta }{\delta \chi }-%
\mathrm{i}\hbar \frac{\delta }{\delta \chi }\cdot \Delta \cdot \frac{\delta 
}{\delta \phi }\right) \\
&&\left. \times \frac{\delta S_{int}\left[ \chi \right] }{\delta \phi
_{A}\left( x\right) }\exp \left( \frac{\mathrm{i}}{\hbar }S_{int}\left[ \phi %
\right] \right) \right\vert _{\chi =\phi }
\end{eqnarray*}
Because  we will restrict ourselves to tree-level amplitudes in what follows, the tadpoles can be effectively omitted
\[
\exp \left( -\mathrm{i}\frac{\hbar }{2}\frac{\delta }{\delta \chi }\cdot
\Delta \cdot \frac{\delta }{\delta \chi }\right) \frac{\delta S_{int}\left[
\chi \right] }{\delta \phi _{A}\left( x\right) }\rightarrow 0,
\]%
and, as above, we can use
\[
\exp \left( -\mathrm{i}\frac{\hbar }{2}\frac{\delta }{\delta \phi }\cdot
\Delta \cdot \frac{\delta }{\delta \phi }\right) \exp \left( \frac{\mathrm{i}%
}{\hbar }S_{int}\left[ \phi \right] \right) =\mathrm{e}^{\frac{\mathrm{i}}{%
\hbar }\mathcal{T}\left[ \phi \right] }.
\]%
Therefore%
\begin{eqnarray*}
\frac{\delta \mathcal{T}\left[ \phi \right] }{\delta \phi _{a}\left(
x\right) } &=&\left. \mathrm{e}^{-\frac{\mathrm{i}}{\hbar }\mathcal{T}\left[
\phi \right] }\exp \left( -\mathrm{i}\hbar \frac{\delta }{\delta \chi }\cdot
\Delta \cdot \frac{\delta }{\delta \phi }\right) \frac{\delta S_{int}\left[
\chi \right] }{\delta \phi _{a}\left( x\right) }\mathrm{e}^{\frac{\mathrm{i}%
}{\hbar }\mathcal{T}\left[ \phi \right] }\right\vert _{\chi =\phi } \\
&=&\left. \mathrm{e}^{-\frac{\mathrm{i}}{\hbar }\mathcal{T}\left[ \phi %
\right] }\frac{\delta S_{int}}{\delta \phi _{a}\left( x\right) }\left[ \chi -%
\mathrm{i}\hbar \Delta \cdot \frac{\delta }{\delta \phi }\right] \mathrm{e}^{%
\frac{\mathrm{i}}{\hbar }\mathcal{T}\left[ \phi \right] }1\right\vert _{\chi
=\phi } \\
&=&\left. \frac{\delta S_{int}}{\delta \phi _{a}\left( x\right) }\left[ \chi
-\mathrm{i}\hbar \Delta \cdot \left( \frac{\delta }{\delta \phi }+\frac{%
\mathrm{i}}{\hbar }\frac{\delta \mathcal{T}\left[ \phi \right] }{\delta \phi 
}\right) \right] 1\right\vert _{\chi =\phi }.
\end{eqnarray*}%
I.e. finally%
\[
\frac{\delta \mathcal{T}\left[ \phi \right] }{\delta \phi _{a}\left(
x\right) }=\left. \frac{\delta S_{int}}{\delta \phi _{a}\left( x\right) }%
\left[ \chi -\mathrm{i}\hbar \Delta \cdot \left( \frac{\delta }{\delta \phi }%
+\frac{\mathrm{i}}{\hbar }\frac{\delta \mathcal{T}\left[ \phi \right] }{%
\delta \phi }\right) \right] 1\right\vert _{\chi =\phi }
\]%
where we have used the algebraic identity
\[
\exp \left( \psi \cdot \frac{\delta }{\delta \phi }\right) F\left[ \phi %
\right] =F\left[ \phi +\psi \right]
\]%
and%
\[
\mathrm{e}^{-\frac{\mathrm{i}}{\hbar }\mathcal{T}\left[ \phi \right] }\frac{%
\delta }{\delta \phi }\mathrm{e}^{\frac{\mathrm{i}}{\hbar }\mathcal{T}\left[
\phi \right] }=\frac{\delta }{\delta \phi }+\frac{\mathrm{i}}{\hbar }\frac{%
\delta \mathcal{T}\left[ \phi \right] }{\delta \phi }.
\]
Expanding now%
\[
\mathcal{T}\left[ \phi \right] =\mathcal{T}^{\left( 0\right) }\left[ \phi %
\right] +\hbar \mathcal{T}^{\left( 1\right) }\left[ \phi \right] +O\left(
\hbar ^{2}\right) 
\]%
we get at the tree level
\[
\frac{\delta \mathcal{T}^{\left( 0\right) }\left[ \phi \right] }{\delta \phi
_{a}\left( x\right) }=\frac{\delta S_{int}}{\delta \phi _{a}\left( x\right) }%
\left[ J\right] 
\]%
where%
\[
J^{a}=\phi^a +\Delta \cdot \frac{\delta \mathcal{T}^{\left( 0\right) }%
\left[ \phi \right] }{\delta \phi^a }
\]%
For non-derivative interaction 
\[
S_{int}\left[ \phi \right] \equiv \int \mathrm{d}x~\left\langle V\left( \phi
\left( x\right) \right) \right\rangle ,~~~~\phi \equiv T^{a}\phi ^{a}
\]%
we can write \ the above formulas in the index-less matrix form,
\begin{eqnarray}
\frac{\delta \mathcal{T}^{\left( 0\right) }}{\delta \phi } =\frac{\partial
V\left( J\right) }{\partial J} ,\,\,\,\,\,\,\,\,\,\,
J =\phi +\Delta \cdot \frac{\delta \mathcal{T}^{(0)}}{\delta
\phi },  \label{B-G compact}
\end{eqnarray}%
where we denoted%
\begin{eqnarray*}
\frac{\delta \mathcal{T}^{\left( 0\right) }}{\delta \phi } &\equiv &T^{a}%
\frac{\delta \mathcal{T}^{\left( 0\right) }}{\delta \phi ^{a}}, \,\,\,
\frac{\partial V\left( J\right) }{\partial J} \equiv T^{a}\frac{\partial
V\left( J\right) }{\partial J^{a}},\,\,\,
\Delta \cdot \frac{\delta \mathcal{T}^{\left( 0\right) }}{\delta \phi }
\equiv \int \mathrm{d}y\Delta \left( x,y\right) \frac{\delta \mathcal{T}%
^{\left( 0\right) }}{\delta \phi \left( y\right) }.
\end{eqnarray*}
The formula (\ref{B-G compact}) is a compact functional form of the Berends-Giele relations. The traditional form can be obtained from (\ref{B-G compact}) by means of expansion
of both sides in powers of $\phi$, stripping the group factors and then going to momentum
representation according to (\ref{amplitudes}) and (\ref{amplitudes 1}).

\subsection{Tree-level cubic BG, the proof}

Explicitly, excluding $\ J$ and expanding $\partial V\left( J\right)
/\partial J$%
\begin{equation}
\frac{\delta \mathcal{T}^{\left( 0\right) }}{\delta \phi }%
=\sum_{k=1}^{\infty }C_{k-1}\left( \phi +\Delta \cdot \frac{\delta \mathcal{%
T}^{\left( 0\right) }}{\delta \phi }\right) ^{2k}  \label{BG_compact_1}
\end{equation}%

Note that the potential $V\left( J\right) $ for the inverse KLT kernel (\ref%
{potential}) satisfies the following identity
\[
\frac{\partial V\left( J\right) }{\partial J}=J^{2}+\left( \frac{\partial
V\left( J\right) }{\partial J}\right) ^{2}.
\]%
Indeed, we have 
\begin{eqnarray*}
J^{2}+\left( \frac{\partial V\left( J\right) }{\partial J}\right) ^{2}
&=&J^{2}+\sum_{i,j=1}^{\infty }C_{i-1}C_{j-1}J^{2(i+j)} 
=J^{2}+\sum_{n=2}^{\infty }\sum_{k=1}^{n-1}C_{k-1}C_{n-k-1}J^{2n}.
\end{eqnarray*}%
Since the Catalan numbers satisfy the recurrent relations
\[
\sum_{k=1}^{n-1}C_{k-1}C_{n-k-1}=C_{n-1},
\]%
and because 
$
C_{0}=1
$, 
we finally get%
\begin{eqnarray*}
J^{2}+\left( \frac{\partial V\left( J\right) }{\partial J}\right) ^{2}
&=&J^{2}+\sum_{n=2}^{\infty }C_{n-1}J^{2n} 
=\sum_{n=1}^{\infty }C_{n-1}J^{2n}=\frac{\partial V\left( J\right) }{%
\partial J}.
\end{eqnarray*}%
Thus, for $\mathcal{T}_{\mathrm{tree}}\left[ \phi \right] $ we can rewrite
the B-G relations (\ref{BG_compact_1}) equivalently as%
\begin{eqnarray}
\frac{\delta \mathcal{T}^{\left( 0\right) }}{\delta \phi } &=&J^{2}+\left( 
\frac{\partial V\left( J\right) }{\partial J}\right) ^{2},\,\,\,
J =\phi +\Delta \cdot \frac{\delta \mathcal{T}^{\left( 0\right) }}{\delta
\phi },
\end{eqnarray}%
or explicitly, using the definition of $J$ and applying once again the
identity (\ref{B-G compact}) (i.e., excluding $J$ and $\partial V\left(
J\right) /\partial J$), we have a new recursion 
\[
\frac{\delta \mathcal{T}^{\left( 0\right) }}{\delta \phi }=\left( \phi
+\Delta \cdot \frac{\delta \mathcal{T}^{\left( 0\right) }}{\delta \phi }%
\right) ^{2}+\left( \frac{\delta \mathcal{T}^{\left( 0\right) }}{\delta \phi 
}\right) ^{2}
\]%
or,%
\begin{eqnarray*}
\frac{\delta \mathcal{T}^{\left( 0\right) }}{\delta \phi } &=&\phi ^{2}+\phi
~\Delta \cdot \frac{\delta \mathcal{T}^{\left( 0\right) }}{\delta \phi }+%
\frac{\delta \mathcal{T}^{\left( 0\right) }}{\delta \phi }\cdot \Delta ~\phi 
+\frac{\delta \mathcal{T}^{\left( 0\right) }}{\delta \phi }\cdot \left(
1+\Delta \Delta \right) \cdot \frac{\delta \mathcal{T}^{\left( 0\right) }}{%
\delta \phi }
\end{eqnarray*}%
More explicitly\footnote{%
Note that the validity of this recursion relation is independent of the
explicit form of the propagator and is shared with all the theories of the general form
\[
\mathcal{L}_{F}=-\frac{1}{2}\left\langle \phi F\left( \square \right) \phi
\right\rangle +V\left( \phi \right) 
\]
with the same potential (\ref{potential}) and arbitrary inverse propagator $%
F\left( \square \right) $.}%
\begin{eqnarray*}
\frac{\delta \mathcal{T}^{\left( 0\right) }}{\delta \phi \left( x\right) }
&=&\phi \left( x\right) ^{2}+\phi \left( x\right) \int \mathrm{d}y\Delta
\left( x,y\right) \frac{\delta \mathcal{T}^{\left( 0\right) }}{\delta \phi
\left( y\right) }+\int \mathrm{d}y\Delta \left( x,y\right) \frac{\delta 
\mathcal{T}^{\left( 0\right) }}{\delta \phi \left( y\right) }\phi \left(
x\right)  \\
&&+\int \mathrm{d}y\mathrm{d}z\left[ \delta \left( x-y\right) \delta \left(
x-z\right) +\Delta \left( x,y\right) \Delta \left( x,z\right) \right] \frac{%
\delta \mathcal{T}^{\left( 0\right) }}{\delta \phi \left( y\right) }\frac{%
\delta \mathcal{T}^{\left( 0\right) }}{\delta \phi \left( z\right) }.
\end{eqnarray*}%
 Taking out the group factors, expanding in the powers of $\phi$ and translating the result into the momentum
representation, we get the cubic tree-level BG relations (\ref{BGBAS}).

\section{Derivation of the all-loop KLT Lagrangian}\label{app:LKLTderiv}

The recursion relations for the contact terms can be written in a compact form as
\begin{equation}
\kappa _{L,n}=\sum\limits_{l=0}^{L}\sum_{k=1}^{n}\kappa _{l,k}\kappa
_{L-l,n-k+1}+\kappa _{L-1,n+1}\label{kapa_recurence}
\end{equation}%
Then, as a consequence,  (derivative of) the $L-$loop Lagrangian\footnote{For simplicity, we suppress the $U(N)$ group structure and treat $\phi$ as a single scalar. The $U(N)$ case can easily be restored by setting $\phi=\phi^a T^a$ and taking the trace of the resulting formulas for the Lagrangians.} generating the contact
terms  
\begin{equation}
\mathcal{L}_{L}^{\prime }\equiv \sum_{n=1}^{\infty }\kappa _{L,n}\phi ^{n-1}
\label{L_L}
\end{equation}%
satisfies the following recursion
\begin{equation*}
\mathcal{L}_{L}^{\prime }=\sum\limits_{l=0}^{L}\mathcal{L}_{l}^{\prime }%
\mathcal{L}_{L-l}^{\prime }+\frac{1}{\phi }\left( \mathcal{L}_{L-1}^{\prime
}-\kappa _{L-1,1}\right) .
\end{equation*}%
More explicitly, solved with respect to $\mathcal{L}_{L}^{\prime }$%
\begin{eqnarray*}
\mathcal{L}_{L}^{\prime } &=&\frac{1}{1-2\mathcal{L}_{0}^{\prime }}\left[
\sum\limits_{l=1}^{L-1}\mathcal{L}_{l}^{\prime }\mathcal{L}_{L-l}^{\prime }+%
\frac{1}{\phi }\left( \mathcal{L}_{L-1}^{\prime }-\kappa _{L-1,1}\right) %
\right]  \\
&=&\frac{1}{\sqrt{1-4\phi ^{2}}}\left[ \sum\limits_{l=1}^{L-1}\mathcal{L}%
_{l}^{\prime }\mathcal{L}_{L-l}^{\prime }+\frac{1}{\phi }\left( \mathcal{L}%
_{L-1}^{\prime }-\kappa _{L-1,1}\right) \right] 
\end{eqnarray*}%
where we used the explicit form for the (derivative of) original
Lagrangian (cf. (\ref{defKLTLag}))%
\begin{equation*}
\mathcal{L}_{0}^{\prime }=\frac{1}{2}\left( 1-\sqrt{1-4\phi ^{2}}\right) 
\end{equation*}%
Finally%
\begin{equation}
\mathcal{L}_{L}^{\prime }\left( \phi \right) =\frac{1}{\sqrt{1-4\phi ^{2}}}%
\left[ \sum\limits_{l=1}^{L-1}\mathcal{L}_{l}^{\prime }\left( \phi \right) 
\mathcal{L}_{L-l}^{\prime }\left( \phi \right) +\frac{1}{\phi }\left( 
\mathcal{L}_{L-1}^{\prime }\left( \phi \right) -\mathcal{L}_{L-1}^{\prime
}\left( 0\right) \right) \right] \label{L_L_recursion1}
\end{equation}%
This is a self-contained recursion for the $L-$ loop (derivative of) contact term Lagrangians with the
initial condition $\mathcal{L}_{0}^{\prime }$ \ which fixes the functions $%
\mathcal{L}_{L}^{\prime }$ uniquely (note that the right hand side depends
only on the functions $\mathcal{L}_{K}^{\prime }$ with $K<L$). 
Several first iterations give
\begin{align}
\mathcal{L}_{1}^{\prime }\!\left( \phi \right)  &=\frac{1}{\sqrt{1-4\phi ^{2}}%
}\frac{1}{\phi }\mathcal{L}_{0}^{\prime }=\frac{1}{2\phi }\!\left(\! \frac{1}{%
\sqrt{1-4\phi ^{2}}}-1 \!\right)\!, \hspace{0.3cm}
\mathcal{L}_{2}^{\prime }\!\left( \phi \right)  =\frac{1}{\left( 1-4\phi
^{2}\right) ^{3/2}}, \label{L_L_solution} \\
\mathcal{L}_{3}^{\prime }\!\left( \phi \right)  &= 8\phi \frac{1-2\phi ^{2}}{%
\left( 1-4\phi ^{2}\right) ^{5/2}}, \hspace{4.1cm}
\mathcal{L}_{4}^{\prime }\!\left( \phi \right)  = \frac{9-16\phi ^{2}}{%
\left( 1-4\phi ^{2}\right) ^{7/2}}, \nonumber\\
\mathcal{L}_{5}^{\prime }\!\left( \phi \right)  &=16\phi \frac{9-56\phi
^{2}+144\phi ^{4}-144\phi ^{6}}{\left( 1-4\phi ^{2}\right) ^{9/2}}, \hspace{1.6cm}
\mathcal{L}_{6}^{\prime }\!\left(\phi\right)=\frac{162-32\phi ^{2}\left( 27-64\phi ^{2}+64\phi
^{4}\right) }{\left( 1-4\phi ^{2}\right) ^{11/2}}. \nonumber
\end{align}

The generating function $\mathcal{L}^{\prime }$ of the all-loop contact terms
(which is a derivative of the all-loop contact term Lagrangian, without the
tree level contribution included) defined as%
\begin{equation*}
\mathcal{L}^{\prime }=\sum\limits_{L=1}^{\infty }\hbar ^{L}\mathcal{L}%
_{L}^{\prime }
\end{equation*}%
satisfies, according to (\ref{L_L_recursion1}), the following functional equation%
\begin{equation*}
\mathcal{L}^{\prime }\left( \phi \right) =\frac{1}{\sqrt{1-4\phi ^{2}}}\left[
\mathcal{L}^{\prime }\left( \phi \right) ^{2}+\frac{\hbar }{\phi }\left( 
\mathcal{L}^{\prime }\left( \phi \right) +\mathcal{L}_{0}^{\prime }\left(
\phi \right) -\mathcal{L}^{\prime }\left( 0\right) \right) \right]. 
\end{equation*}%
Solution of the latter reads
\begin{equation*}
\mathcal{L}^{\prime }=\frac{1}{2}\left( 1-\frac{\hbar }{\phi }\right) \left[
1-\sqrt{1+4\frac{\frac{\hbar }{\phi }\mathcal{L}^{\prime }\left( 0\right)
-\phi ^{2}}{\left( 1-\frac{\hbar }{\phi }\right) ^{2}}}\right] -\mathcal{L}%
_{0}^{\prime },
\end{equation*}%
which is the formula (\ref{KLTLagL}). The consistency condition, which fixes the sign of
the square root 
\begin{equation*}
\lim_{\phi \rightarrow 0}\mathcal{L}^{\prime }=\mathcal{L}^{\prime }\left(
0\right) 
\end{equation*}%
is then satisfied identically. Therefore, $\mathcal{L}^{\prime }\left(
0\right)$ cannot be fixed from the above solution in a closed form. Nevertheless, parameterizing $\mathcal{L}^{\prime }\left(0\right)$, i.e. writing
\[\mathcal{L}^{\prime }\left(0\right)=\sum_{k=1}^{\infty}\kappa_{2k,1}\hbar^{2k}\]
the parameters $\kappa_{L,1}$ can be fixed step by step from the requirement of absence of poles at $\phi=0$ in $\mathcal{L}^{\prime }_L(\phi)$ calculated from the expansion of $\mathcal{L}^{\prime }$ in powers of $\hbar$.
That is, defining
\[
\mathcal{L}_{L}^{\prime }\left(\phi,\{ \kappa _{l,1}\}\right) =\frac{1}{L!}\left. 
\frac{\partial ^{L}}{\partial \hbar ^{L}}\mathcal{L}^{\prime }\right\vert
_{\hbar =0}
\]
we get e.g.
\begin{eqnarray}
\mathcal{L}_{1}^{\prime }\left(\phi,\{ \kappa _{l,1}\}\right)  &=&\frac{1}{2\phi }\left( \frac{1}{%
\sqrt{1-4\phi ^{2}}}-1\right)  \nonumber\\
\mathcal{L}_{2}^{\prime }\left(\phi,\{ \kappa _{l,1}\}\right)  &=&\frac{1}{\left( 1-4\phi
^{2}\right) ^{3/2}} \nonumber\\
\mathcal{L}_{3}^{\prime }\left(\phi,\{ \kappa _{l,1}\}\right) &=&\frac{1-\left( 1-4\phi ^{2}\right)^2 \kappa _{2,1}%
}{\phi \left( 1-4\phi ^{2}\right) ^{5/2}}=\frac{1-\kappa _{2,1}}{\phi }%
+O(\phi ) \nonumber\\
\mathcal{L}_{4}^{\prime }\left(\phi,\{ \kappa _{l,1}\}\right) &=&\frac{1+\phi ^{2}-\left( 1-4\phi ^{2}\right)^2
\kappa _{2,1}}{\phi ^{2}\left( 1-4\phi ^{2}\right) ^{7/2}}=\frac{1-\kappa
_{2,1}}{\phi ^{2}}+O\left( 1\right) \nonumber \\
\mathcal{L}_{5}^{\prime }\left(\phi,\{ \kappa _{l,1}\}\right) &=&\frac{1-\kappa _{2,1}+\phi ^{2}\left[
3+(6-32\phi ^{4})\kappa _{2,1}-\left( 1-4\phi ^{2}\right) ^{4}\kappa _{4,1}%
\right] }{\phi ^{3}\left( 1-4\phi ^{2}\right) ^{9/2}} \nonumber\\
&=&\frac{1-\kappa _{2,1}}{\phi ^{3}}+\frac{21-12\kappa _{2,1}-\kappa _{4,1}}{%
\phi }+O\left( \phi \right) \nonumber \\
&&\ldots \label{L_L_parametrized}
\end{eqnarray}%
Consistency with (\ref{L_L}) requires the vanishing of the poles, i.e.,%
\[
\kappa _{2,1}=1,~~\kappa _{4,1}=9,\,\,\,\ldots 
\]%
and similarly in higher orders $\hbar ^{L}$, the corresponding iterative calculation gives 
\[
\kappa _{2k,1}=1,\,9,\,162,\,3861,\,107406,\,3296538,\,108314820,\,\,\ldots 
\]%
which coincides with the results of the iteration based on the original recursion relations (\ref{kapa_recurence}).
Inserting these numbers back into (\ref{L_L_parametrized}), we reproduce the solutions (\ref{L_L_solution}) of the recurrence (\ref{L_L_recursion1}).

\section{Feynman diagrams for $m^{\alpha'}_{\xRed{2},1}$ }\label{app:m21Feyn}
Here we list the Feynman diagrams contributing to the two-loop tadpole $m_{\xRed{2},1}^{\alpha'}$ in \eqref{BG2loop1}. With the exception of the two-loop contact term, all diagrams can be derived from the effective one-loop Lagrangian $\mathcal{L}_{\text{KLT}}^{\xRed{(1)}}$ as defined below \eqref{KLTLag1Loop}:
\begin{align}
    \begin{matrix}
         \hspace{0.1cm}\vspace{-0.05cm}\includegraphics[width=1.7cm]{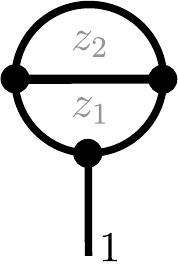}
    \end{matrix} = \frac{1}{t_{1z_1}^2 t_{z_1z_2}t_{1z_2}}, \hspace{0.6cm}
    \begin{matrix}
         \hspace{0.1cm}\vspace{-0.05cm}\includegraphics[width=1.7cm]{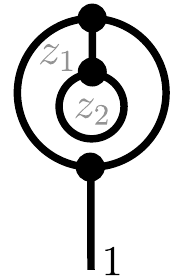}
    \end{matrix} &= \frac{1}{t_{1z_1}^2 t_{z_1z_1}t_{z_1z_2}}, \hspace{0.5cm}
    \begin{matrix}
         \hspace{0.1cm}\vspace{-0.05cm}\includegraphics[width=1.0cm]{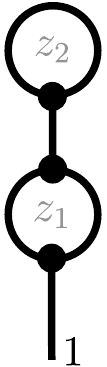}
    \end{matrix} = \frac{1}{t_{1z_1}^2 t_{11}t_{1z_2}},
    \nonumber
    \\
    \begin{matrix}
         \hspace{0.1cm}\vspace{-0.05cm}\includegraphics[width=3.5cm]{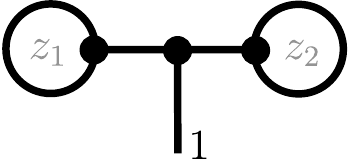}
    \end{matrix} = \frac{1}{t_{1z_1}t_{11}^2t_{1z_2}}, \hspace{0.6cm}
    \begin{matrix}
         \hspace{0.1cm}\vspace{-0.05cm}\includegraphics[width=2.0cm]{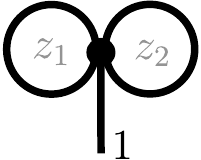}
    \end{matrix} &= \frac{1}{t_{1z_1}t_{1z_2}}, \hspace{0.65cm}
    \begin{matrix}
         \hspace{0.1cm}\vspace{-0.05cm}\includegraphics[width=1.6cm]{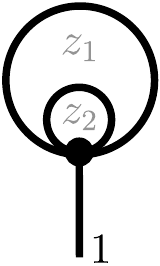}
    \end{matrix} = \frac{1}{t_{1z_1}t_{z_1z_2}},
    \nonumber
    \\
    \begin{matrix}
         \hspace{0.1cm}\vspace{-0.05cm}\includegraphics[width=0.95cm]{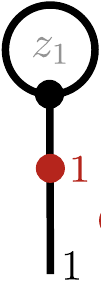}
    \end{matrix} = \frac{1}{t_{1z_1}t_{11}}, \hspace{1.8cm}
    \begin{matrix}
         \hspace{0.1cm}\vspace{-0.05cm}\includegraphics[width=1.55cm]{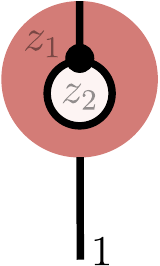}
    \end{matrix} &= \frac{1}{t_{z_1z_1}t_{z_1z_2}}, \hspace{0.9cm}
    \begin{matrix}
         \hspace{0.1cm}\vspace{-0.05cm}\includegraphics[width=1.0cm]{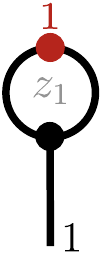}
    \end{matrix} = \frac{1}{t_{1z_1}^2},
    \nonumber
    \\
    \begin{matrix}
         \hspace{0.1cm}\vspace{-0.05cm}\includegraphics[width=0.4cm]{figures/KLTVert2loop1pt.pdf}
    \end{matrix} &= 2.
\end{align}

\newpage

\bibliography{mainbib}

\end{document}